\documentclass[11pt,english,a4paper]{article}
\usepackage{amsfonts, amssymb, a4, setspace, lscape,color}
\usepackage{amsmath}
\usepackage{bbm}
\usepackage{newinutile}
\usepackage{graphicx,calc}
\usepackage{cite}
\usepackage{amsthm}
\usepackage{babel}

\usepackage{mathrsfs}

\newcommand{\rr}[1]{{\textrm{#1}}}
\newcommand{\bb}[1]{{\mathbb{#1}}}
\newtheorem{Remark}{Remark}[section]

\begin{document}

\title{Effects of an impermeable wall in dissipative dynamics of saturated porous media}
\author{P.\ Artale Harris}
\affiliation{Dipartimento di Scienze di Base e Applicate per l'Ingegneria, 
              Sapienza Universit\`a di Roma, 
              via A.\ Scarpa 16, I--00161, Roma, Italy.}
\email{pietro.artale.h@gmail.com}
\maketitle

\begin{abstract}
A phase transition model for porous media in consolidation is studied. The model is able to describe the phenomenon of fluid--segregation during the consolidation process, i.e., the coexistence of two phases
differing from fluid content inside the porous medium under static load. Considering pure Darcy dissipation, the dynamics is described by a Cahn--Hilliard--like system of partial differential equations (PDE). The goal, here, is
to study the dynamics of the formation of stationary fluid--rich bubbles. The evolution of the strain and fluid density profiles of the porous medium is analyzed in two physical situation: fluid free to flow through the boundaries of the medium and fluid flow prevented at one of the two boundaries. Morover, an analytic result on the position of the interface between the two phases is provided.
\end{abstract}

%%%%%%%% Il lavoro
\section{Introduction}
\label{s:introduzione}
In the context of soil consolidation, one can consider the problem of fluid--segregation. Terzaghi's and Biot's theories \cite{Terzaghi} and \cite{biot41, biot55} admit only one sationary state. On the other hand the so--called Mandel--Cryer effect, see \cite{Mandel} and \cite{Cryer63}, is capable to predict fluid--segregation for special geometries of the porous material and for short--time only. 

In the recent literature an extension of the classical Biot theory has been formulated via a nonlinear poromechanical model within the
framework of second gradient theories \cite{CIS2009,CIS2011,CIS2010,CIS2013}. This model is able to describe the occurrence of a second stationary state richer in fluid. Thus, fluid--segregation is described by considering a kind of phase transition.

The model in \cite{CIS2009,CIS2011,CIS2010,CIS2013} is based on the introduction of a quartic energy potential, depending on the strain $\varepsilon$ and the variation of the fluid mass density $m$. This potential is an extension to that of Biot \cite{biot41} and, beside the new fluid--rich phase, it provides the same Biot's standard phase. The dissipation process which characterizes the dynamics of consolidation is described considering a pure Darcy dissipation rather than a pure Stokes dissipation. It is possible to prove that Darcy dissipation implies the behavior of the porous system to be governed by a kind of Cahn--Hilliard set of partial differential equations (PDE),  whilst pure Stokes dissipation yields a kind of Allen--Cahn set of equations \cite{CIS2013}. The effect of the external world on the consolidating porous medium can be coded into the boundary conditions of the PDE problem describing the evolution of the system.  The main mathematical difference between the two PDE cases is that the Allen--Cahn--like equation is a second--order PDE, while the Cahn--Hilliard equation a fourth--order one. Thus, in the latter case two additional boundary conditions have to be prescribed. 

Here we describe the effects of an impermeable wall on the consolidation process. The main appropriate equation in this context is the Cahn--Hilliard one; in fact in this case we can prescribe boundary conditions on the chemical--potential or on its derivative (seepage velocity), while in the Allen--Cahn case, being this choice impossible, the chemical potential is fixed to be zero in the boundaries, thus letting the fluid be free to flow. 

In \cite{ACS2014}, as an extension to the model  \cite{CIS2009,CIS2011,CIS2010,CIS2013}, the authors introduced an impermeable wall in one of the boundaries; in order to obtain the fluid--segregation inside the porous medium, the fluid--poor and the fluid--rich phases at the boundaries of the medium itself were prescribed thanks to Dirichlet boundary conditions. The numerical analysis performed there allowed to learn how the segregation occurs: independently of the presence of the wall, the dynamics is divided in two steps, the formation of the interfaces and the motion towards the stationary profile. 

Once the fluid--segregation dynamics is obtained by a prescription of the boundary values,
a set of more realistic boundary conditions has to be taken into account. In the present work we  discuss the effects of the impermeable wall by choosing a set of Neumann homogenous boundary conditions. We find that in the case of impermeable wall the dynamics ends with the formation of an interface between the
fluid--poor and the fluid--rich phases, while the dynamics without the wall reaches the standard Biot phase. Regarding the study of the
stationary problem, we are able to prove analytically that the position of the interface in this case of Neumann homegenous boundary conditions is the same of the Dirichlet case,
a result firstly discussed in \cite{CIS2012}. For a detailed discussion regarding the more natural situation described by these boundary conditions we remand to Subsection \ref{discussioneneumann}. 

The paper is organized as follows. In Section \ref{s:intro} we resume the model of \cite{CIS2009,CIS2011,CIS2010,CIS2013} and we discuss the physical meaning of the boundary conditions under consideration. In Section \ref{s:stazionario} we discuss numerical simulations of the stationary problem and we prove an analytic result regarding the position of the interface between the two phases. In Section \ref{section:darcyneumann} we discuss the dynamical problem, pointing out the role of the impermeable wall in one of the boundaries of the porous material. Finally, in Section \ref{s:conclusioni} we resume the main results of this work.

\section{The model}
\label{s:intro}

In this Section we resume the model of \cite{CIS2009,CIS2011,CIS2010,CIS2013} with the extension made in \cite{ACS2014}. Morover we present a discussion on the meaning of the boundary conditions taken into account and we apply the model to a special choice of fourth--order overall potential energy.

\subsection{Equations of motion}

We resume the one dimensional poromechanical model introduced in \cite{CIS2013}, where the authors have derived 
the equations of motion by using a
variational approach much similar to that developed in~\cite{sciarra2008}; for more details we remand to \cite{CIS2013}. 

%tienila per dopo
%ùIn \cite{ACS2014} the authors, considering a pure Darcy dissipation mechanism, introduced an impermeable wall in the dynamics of the porous medium.
%
%We shall summarize, here, the main results of 	\cite{ACS2014}. The equations governing 
%the behavior of the porous system are deduced prescribing 
%the conservative part of
%the constitutive law through a suitable potential energy density 
%$\Phi$ and the dissipative 
%contributions through purely Darcy term. 

Let $B_\textrm{s}:=[\ell_1,\ell_2]\subset\mathbb{R}$, with 
$\ell_1,\ell_2\in\mathbb{R}$, 
and $B_\textrm{f}:=\mathbb{R}$ be the \textit{reference}
configurations
for the solid and fluid components, see~\cite{Coussy}. 
The \textit{solid placement} 
$\chi_\textrm{s}:B_\textrm{s}\times\mathbb{R}\to\mathbb{R}$ is a $C^2$ function such that 
the map $\chi_\textrm{s}(\cdot,t)$, 
associating to each $X_\textrm{s}\in B_\textrm{s}$
the position occupied at time $t$ by the particle labeled 
by $X_\textrm{s}$ in the reference configuration $B_\textrm{s}$,
is a $C^2$--diffeomorphism.
The \textit{fluid placement} map 
$\chi_\textrm{f}:B_\textrm{f}\times\mathbb{R}\to\mathbb{R}$
is defined analogously.
The \textit{current configuration} $B_t:=\chi_\textrm{s}(B_\textrm{s},t)$ at time 
$t$ is the set of positions 
of the superposed solid and fluid particles.

Consider the $C^2$ function
$\phi:B_\textrm{s}\times\mathbb{R}\to B_\textrm{f}$ 
such that 
$\phi(X_\textrm{s},t)$ is 
the fluid particle that at time $t$ occupies the 
same position of the solid particle $X_\textrm{s}$; 
assume, also, that $\phi(\cdot,t)$ is a $C^2$--diffeomorphism 
mapping univocally a solid particle 
into a fluid one.
The three fields
$\chi_\textrm{s}$, $\chi_\textrm{f}$, and $\phi$ are 
not at all independent\footnote{indeed, by definition, we immediately have that 
$\chi_\textrm{f}(\phi(X_\textrm{s},t),t)=\chi_\textrm{s}(X_\textrm{s},t)$ 
for any $X_\textrm{s}\in B_\textrm{s}$ and $t\in\mathbb{R}$}. 

The \textit{Lagrangian velocities} are the two maps
$u_\alpha:B_\alpha\times\mathbb{R}\to\mathbb{R}$
defined by setting
%\begin{equation}
%\label{vel-lagr}
$u_\alpha(X_\alpha,t):=\partial\chi_\alpha/\partial t$
%\end{equation}
for any $X_\alpha\in B_\alpha$, where $\alpha=\textrm{s},\textrm{f}$.
We also consider the \textit{Eulerian velocities}
$v_\alpha:B_t\times\mathbb{R}\to\mathbb{R}$ associating
with each point $x\in B_t$ and for each time $t\in\mathbb{R}$ the velocities
of the solid and fluid particle occupying the place $x$ at time $t$;
more precisely we set
$v_\alpha(x,t):=u_\alpha(\chi^{-1}_\alpha(x,t),t)$.

Since the reference configuration $B_\textrm{s}$ of the solid component 
is known {\em a priori}, we express the dynamical observables 
in terms of the fields $\chi_\textrm{s}$ and $\phi$ which are defined 
on $B_\textrm{s}$.

Assume that the 
fluid component of the system is acted upon by dissipative
forces. 
We consider the independent variations 
$\delta\chi_\textrm{s}$ 
and $\delta\phi$ of the two fields $\chi_\textrm{s}$ and $\phi$ and
denote by $\delta W$ the corresponding elementary 
\textit{virtual work}
made by the dissipative forces acting on the fluid component.

The possible motions of the system, see for instance 
\cite[Chapter~5]{Blanchard},
in an interval of time $(t_1,t_2)\subset\mathbb{R}$
are those such that the fields $\chi_\textrm{s}$ and $\phi$ satisfies
the variational principle
\begin{equation}
\label{vardiss}
\delta 
 \int_{t_1}^{t_2}\textrm{d}t
 \int_{B_\textrm{s}}\textrm{d}X_\textrm{s}
 \,
 {L}(
            \dot\chi_\textrm{s}(X_\textrm{s},t),
            \dots,
            \phi(X_\textrm{s},t))
=
-\int_{t_1}^{t_2}\delta W\,\textrm{d}t,
\end{equation}
where the \textit{Lagrangian density} $L$ is equal to the minus overall potential energy associated to both the internal and external conservative forces; here we do not consider inertial effects. Thus, the variation of the the action integral in 
correspondence of a possible motion is equal to the 
integral over time of minus the virtual work of the dissipative 
forces corresponding to the considered variation of the fields. 
%The way in which dissipation has to be 
%introduced in saturated porous media 
%models is still under debate.
%In particular, according to the effectiveness of the hypothesis 
%of separation of scales, between the local and macroscopic level,
%Darcy's or Stokes' effects are accounted for. We refer 
%the interested reader to \cite{CIS2013} for a detailed discussion of this 
%issue. 

In this paper we shall 
consider the so--called Darcy effect, i.e., the dissipation 
due to forces proportional to the velocity of the 
fluid component measured with respect to the solid, \cite{darcy}.

A natural expression for the virtual work of the dissipation forces 
acting on the fluid component and taking into account the Darcy 
effect is 
\begin{equation}
\label{lv01}
 \delta W:=
 -
 \int_{B_t}
 D[v_\rr{f}(x,t)-v_\rr{s}(x,t)]
 [\delta\chi_\rr{f}(\chi^{-1}_\rr{f}(x,t),t)
 -
 \delta\chi_\rr{s}(\chi^{-1}_\rr{s}(x,t),t)]
\,\rr{d}x
\end{equation}
where $\delta\chi_\rr{f}$ is the variation of the field $\chi_\rr{f}$
induced by the independent variations $\delta\chi_\rr{s}$ and 
$\delta\phi$. $D>0$ is the inverse of the permeability $k$ of the porous medium. For more details on equation \eqref{lv01}, we refer to \cite{CIS2013} and \cite{ACS2014}.

It is reasonable to assume that the potential energy density 
depends on the space and time variable only via two 
physically relevant functions: 
the strain of the solid and a properly normalized fluid mass 
density~\cite{CIS2013}, i.e.,
\begin{equation}
\label{deformazione}
\varepsilon(X_\textrm{s},t):=[(\chi'_\textrm{s}(X_\textrm{s},t))^2-1]/2
\;\;\textrm{ and }\;\;
m_\textrm{f}(X_\textrm{s},t)
 :=\varrho_{0,\textrm{f}}(\phi(X_\textrm{s},t))
   \phi'(X_\textrm{s},t)
\end{equation}
where $\varrho_{0,\textrm{f}}:B_\textrm{f}\to\mathbb{R}$ 
is a fluid reference \textit{density}.
In other words,  we assume that 
the potential energy density $\Phi$ is a function of the 
fields 
$m_\textrm{f}$ and $\varepsilon$ and on their space
derivative
$m'_\textrm{f}$ and $\varepsilon'$.

By a standard variational computation, see \cite[equation (24)]{CIS2013},
one gets the equation of motion. In this framework, we are interested in the 
geometrically linearized version of such equations: 
we assume $\varrho_{0,\textrm{f}}$ to be constant and 
introduce the \textit{displacement fields} 
$u(X_\textrm{s},t)$ and $w(X_\textrm{s},t)$ by setting
\begin{equation}
\label{gd00}
\chi_\textrm{s}(X_\textrm{s},t)=X_\textrm{s}+u(X_\textrm{s},t)
\;\textrm{ and }\;
\phi(X_\textrm{s},t)=X_\textrm{s}+w(X_\textrm{s},t)
\end{equation}
for any $X_\textrm{s}\in B_\textrm{s}$ and $t\in\mathbb{R}$. We then 
assume that $u$ and $w$ are small, together with 
their space and time derivatives, and write 
\begin{equation}
\label{gd01}
m_\textrm{f}=\varrho_{0,\textrm{f}}(1+w'),\;
m:=m_\textrm{f}-\varrho_{0,\textrm{f}}=\varrho_{0,\textrm{f}}w',\;
\varepsilon\approx u',\;
\end{equation}
where $\approx$ means that all the terms of order larger than one 
have been neglected.

We have introduced above the field $m$. In the following we shall imagine 
$\Phi$ as a function of $m,\,\varepsilon$ and $m'\,\varepsilon'$ and the equations of motion and 
the boundary conditions will be written in terms of these fields. 
We get the equations of motion \cite{ACS2014}
\begin{equation}
\label{gd02}
 \frac{\partial\Phi}{\partial\varepsilon}
 -
 \Big(
 \frac{\partial\Phi}{\partial\varepsilon'}
 \Big)'
 =0
\;\;\textrm{ and }\;\;
 \varrho_{0,\rr{f}}^2
 \Big[
      \frac{\partial\Phi}{\partial m}
      -
      \Big(
      \frac{\partial\Phi}{\partial m'}
      \Big)'
 \Big]''
 =
 D\dot{m},
\end{equation}
and the associated boundary conditions that are compatible with the choices of
Dirichlet and Neumann boundary conditions:	
\begin{equation}
\label{gd02bc}
 \Big\{\left(
 \frac{\partial\Phi}{\partial\varepsilon'}
 \delta\varepsilon
 +
 \frac{\partial\Phi}{\partial m'}
 \delta m\right)
 +
 \Big[
 \Big(\frac{\partial\Phi}{\partial m}
      -
      \Big(
      \frac{\partial\Phi}{\partial m'}
      \Big)'
 \Big)
 \varrho_{0,\rr{f}}
 \Big]
 \delta w
 \Big\}_{\ell_1}^{\ell_2}
 \!\!\!\!
 =0
\end{equation}

Recalling that in our approximation $m=\varrho_{0,\textrm{f}}w'$, see 
the second among equations (\ref{gd01}), we have that 
the equations (\ref{gd02}) are evolution equations for the 
fields $m$ and $\varepsilon$.

The second between the equations of motion (\ref{gd02}), thanks to a suitable choice of $\Phi$, see Section \ref{sec.phi}, will become
is a Cahn--Hilliard--like equation for the field $m$ 
with driving field still depending 
parametrically on $\varepsilon$ \cite{CIS2013}.

\subsection{The Zero chemical potential problem}
\label{ss:zcp}

A set of boundary conditions implying that (\ref{gd02bc}) 
are satisfied is
\begin{equation}
\label{gd03bc}
\Big(
 \frac{\partial\Phi}{\partial\varepsilon'}
 \delta\varepsilon
 +
 \frac{\partial\Phi}{\partial m'}
 \delta m
\Big)_{\ell_1,\ell_2}
 \!\!\!\!
=
 \Big[\frac{\partial\Phi}{\partial m}
      -
      \Big(
      \frac{\partial\Phi}{\partial m'}
      \Big)'
 \Big]_{\ell_1,\ell_2}
 \!\!\!\!
=0
\end{equation}
where the notation above means that the functions in brackets are 
evaluated both in $\ell_1$ and $\ell_2$.
With this choice it is possible to fix the boundary conditions 
directly on fields $m$ and $\varepsilon$ (and derivatives).

The first equation (\ref{gd03bc}) is the additional 
boundary condition due to the presence of the gradient terms in the 
potential energy density $\Phi$. This equation specifies essential boundary 
conditions on the derivatives of the displacement fields or natural 
boundary conditions on the so called double forces, see \cite{germain73} and the next Subsection \ref{discussioneneumann}.
The generalized essential boundary conditions can be read as a 
prescription on the derivative of the independent fields $\chi_\rr{s}$ 
and $\phi$, see equations (\ref{deformazione}); whilst 
the extended natural boundary conditions prescribe, on one hand, 
the additional forces which the solid continuum is able to balance at
the boundary and, on the other, the wetting properties of the fluid
which fills the pores \cite{Seppecher89}.

The second equation (\ref{gd03bc}) provide 
natural boundary conditions prescribing 
the chemical potential of the fluid, so that the fluid is free to flow trough both the two boundaries. For a more detailed discussion on this boundary condition see \cite{ACS2014} and \cite{baek}. 

Finally, we get the PDE problem
\begin{equation}
\label{problema-d}
\left\{
\begin{array}{l}
{\displaystyle
 \frac{\partial\Phi}{\partial\varepsilon}
 -
 \Big(
 \frac{\partial\Phi}{\partial\varepsilon'}
 \Big)'
 =0
\;\;\textrm{ and }\;\;
 \varrho_{0,\rr{f}}^2
 \Big[
      \frac{\partial\Phi}{\partial m}
      -
      \Big(
      \frac{\partial\Phi}{\partial m'}
      \Big)'
 \Big]''
 =
 D\dot{m}
\vphantom{\bigg\{_\big\{}
}
\\
{\displaystyle
\Big(
 \frac{\partial\Phi}{\partial\varepsilon'}
 \delta\varepsilon
 +
 \frac{\partial\Phi}{\partial m'}
 \delta m
\Big)_{\ell_1,\ell_2}
=
 \Big[\frac{\partial\Phi}{\partial m}
      -
      \Big(
      \frac{\partial\Phi}{\partial m'}
      \Big)'
 \Big]_{\ell_1,\ell_2}
=0
}
\\
\end{array}
\right.
\end{equation}
that will be called the \textit{zero chemical potential} problem.

\subsection{The one--side impermeable problem}
\label{s:osi}
A  very interesting situation in 
applications is the one in which one of the two 
boundaries is not permeable to the fluid. 

This physical situation can be mathematically implemented assuming that, at any time
$t$, the seepage velocity, 
say $v=u_{\rr f}-u_{\rr s}=-\dot w$, to identically vanish and therefore the boundary to be
impermeable to the fluid flow, see \cite{ACS2014}. Considering the bulk equation (\ref{gd02})$_2$ 
this boundary condition can therefore be rephrased as follows:
\begin{equation}
\label{vel000}
v=-\frac{\varrho_{0,\rr{f}}}{D}\Big(
     \frac{\partial\Phi}{\partial m}
     -
     \Big(\frac{\partial\Phi}{\partial m'}\Big)'
\Big)'=0
\end{equation}

% Moreover, by assuming that the quantity (\ref{vel000}) is 
% equal to zero at one of the boundaries of the system, by the 
% second of (\ref{gd02}), we implicitely assume that the field $w$ 
% is contant in time at such a boundary. In other words we 
% assume that the boundary condition $\delta w=0$ is satisfied at 
% this boundary. 
Thus, it follows that a set of boundary conditions 
implying that (\ref{gd02bc}) are satisfied is
\begin{equation}
\label{vel010}
\begin{array}{rcl}
{\displaystyle
 \Big(
  \frac{\partial\Phi}{\partial\varepsilon'}
  \delta\varepsilon
  +
  \frac{\partial\Phi}{\partial m'}
  \delta m
 \Big)_{\ell_1,\ell_2}
  \!\!\!\!
}
&\!\!=&\!\!
{\displaystyle
 \Big[\Big(\frac{\partial\Phi}{\partial m}
      -
      \Big(
      \frac{\partial\Phi}{\partial m'}
      \Big)'
      \Big)'
 \Big]_{\ell_2}
 \!\!\!\!
 \vphantom{\bigg\{_\big\}}
}
\!\!=\!\!
{\displaystyle
 \Big[\frac{\partial\Phi}{\partial m}
      -
      \Big(
      \frac{\partial\Phi}{\partial m'}
      \Big)'
 \Big]_{\ell_1}
 \!\!\!\!
=0
}\\
\end{array}
\end{equation}

Finally we get the PDE problem
\begin{equation}
\label{problema-d-osi}
\left\{
\begin{array}{l}
{\displaystyle
 \frac{\partial\Phi}{\partial\varepsilon}
 -
 \Big(
 \frac{\partial\Phi}{\partial\varepsilon'}
 \Big)'
 =0
\;\;\textrm{ and }\;\;
 \varrho_{0,\rr{f}}^2
 \Big[
      \frac{\partial\Phi}{\partial m}
      -
      \Big(
      \frac{\partial\Phi}{\partial m'}
      \Big)'
 \Big]''
 =
 D\dot{m}
\vphantom{\bigg\{_\big\{}
}
\\
{\displaystyle
\Big(
 \frac{\partial\Phi}{\partial\varepsilon'}
 \delta\varepsilon
 +
 \frac{\partial\Phi}{\partial m'}
 \delta m
\Big)_{\ell_1,\ell_2}
\!\!
\!\!
\!
=
 \Big[\Big(\frac{\partial\Phi}{\partial m}
      -
      \Big(
      \frac{\partial\Phi}{\partial m'}
      \Big)'
      \Big)'
 \Big]_{\ell_2}
\!\!
\!\!
\!
=
 \Big[\frac{\partial\Phi}{\partial m}
      -
      \Big(
      \frac{\partial\Phi}{\partial m'}
      \Big)'
 \Big]_{\ell_1}
\!\!
\!\!
\!
=0
}
\\
\end{array}
\right.
\end{equation}
that will be called the \textit{one--side impermeable} problem.

\subsection{Neumann boundary conditions}
\label{discussioneneumann}

In order to explain the subsequent choice of homogeneous Neumann boundary conditions (see Sections \ref{s:stazionario} and \ref{section:darcyneumann}), we briefly resume the main concepts of second--gradient theories in the one--dimensional case. The external working at the boundaries for a biphasic second gradient porous medium is 	\cite{sciarra2007,sciarra2008}
\begin{eqnarray}
&&\mathcal{P}^{\mbox{ext}}(v_s,v_f)=\left[t_s v_s+t_f(v_f-v_s)+\tau v'_s+\tau_f(v'_f-v'_s)\right]_{\ell_1}^{\ell_2}
\label{9sciarra2008}
\end{eqnarray}
Here $t_\alpha$ are the tractions and $\tau_\alpha$ are the double forces, $\alpha=s,f$. According to \cite{germain73} and \cite{dellisolaseppecher97}, in the one dimensional case, the double force $\tau_\alpha$ can be regarded as the rate of dilatancy.

Recall now the first equation \eqref{gd02bc}; this condition can be satisfied in two cases, \textit{i.e.} by setting $\delta\varepsilon=\delta m=0$ or
$
\partial\Phi/\partial\varepsilon'=\partial\Phi/\partial m'=0$. The first choice is that of \cite{CIS2013,ACS2014} and it was obtained thanks to Dirichlet boundary conditions, while the second option can be obtained by choosing Neumann homogenous boundary conditions $\varepsilon'(\ell_1)=\varepsilon'(\ell_2)=m'(\ell_1)=m'(\ell_2)=0$, thus setting the double forces equal to zero in both the two boundaries. The last one is the choice of the present paper, so that here we do not allow dilatancy  at the boundaries of the porous material. 

Neumann boundary conditions in this context are interesting not only for a more realistic scenario with respect to Dirichlet boundary conditions, but also because they introduce both analytic and numerical complications. In fact, in this case the solution is not unique, so that a various set of stationary solutions has to be studied, as we will see in Section \ref{s:stazionario}; regarding this point, an analytic result about the position of the interface (one of the stationary solutions) between the two phases (poor and rich in fluid content) has been proven in Subsection \ref{sss:kink1}.

\subsection{Overall potential energy}
\label{sec.phi}
We specialize the Porous Medium model we are studying by choosing the 
second gradient part of the dimensionless potential energy, that is we assume 
\begin{equation}
\label{sec010}
\Phi(m',\varepsilon',m,\varepsilon)
 :=
 \frac{1}{2}[k_1(\varepsilon')^2+2k_2\varepsilon' m'+k_3(m')^2]
 +
 \Psi(m,\varepsilon),
\end{equation}
with $k_1,k_3>0$, $k_2\in\bb{R}$ such that $k_1k_3-k_2^2\ge0$.
These parameters provide energy penalties for 
the formation of interfaces; they have the physical dimensions of
squared lengths and, according with the above mentioned 
conditions, provide a well--grounded identification of the intrinsic
characteristic lengths of the one--dimensional porous continuum.

Moreover, 
we consider the following expression for the total potential 
energy density in the perspective of describing the transition between
a fluid--poor and a fluid--rich phase 
\begin{equation}
\label{sec015}
\Psi(m,\varepsilon)
 :=
 \frac{\alpha}{12}m^2(3m^2\!-8b\varepsilon m+6b^2\varepsilon^2)
 +
 \Psi_\rr{B}(m,\varepsilon)
\end{equation}
where 
\begin{equation}
\label{sec020}
\Psi_\rr{B}(m,\varepsilon):=
 p\varepsilon+\frac{1}{2}\varepsilon^2+\frac{1}{2}a(m-b\varepsilon)^2
\end{equation}
is the Biot potential energy density~\cite{biot41},
$a>0$ is the ratio between the fluid and the solid rigidity, 
$b>0$ is a coupling between the fluid and the solid component, 
$p>0$ is the external pressure,
and
$\alpha>0$ is a material parameter responsible for the showing 
up of an additional equilibrium.

The choice \eqref{sec010} is due to the fact that we want to study diffusive interfaces, so that a second--gradient model is needed. For more details on this particular choice and second--gradient theories we remand to \cite{CIS2013} and \cite{sciarra2007,sciarra2008}.

\section{The stationary problem}
\label{s:stazionario}
In the above Section we have introduced two different physically 
interesting problems for the porous medium we are studying by 
specifying two different sets of boundary conditions. 
The stationary solution of those two problems are the solution of 
the two differential equations
\begin{equation}
\label{stazionario000}
 \frac{\partial\Phi}{\partial\varepsilon}
 -
 \Big(
 \frac{\partial\Phi}{\partial\varepsilon'}
 \Big)'
 =0
\;\;\textrm{ and }\;\;
 \Big[
      \frac{\partial\Phi}{\partial m}
      -
      \Big(
      \frac{\partial\Phi}{\partial m'}
      \Big)'
 \Big]''
 =
0
\end{equation}
with the boundary conditions provided respectiveley in (\ref{problema-d}) 
and (\ref{problema-d-osi}) for the zero chemical potential and 
the one--side impermeable problems.

It is easy to show that in both cases the boundary conditions 
imply that, see also \cite[equation~(40) and related discussion]{CIS2013},
$\partial\Phi/\partial m-(\partial\Phi/\partial m')'=0$, so that 
for both the problems the stationary solutions are the 
solutions of the PDE problem
\begin{equation}
\label{stazionario020} 
\left\{
\begin{array}{l}
{\displaystyle
 \frac{\partial\Phi}{\partial\varepsilon}
 -
 \Big(
 \frac{\partial\Phi}{\partial\varepsilon'}
 \Big)'
 =0
\;\;\textrm{ and }\;\;
 \Big[
      \frac{\partial\Phi}{\partial m}
      -
      \Big(
      \frac{\partial\Phi}{\partial m'}
      \Big)'
 \Big]''
 =
0
\vphantom{\bigg\{_\big\{}
}
\\
{\displaystyle
\Big(
 \frac{\partial\Phi}{\partial\varepsilon'}
 \delta\varepsilon
 +
 \frac{\partial\Phi}{\partial m'}
 \delta m
\Big)_{\ell_1,\ell_2}
=0
}
\\
\end{array}
\right.
\end{equation}

\subsection{Constant and interfaces stationary solutions}
\label{sottosezione}
Here we discuss the stationary Problem \eqref{stazionario020}, with the choiche \eqref{sec010}, \eqref{sec015} and  with homegenous Neumann natural boundary conditions, so that system \eqref{stazionario020} becomes
\begin{equation}
\label{problema-staz01}
\left\{
\begin{array}{l}
-(2/3)\alpha b m^3+\alpha b^2m^2\varepsilon+p+\varepsilon-ab(m-b\varepsilon)
-k_1\varepsilon''-k_2m''=0,
\\
 \alpha m^3-2\alpha b m^2 \varepsilon+\alpha b^2m\varepsilon^2
 +a(m-b\varepsilon)
 -k_2\varepsilon''-k_3m'' 
=0,
\\
{\displaystyle
\varepsilon'(\ell_1)=m'(\ell_1)=\varepsilon'(\ell_2)=m'(\ell_2)
=0.
}
\\
\end{array}
\right.
\end{equation}
The problem with these boundary conditions is interesting because here we do not force the formation of the interface by prescribing the two phases at the boundaries as in the Dirichlet boundary conditions case discussed in \cite{ACS2014}. Equation \eqref{problema-staz01} has not a unique solution, and the interface is one of them. The reason of this choiche, that is a physically natural choice (see Subsection \ref{discussioneneumann}), is that we want to understand (in particular in the dynamic case) if and how the fluid bubble inside the porous medium is formed. 
\begin{figure}[h!]
\vskip-0.13cm
\begin{picture}(200,400)(15,0)
\put(30,300)
{
\resizebox{5.cm}{!}{\rotatebox{0}{\includegraphics{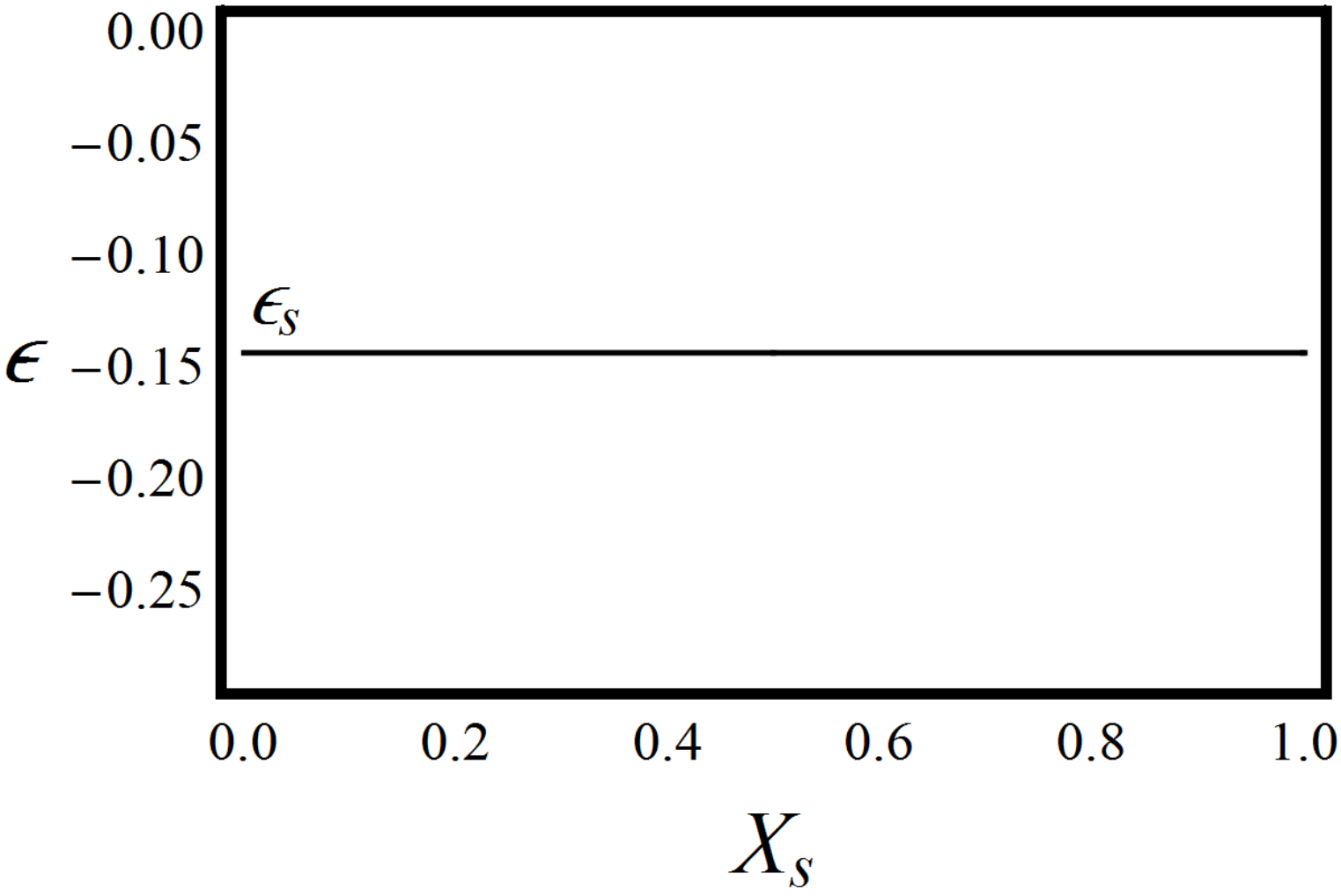}}}
}
\put(245,300)
{
\resizebox{5.cm}{!}{\rotatebox{0}{\includegraphics{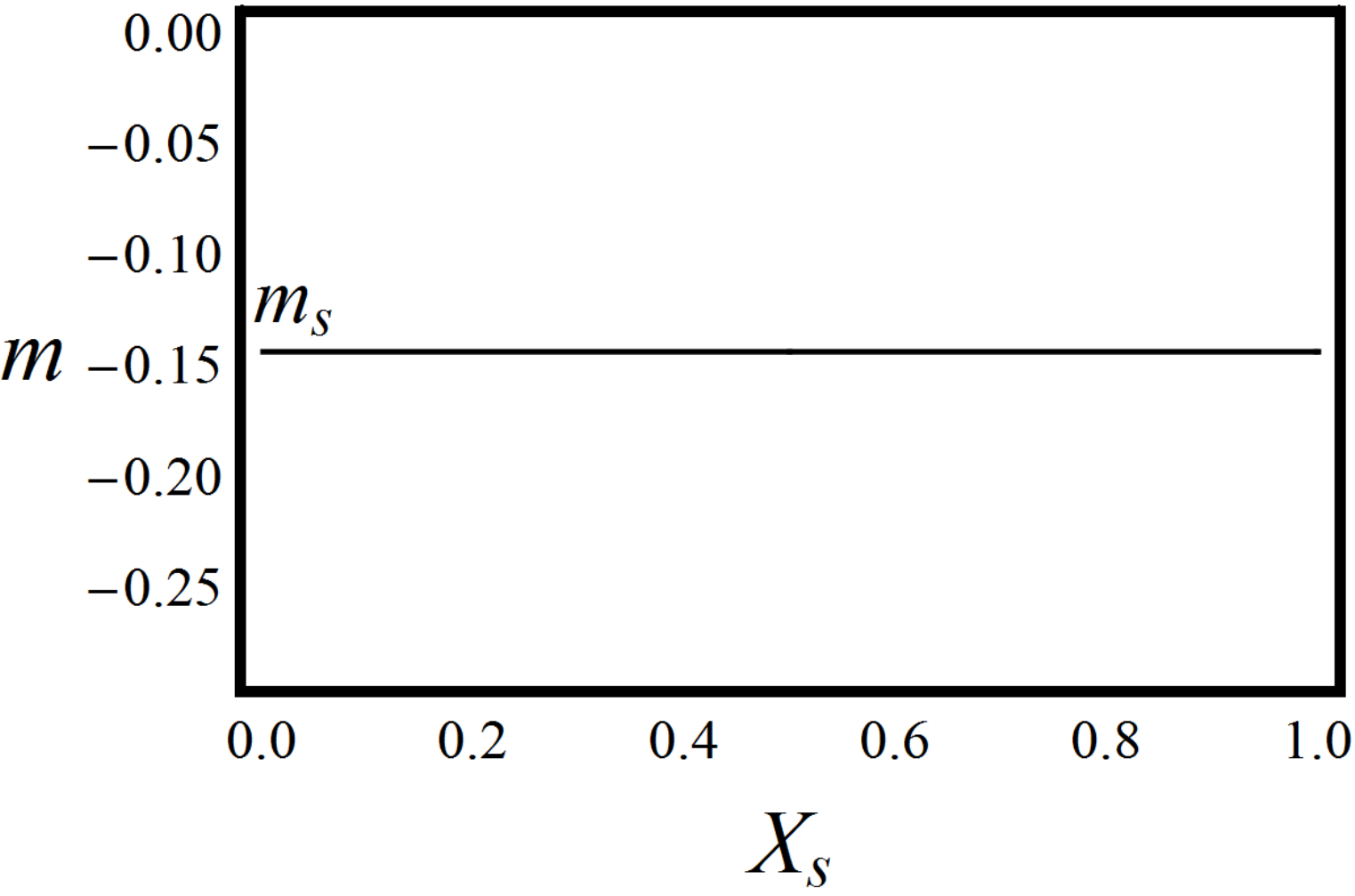}}}
}
\put(30,200)
{
\resizebox{5.cm}{!}{\rotatebox{0}{\includegraphics{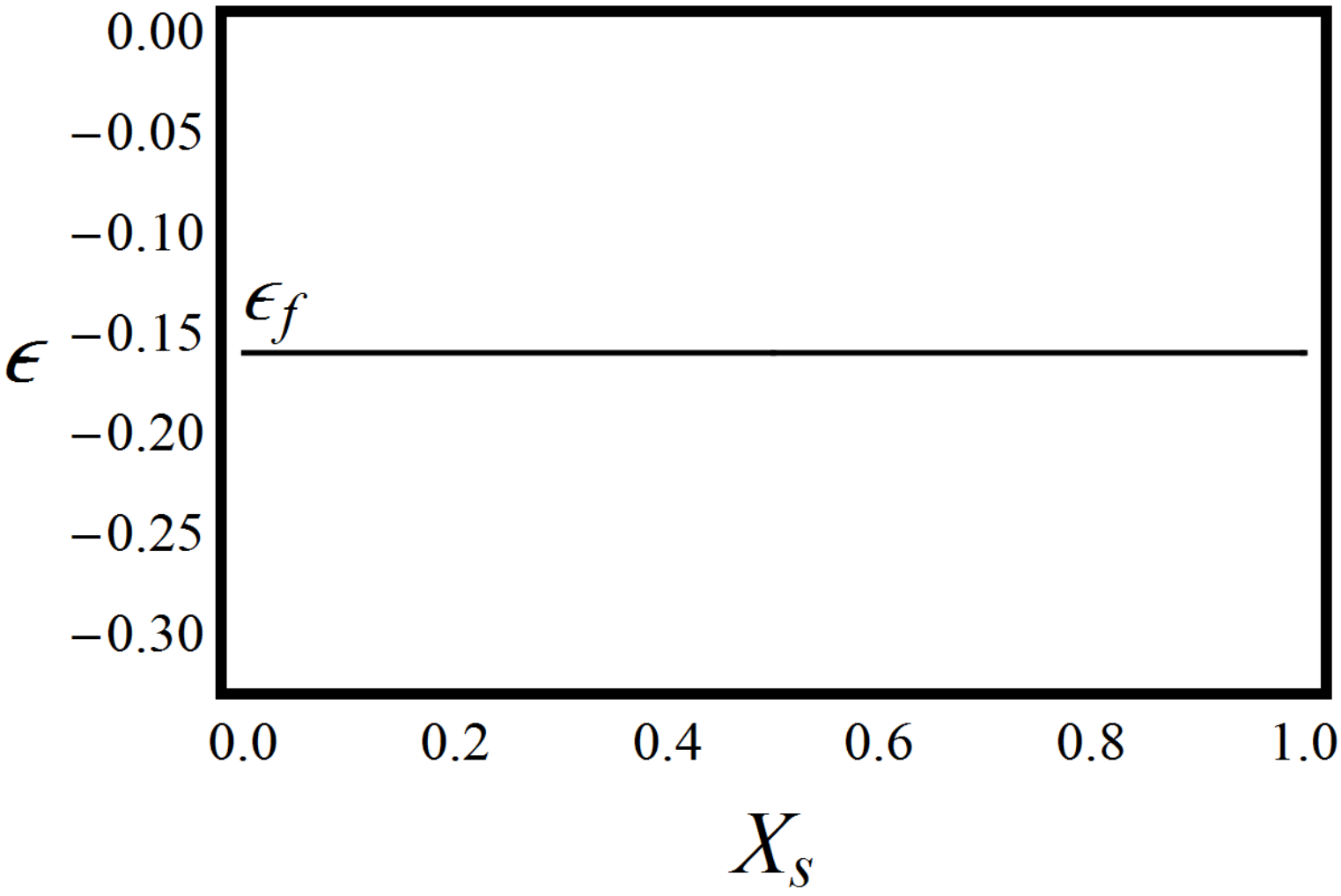}}}
}
\put(245,200)
{
\resizebox{5.cm}{!}{\rotatebox{0}{\includegraphics{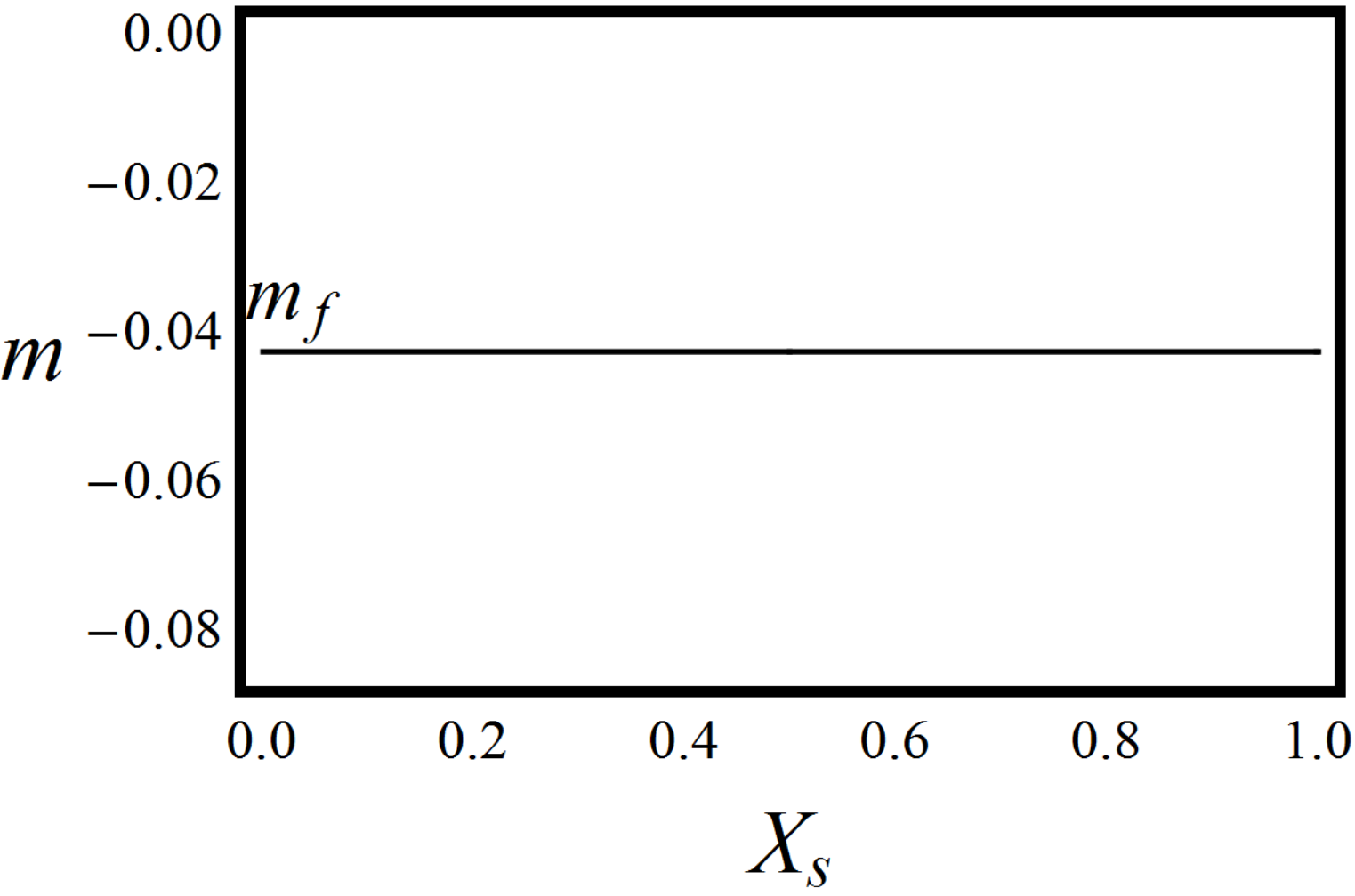}}}
}

\put(30,100)
{
\resizebox{5.cm}{!}{\rotatebox{0}{\includegraphics{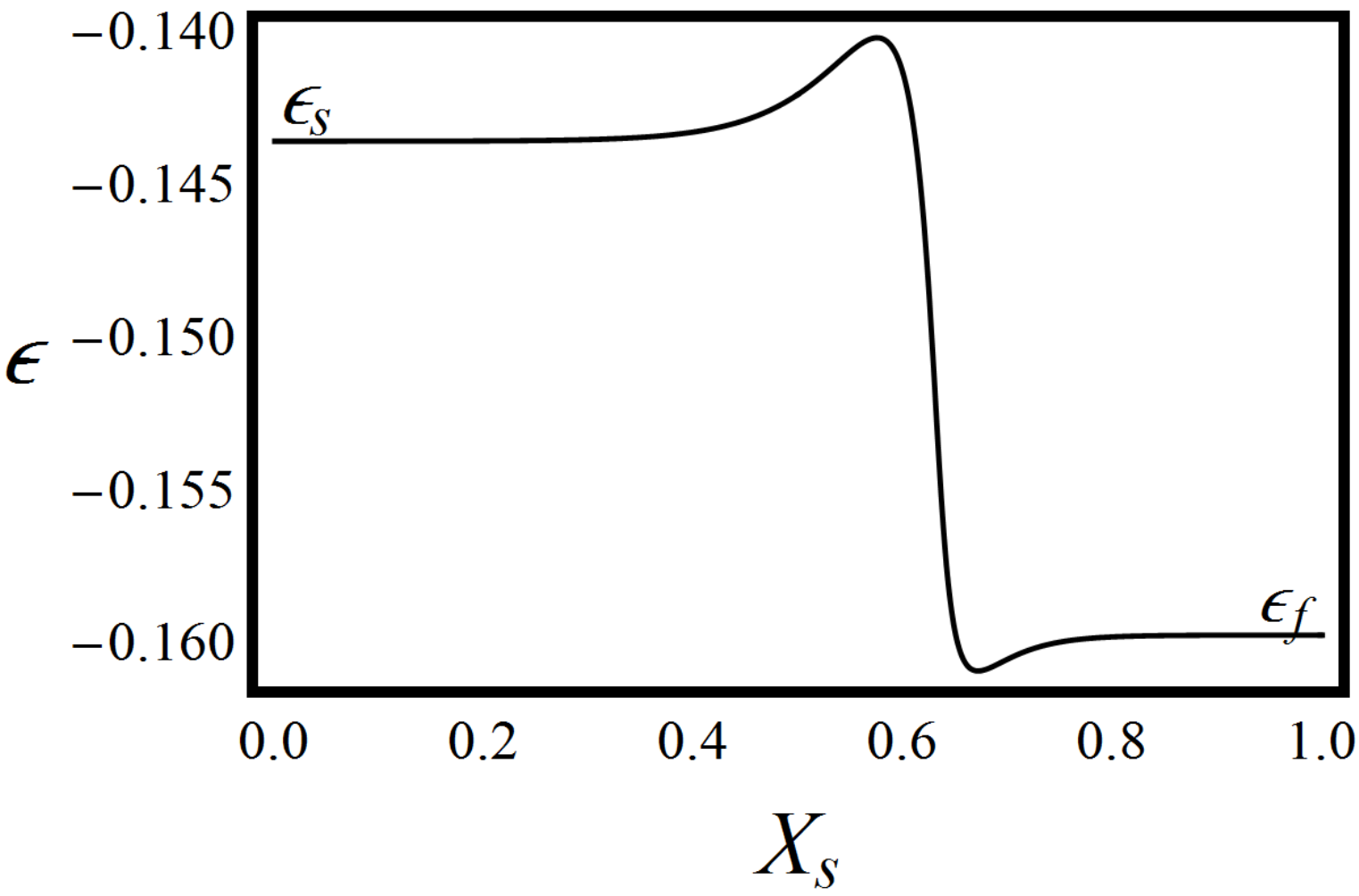}}}
}
\put(245,100)
{
\resizebox{5.cm}{!}{\rotatebox{0}{\includegraphics{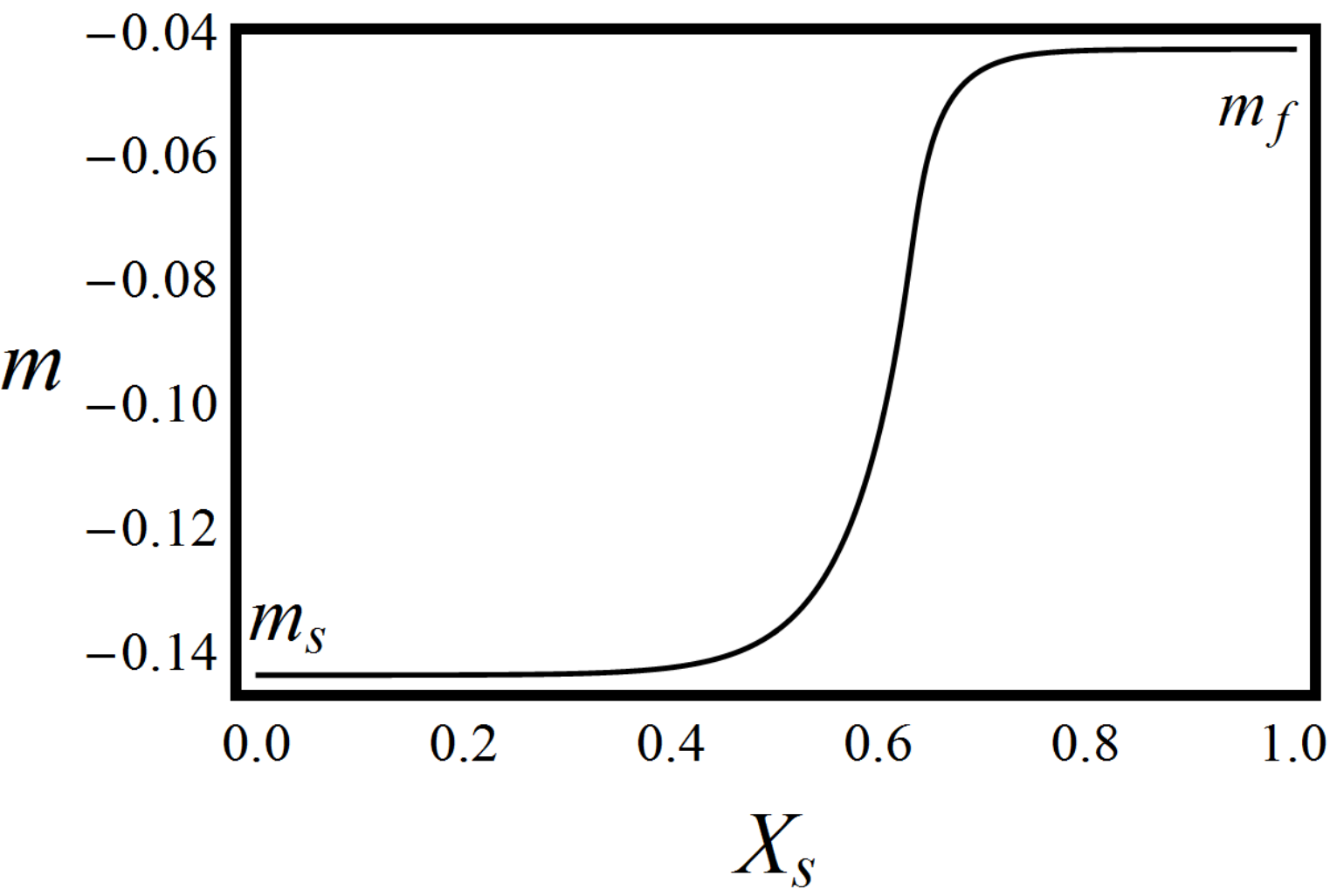}}}
}
\put(30,0)
{
\resizebox{5.cm}{!}{\rotatebox{0}{\includegraphics{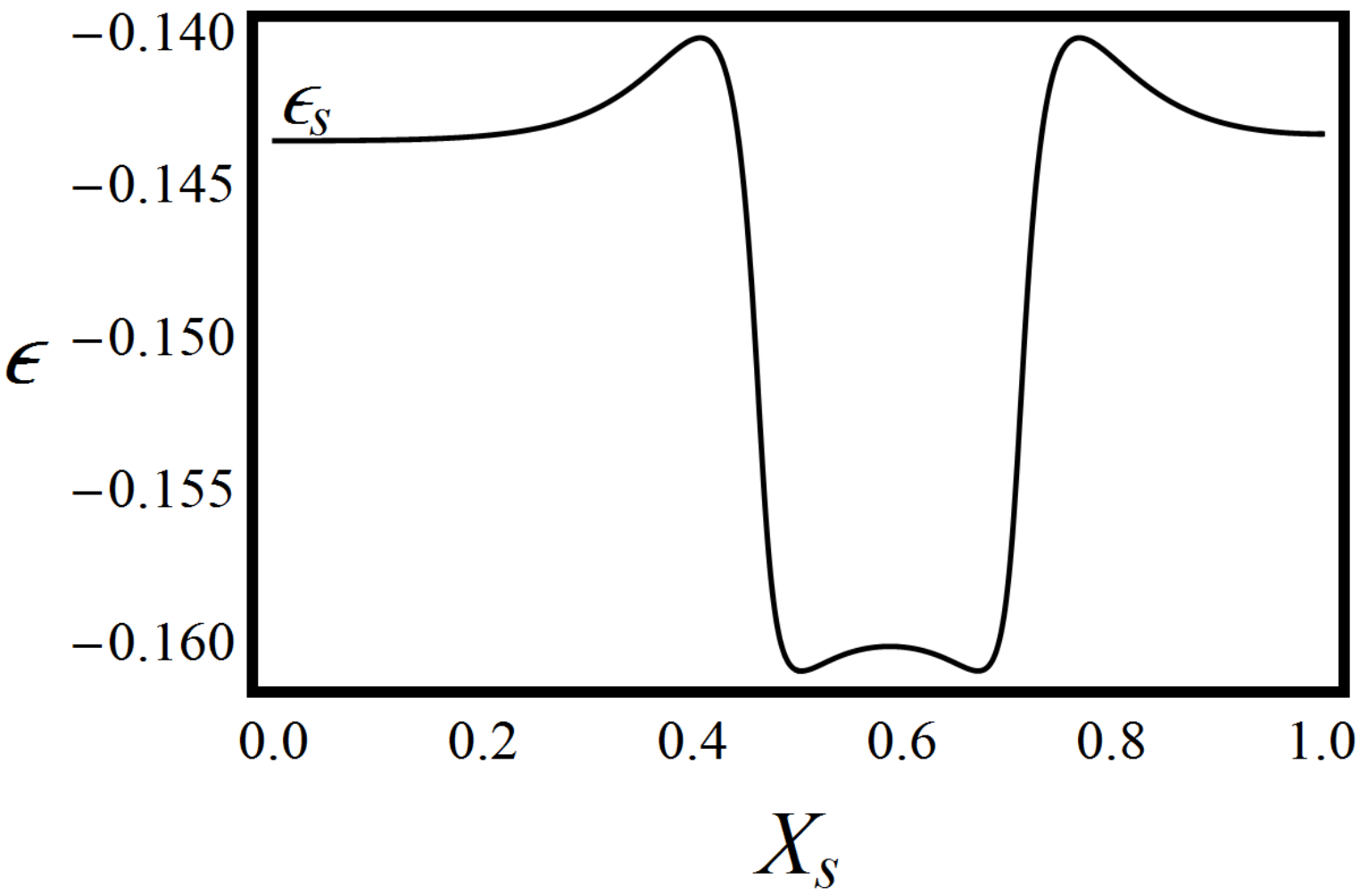}}}
}
\put(245,0)
{
\resizebox{5.cm}{!}{\rotatebox{0}{\includegraphics{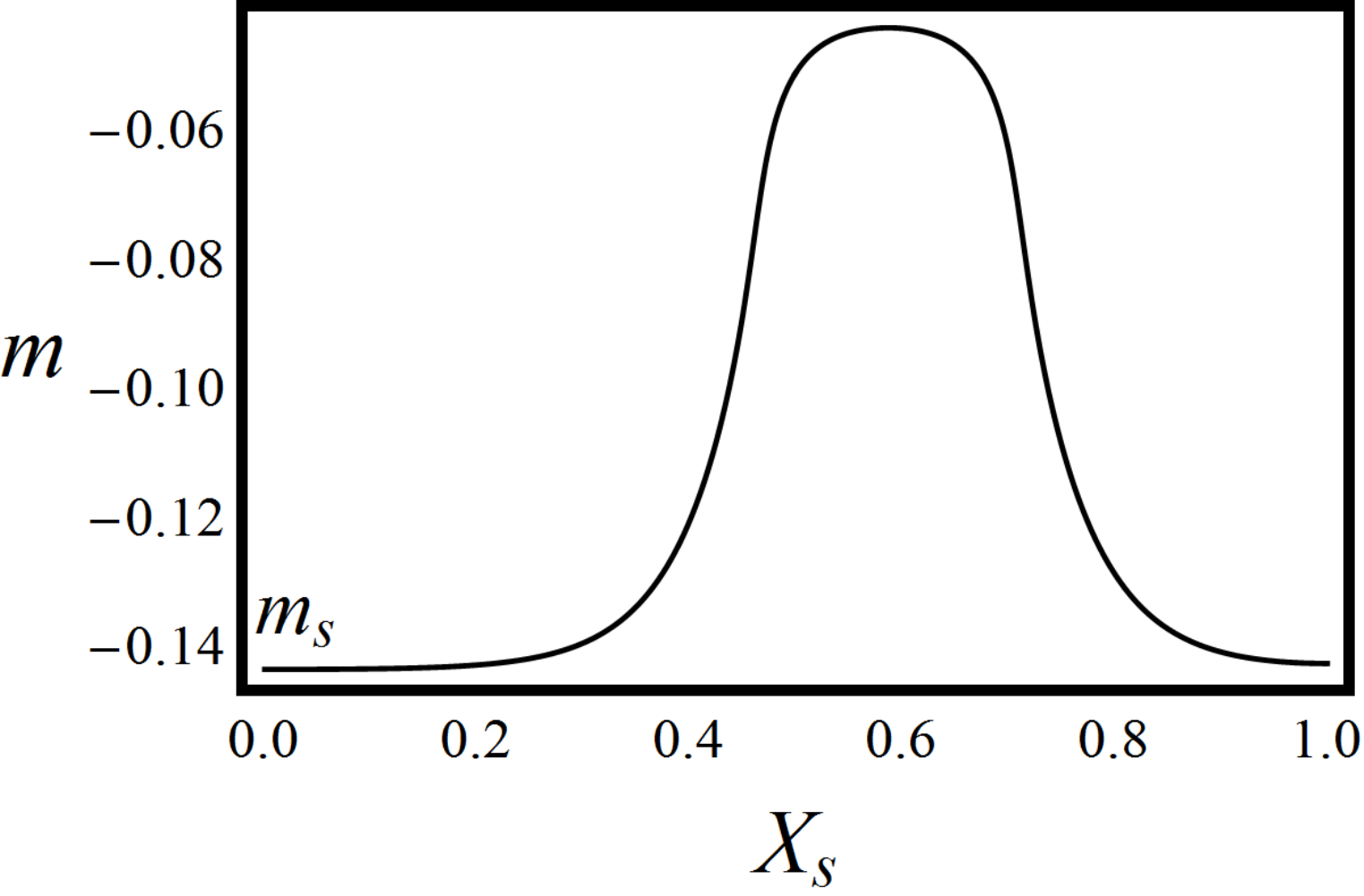}}}
}
\end{picture}
\vskip .1 cm
\centering
\caption{
Four types of solutions $\varepsilon(X_s)$ (left) and $m(X_s)$ (right) of the stationary problem \eqref{problema-staz01}. We used Neumann boundary conditions $m'(0)=\varepsilon'(0)=\varepsilon(1)=m'(1)=0$ on the finite interval $[0,1]$, at the coexistence pressure for $a=0.5,\,b=1,\,\alpha=100,\,k_1=k_2=k_3=10^{-3}$. The first and the second solutions represent the two phases, fluid--poor $(\varepsilon_s,m_s)$ and fluid--rich $(\varepsilon_f,m_f)$ respectively, while the third and the fourth one are interface--type solutions, one single interface and double interface respectively. 
}
\label{figurastazionario1}
\end{figure}

In Figure \ref{figurastazionario1} are depicted
four types of solution of problem \eqref{problema-staz01}. We found numerically these solutions with $k_1=k_2=k_3=10^{-3}$ 
via the finite difference
method powered with the Newton-–Raphson algorithm. The use of different initial guess in the Newton–-Raphson algorithm has allowed us to find numerically the different stationary solutions. The solutions reported here are those of main interest since they represent the
two phases of the system, fluid--poor and fluid--rich phases respectively, the single interface solution, which as we will prove in Subsection \ref{sss:kink1} is practically the same of \cite{CIS2013,ACS2014} (see Remark \ref{remark2} in \ref{sss:kink1}), and a double interface solution, that is characteristic of dynamics with Neumann homogeneous boundary conditions.

\subsection{General result on kink localization}
\label{sss:kink1}
Here we prove that the interface--like solution to Problem \eqref{problema-staz01} is the same as that of the Problem with Dirichlet boundary conditions, see \cite{CIS2012,CIS2013,ACS2014}. In fact, in \cite{CIS2012}, it has been shown that, for $k_1k_3-k_2^2=0$ (\textit{degenerate} case), the 
problem of finding a solution to the stationary problem \eqref{stazionario020}
can be reduced to the computation of a definite integral via a rephrasing of the problem as a one dimensional conservative mechanical 
system.

A rotation of the Cartesian reference system can be done; in particular by setting
\begin{equation}
\label{secondo07}
x:=\frac{m+k\varepsilon}{\sqrt{1+k^2}}
\;\;\;\textrm{ and }\;\;\;
y:=\frac{-k m+\varepsilon}{\sqrt{1+k^2}}
\end{equation}
in the plane $m$--$\varepsilon$,
where $k:=k_1/k_2=k_2/k_3$. Defining
\begin{equation}
\label{secondo08}
V(x,y)
=
\Psi(m(x,y),\varepsilon(x,y)),
\end{equation}
it can be shown that
$m(x)$ and $\varepsilon(x)$ are solutions of  (\ref{problema-staz01}) if and only if the corresponding fields $x(\xi)$ and $y(\xi)$ satisfy
\begin{equation}
\label{secondo12}
k_3(1+k^2)
x''=\frac{\partial V}{\partial x}(x,y)
\;\;\;\textrm{ and }\;\;\;
\frac{\partial V}{\partial y}(x,y)=0.
\end{equation}
The root locus of
the \textit{constraint curve}
$\partial V(x,y)/\partial y=0$ is made of
a certain number of maximal components
such that each of them is the graph of a
function $x\in\mathbb{R}\to y(x)\in\mathbb{R}$;
for each of them the first between the two equations
(\ref{secondo12}) becomes
a one--field one--dimensional problem.
Since
the function $V$ is obtained by
rotating the coordinate axes,
then at the coexistence pressure it
has the two absolute minimum points
$(x_\rr{s},y_\rr{s})$
and
$(x_\rr{f},y_\rr{f})$
corresponding, respectively, to the standard and to the fluid--rich phases.
Since $(m_\rr{s},\varepsilon_\rr{s})$
and $(m_\rr{f},\varepsilon_\rr{f})$
satisfy the equations $\partial\Psi(m,\varepsilon)/\partial m=0$ and
$\partial\Psi(m,\varepsilon)/\partial\varepsilon=0$, we have that the two points
$(x_\rr{s},y_\rr{s})$ and
$(x_\rr{f},y_\rr{f})$ are solutions of the constraint equation
$\partial V(x,y)/\partial y=0$
and hence they belong to the constraint curve.

In our case of homogenous Neumann boundary conditions, we have to study the system
\begin{eqnarray}
\label{equazione}
\begin{cases}
ku_{xx}=-V'(u)\\
u_{x}(0)=0,\;u_{x}(L)=0
\end{cases}
\;\;,
\end{eqnarray}
for the field $u(x)$ on the space interval $[0,L]$, for some $k,L>0$. Here $V:\mathbb{R}\rightarrow\mathbb{R}$ is a negative $C^2(\mathbb{R})$
function with two single isolated local maxima $a$ and $b$ (assume $a < b$), 
a single local minimum $c\in(a,b)$, and such that 
$V(a)=V(b)=0$, $V''(a)< 0$, $V''(b) < 0$, and $V''(c)>0$.

%%% Figura
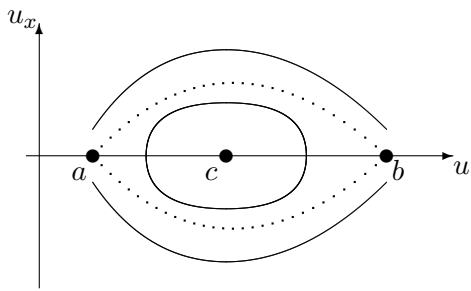
\begin{figure}[h!]
\begin{picture}(200,120)(-140,-10)
\thinlines
\put(-10,0){\vector(0,1){100}}
\put(-15,50){\vector(1,0){160}}
\put(10,50){\circle*{5}}
\put(60,50){\circle*{5}}
\put(120,50){\circle*{5}}
\thicklines
\qbezier[30](10,50)(60,105)(120,50)
\qbezier[30](10,50)(60,-5)(120,50)
\thinlines
\qbezier(10,60)(30,90)(60,90)
\qbezier(60,90)(90,90)(120,60)
\qbezier(10,40)(30,10)(60,10)
\qbezier(60,10)(90,10)(120,40)

\qbezier(30,50)(30,70)(60,70)
\qbezier(60,70)(90,70)(90,50)
\qbezier(30,50)(30,30)(60,30)
\qbezier(60,30)(90,30)(90,50)

\put(145,43){${u}$}
\put(-22,100){${u_x}$}
\put(2,41){${a}$}
\put(52,41){${c}$}
\put(122,41){${b}$}
\end{picture}
%\vskip 2 cm
\caption{Phase portrait of the 
stationary equation \eqref{equazione}.
Dotted lines represent asymptotic solutions.
Solid disks represent constant solutions. 
}
\label{f:qual}
\end{figure}

We remark that 
the equation above trivially satisfies the following 
\emph{energy conservation} principle
\begin{eqnarray}
\frac{1}{2}ku_{x}^2+V(u)=E.
\label{energycons}
\end{eqnarray}
By a standard Weirstrass qualitative analysis it follows that the 
solutions of the problem \eqref{equazione} have negative energy 
and, in the phase portrait depicted in figure~\ref{f:qual}, 
correspond to the closed phase lines lying between the 
two dotted separatrix components.
For any $V(c)<E<0$, 
these solutions are periodic with period $T_k(E)$ such that
\begin{equation}
\frac{T_k(E)}{2}=\int_{u_-(E)}^{u_+(E)}\frac{\sqrt{k}\rr{d}s}{\sqrt{2(E-V(s))}}
\end{equation}
where $u_\pm(E)$ are the solutions 
of the equation $V(u)=E$, with $u_-(E)<u_+(E)$. 

We remark that, for any fixed $k$, from 
\cite[Section $5$]{CI}
it follows that 
\begin{equation}
T_k(E)\overset{E\rightarrow 0^-}{\longrightarrow}+\infty,
\quad T_k(E)\overset{E\rightarrow V(c)^+}{\longrightarrow}
\frac{\sqrt{k}\,2\pi}{\sqrt{V''(c)}},
\end{equation} 
and
\begin{eqnarray}
\label{crescente}
\frac{\rr{d}}{\rr{d}E}T_k(E)>0.
\end{eqnarray}
Hence, if $L>\sqrt{k}\pi/\sqrt{V''(c)}$, there exists at least a solution of 
\eqref{equazione} and
a unique energy value $E_k(L)$ 
to which corresponds a monotone increasing 
solution connecting $u_-(E_k(L))$ and $u_+(E_k(L))$.
The energy level $E_k(L)$ is defined by
the equality 
\begin{eqnarray}
\int_{u_-(E_k(L))}^{u_+(E_k(L))}\frac{\sqrt{k}\rr{d}s}{\sqrt{2(E_k(L)-V(s))}}=L.
\label{Lfixed}
\end{eqnarray}
This solution is called \textit{connection} on the interval $[0,L]$. 

From now on fix $L>\pi/\sqrt{V''(c)}$ and always assume 
$k\le 1$ so that the existence of the connection is ensured. 
We then skip $L$ from the notation and 
summarize our resuls as follows.
We define the \textit{interface position} $x_k$
as the (unique) point where the connection profile attains
the value $(u_-(E_k)+u_+(E_k))/2$, namely,
\begin{equation}
\label{interk}
x^i_k
=
\int_{u_-(E_k)}^{\frac{u_-(E_k)+u_+(E_k)}{2}}
\frac{\sqrt{k}\rr{d}s}
             {\sqrt{2[E_k-V(s)]}}.
\end{equation} 
and prove that 
\begin{equation}
\label{interfaccia00}
\lim_{k\to 0}
 [\sqrt{V''(a_k)}\,x^i_k-\sqrt{V''(b_k)}\,(L-x^i_k)]=0.
\end{equation}
Moreover, we shall prove 
\begin{equation}
\label{ugualifasi}
\lim_{k\rightarrow 0}u_-(E_k) = a
\;\;\textrm{ and }\;\;
\lim_{k\rightarrow 0}u_+(E_k) = b 
\end{equation}
so that in the limit $k\to0$, 
the interface position $x^i_k$ tends to 
\begin{equation}
\label{interfaccia01}
x^i_0:=\frac{L\sqrt{V''(b)}}
              {\sqrt{V''(a)}+\sqrt{V''(b)}}.
\end{equation}
This result is analogous to the case of Dirichlet boundary 
conditions of \cite{CIS2012}.

The first step in the proof of \eqref{interfaccia00}is to show that 
the connection energy level tends to $0$ for $k\to0$, i.e., 
\begin{eqnarray}
\label{limiteezero}
\lim_{k\rightarrow 0}E_k=0.
\end{eqnarray}
Indeed, consider $k'<k<1$ and note that by the definition 
\eqref{Lfixed} of the energy levels $E_k$ and $E_{k'}$ 
we have
\begin{eqnarray}
\int_{u_-(E_k)}^{u_+(E_k)}\frac{\sqrt{k}\rr{d}s}{\sqrt{2(E_k-V(s))}}=\int_{u_-(E_{k'})}^{u_+(E_{k'})}\frac{\sqrt{k'}\rr{d}s}{\sqrt{2(E_{k'}-V(s))}}, 
\end{eqnarray}
so that
\begin{eqnarray}
\int_{u_-(E_k)}^{u_+(E_k)}\frac{\rr{d}s}{\sqrt{2(E_k-V(s))}}
<\int_{u_-(E_{k'})}^{u_+(E_{k'})}\frac{\rr{d}s}{\sqrt{2(E_{k'}-V(s))}},
\end{eqnarray}
Hence, by using \eqref{crescente} with $k=1$ it follows that 
$E_{k'}>E_k$. We have thus proved that $E_k$ is a decreasing function 
of $k\in(0,1]$.
Now, if by absurdity it were  
$\lim_{k\rightarrow 0}E_k<0$, 
then it would be 
\begin{eqnarray}
\lim_{k\to0}
\sqrt{k}
\int_{u_-(E_k)}^{u_+(E_k)}\frac{\rr{d}s}{\sqrt{2(E_k-V(s))}}
=0
\end{eqnarray}
since the integral above would be bounded for $k\in(0,1)$, and this 
is absurd in view of the definition \eqref{Lfixed} of the energy level 
$E_k$.
Thus, we have estabilished \eqref{limiteezero}.

The next step to get \eqref{interfaccia00} is proving that for every $u_1,u_2\in[u_-(E_k),u_+(E_k)]$,
\begin{eqnarray}
\lim_{k\rightarrow 0}\int_{u_1}^{u_2}\frac{\sqrt{k}\rr{d}s}{\sqrt{2(E_k-V(s))}}=0.
\label{eqthm1}
\end{eqnarray}
For sufficiently small $k$, by \eqref{limiteezero}, we have that
\begin{eqnarray}
E_k>\max\left\lbrace V(u_1),V(u_2)\right\rbrace,
\end{eqnarray}
so that there exists $\delta>0$ such that
\begin{equation}
\frac{1}{\sqrt{2(E_k-V(s))}}\leq\delta.
\end{equation}
Thus
\begin{eqnarray}
\lim_{k\rightarrow 0}\int_{u_1}^{u_2}\frac{\rr{d}s}{\sqrt{2(E_k-V(s))}}=\int_{u_1}^{u_2}\frac{\rr{d}s}{\sqrt{2(-V(s))}}<\infty.
\label{limitefinito1}
\end{eqnarray}
By \eqref{limitefinito1} we get \eqref{eqthm1}. The proof of \eqref{interfaccia00} is now just an extension of the proof of Theorem $2$ of\cite{CIS2012}.

From \eqref{limiteezero}, \eqref{eqthm1} and \cite{CIS2012} we get \eqref{ugualifasi}, where $a$ and $b$ correspond to the fluid--poor and fluid--rich phases of \eqref{problema-staz01} so that we can state that the interface position is the same in both Dirichlet and Neumann--Neumann homogeneous boundary conditions. Actually the authors of \cite{CIS2012} determined the position of the interface between the fluid--poor and the fluid--rich phases that turned out to be $X_s=0.6164$ in agreement with the numerical simulations, see the third row of Figure \ref{figurastazionario1}.
\begin{Remark}
\label{remark2}
At the end of Subsection \ref{sottosezione} we said that the solution are ``practically'' the same becouse the zero energy level can not be reached, in this case of Neumann boundary conditions, as for a qualitative analysis point of view, we are considering trajectories starting and ending with velocity zero. The values $a$ and $b$ are only limit (for $k\to0$) values.
\end{Remark}
\section{The not stationary problem}
\label{section:darcyneumann}
In this section we discuss the not stationary solution of 
the problems (\ref{problema-d}) and (\ref{problema-d-osi}) with homogenous Neumann natural boundary conditions. This choice appears to be very natural if one does not want to prescribe the two phases at the boundaries via Dirichlet boundary conditions. Now, the physical problem of main interest is that of asking: ``is there the possibility of the spontaneous formation of a bubble inside a porous medium in the neighborhood of the impermeable wall?'', ``how does this phenomenon happen? and for which initial conditions?''
\\\indent
As we will show below, in order to answer to these questions and to understand this phenomenon we have studied numerically Problems (\ref{problema-d}) and (\ref{problema-d-osi}) with piecewise initial conditions.
\\\indent
The stationary solutions are several (see Section \ref{s:stazionario}), so that the dynamics with or without an impermeable wall, starting with the same initial condition, could possibly lead to different final states. 

The equations of motion describing the evolution of the system 
in the two cases, with the choiche \eqref{sec010}, \eqref{sec015} and  with homegenous Neumann natural boundary conditions  read
\begin{equation}
\label{num05}
\left\{
\begin{array}{l}
-(2/3)\alpha b m^3+\alpha b^2m^2\varepsilon+p+\varepsilon-ab(m-b\varepsilon)
-k_1\varepsilon''-k_2m''=0
\\
 \varrho^2_{0,\rr{f}}
 ( 
  \alpha m^3-2\alpha b m^2 \varepsilon+\alpha b^2m\varepsilon^2
  +a(m-b\varepsilon)
  -k_2\varepsilon''-k_3m'' 
 )''
 =D\dot{m}
\\
\end{array}
\right.
\end{equation}
with the boundary conditions
\begin{equation}
\label{num01}
\left\{
\begin{array}{l}
{\displaystyle
\varepsilon'(\ell_1)=m'(\ell_1)=\varepsilon'(\ell_2)=m'(\ell_2)
=0
}
\\ \\
( 
 \alpha m^3-2\alpha b m^2 \varepsilon+\alpha b^2m\varepsilon^2
 +a(m-b\varepsilon)
 -k_2\varepsilon''-k_3m'' 
)_{\ell_1,\ell_2}
=0
\end{array}
\right.
\end{equation}
for the zero chemical potential problem (\ref{problema-d}) 
and 
\begin{equation}
\label{num02}
\left\{
\begin{array}{l}
{\displaystyle
\varepsilon'(\ell_1)=m'(\ell_1)=\varepsilon'(\ell_2)=m'(\ell_2)
=0
}
\\ \\
( 
 3\alpha m^2m'
 -4\alpha b mm' \varepsilon
 -2\alpha b m^2 \varepsilon'
 +\alpha b^2m'\varepsilon^2
\\
\phantom{(3\alpha m^2m'}
 +2\alpha b^2m\varepsilon\varepsilon'
 +a(m'-b\varepsilon')
 -k_2\varepsilon'''-k_3m''' 
)_{\ell_2}
=0
\\ \\
( 
 \alpha m^3-2\alpha b m^2 \varepsilon+\alpha b^2m\varepsilon^2
 +a(m-b\varepsilon)
 -k_2\varepsilon''-k_3m'' 
)_{\ell_1}
=0
\end{array}
\right.
\end{equation}
for the one--side impermeable problem 
(\ref{problema-d-osi}).

The dissipative evolution of the elastic strain $\varepsilon$ and 
the fluid mass density variations $m$, relative to the zero chemical 
potential boundary condition and the one--side impermeability condition, 
are discussed for four type of different initial conditions, say:
three different piecewise initial condition and a piecewise condition with a continuos junction.  
\\\indent
In the one--side impermeble case we were able to obtain different final states corresponding to different stationary solutions (with respect to those of the evolution without the impermeable wall) of the problem, see figure \ref{figurastazionario1}. In the case of a piecewise initial condition, depending on the fluid--poor portion of the medium $s$ and the water level $h$ of the $m$ profile (and corresponding $\varepsilon$ profile) of the initial state, we obtained these different solutions, as reported in Table \ref{tabella1}.
\vskip 1cm
\begin{table}[h!]
\begin{center}
\begin{tabular}{|l|l|l|l||l|l|l|r|}
\hline
n.&$s$&$h$&solution type&n.&$s$&$h$&solution type\\
\hline
1&50&-0.093& double interface&7&200&-0.093& interface\\
\hline
2&50&-0.092&double interface&8&200&-0.068&interface\\
\hline
3&50&-0.091&double interface&9&350&-0.090&interface\\
\hline
4&50&-0.090&double interface&10&500&-0.093& interface\\
\hline
5&50&-0.089&interface&11&500&-0.091&interface\\
\hline
6&100&-0.068&interface&12&500&-0.090&interface\\
\hline
\end{tabular}
\end{center}
\caption{fluid--poor portion of the medium $s$, water level $h$ of the initial $m$ profile and final state of the dynamics with impermeable condition.}
\label{tabella1}
\end{table}
The choice of these piecewise conditions is that it is straightforward to expect an accumulation of fluid in the neighborhood of the impermeable wall as the level $h$ grows (with $s$ remaining fixed), so that the fluid could remain trapped into the porous medium, thus giving a dynamics towards a stationary solution of interface type.   
On the other hand, it is also possible that, if the fluid level $h$ is too high, also the dynamics without the impermeable wall can lead to a not constant stationary solution (see Figure \ref{dinamica1} and discussion below), so that it is important to study thoroughly the effects of the more (or the less) amount of fluid in the initial state, in order to obtain different final states for the dynamics with the two different natural boundary conditions.

\subsection{Piecewise initial conditions}
\begin{figure}[h!]
\vskip 1.80cm
\begin{picture}(200,400)(15,0)
\put(-40,340)
{
\resizebox{5.5cm}{!}{\rotatebox{0}{\includegraphics{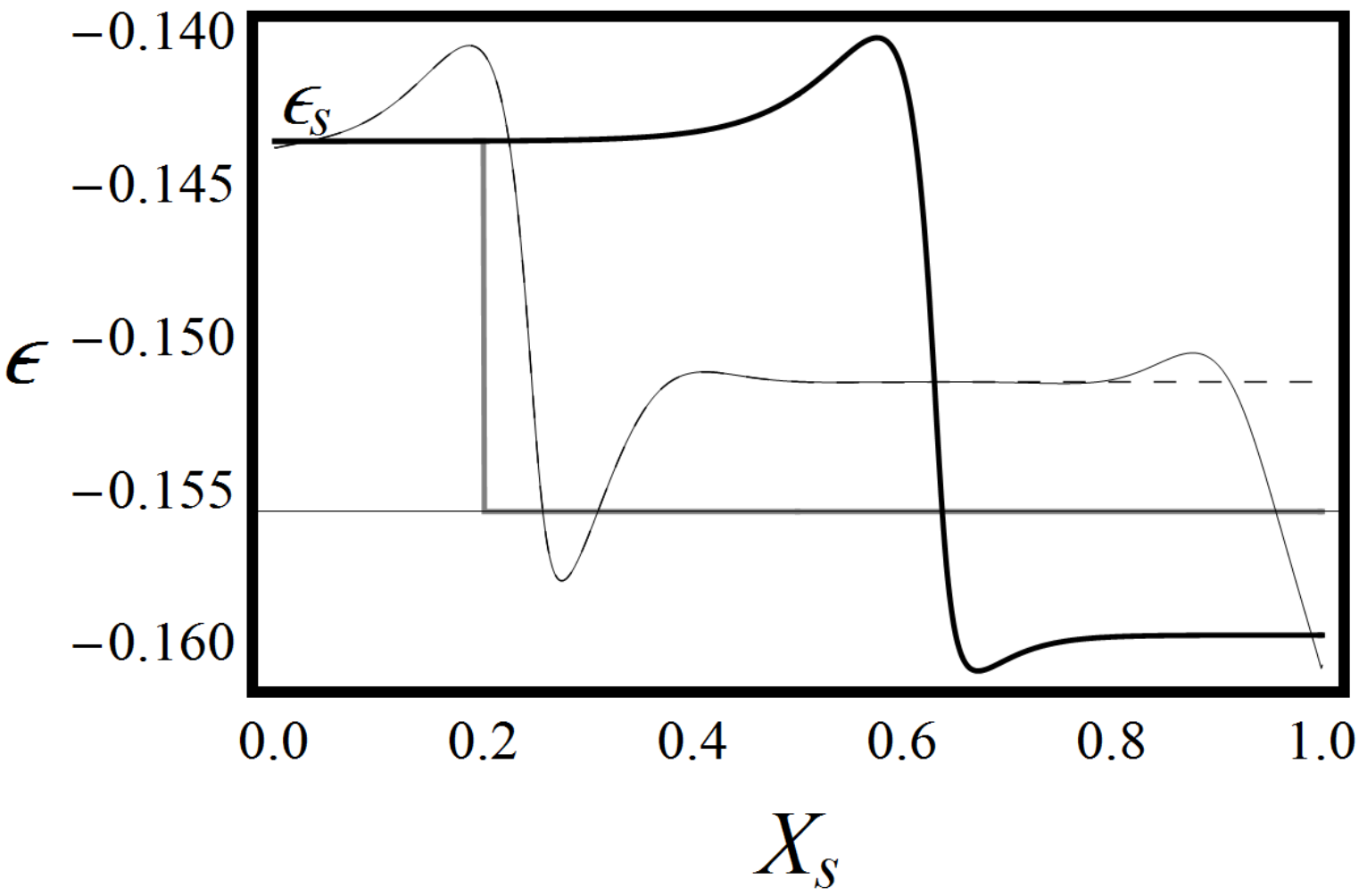}}}
}
\put(135,340)
{
\resizebox{5.5cm}{!}{\rotatebox{0}{\includegraphics{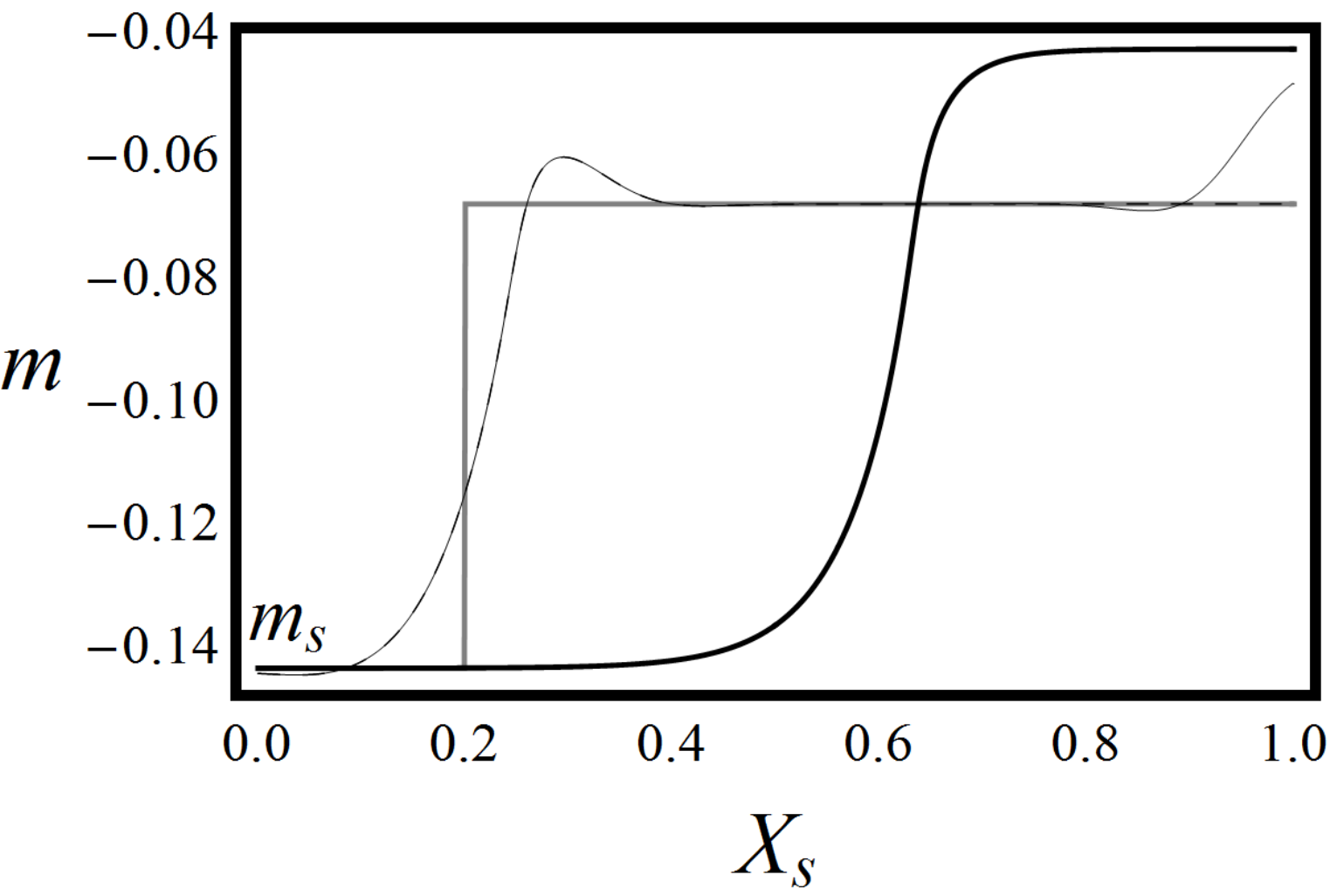}}}
}
\put(310,340)
{
\resizebox{5.5cm}{!}{\rotatebox{0}{\includegraphics{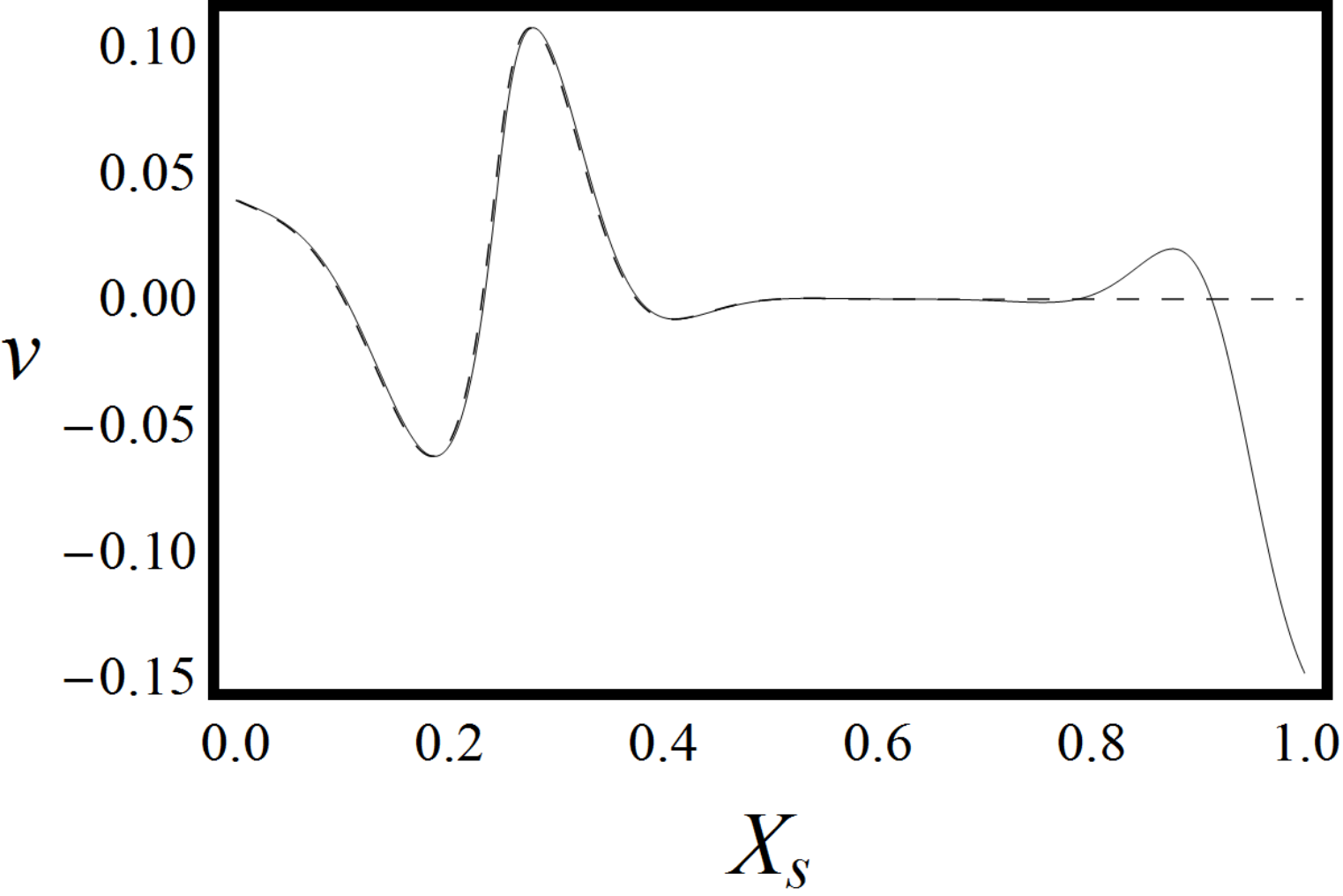}}}
}
\put(-40,220)
{
\resizebox{5.5cm}{!}{\rotatebox{0}{\includegraphics{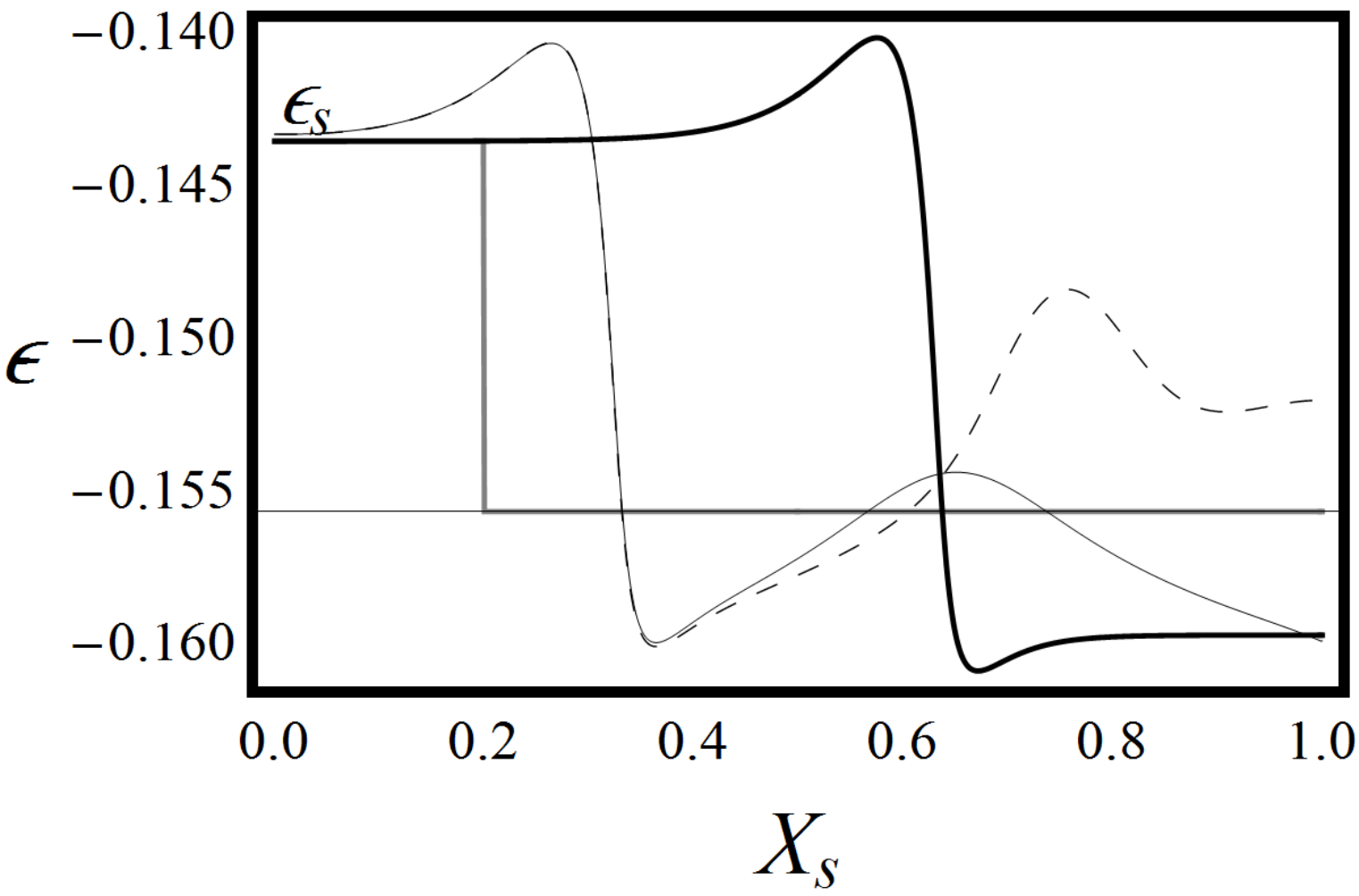}}}
}
\put(135,220)
{
\resizebox{5.5cm}{!}{\rotatebox{0}{\includegraphics{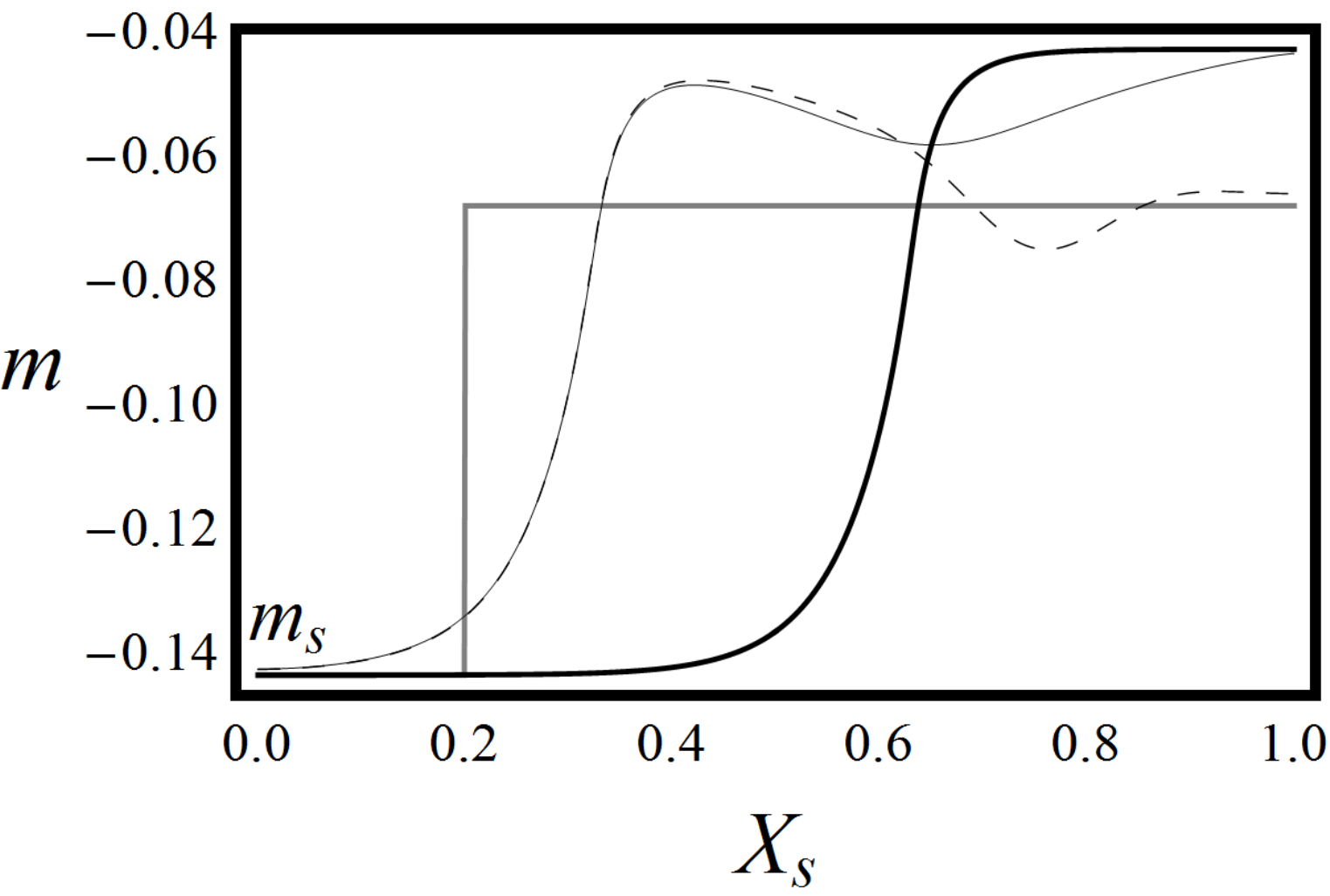}}}
}
\put(310,220)
{
\resizebox{5.5cm}{!}{\rotatebox{0}{\includegraphics{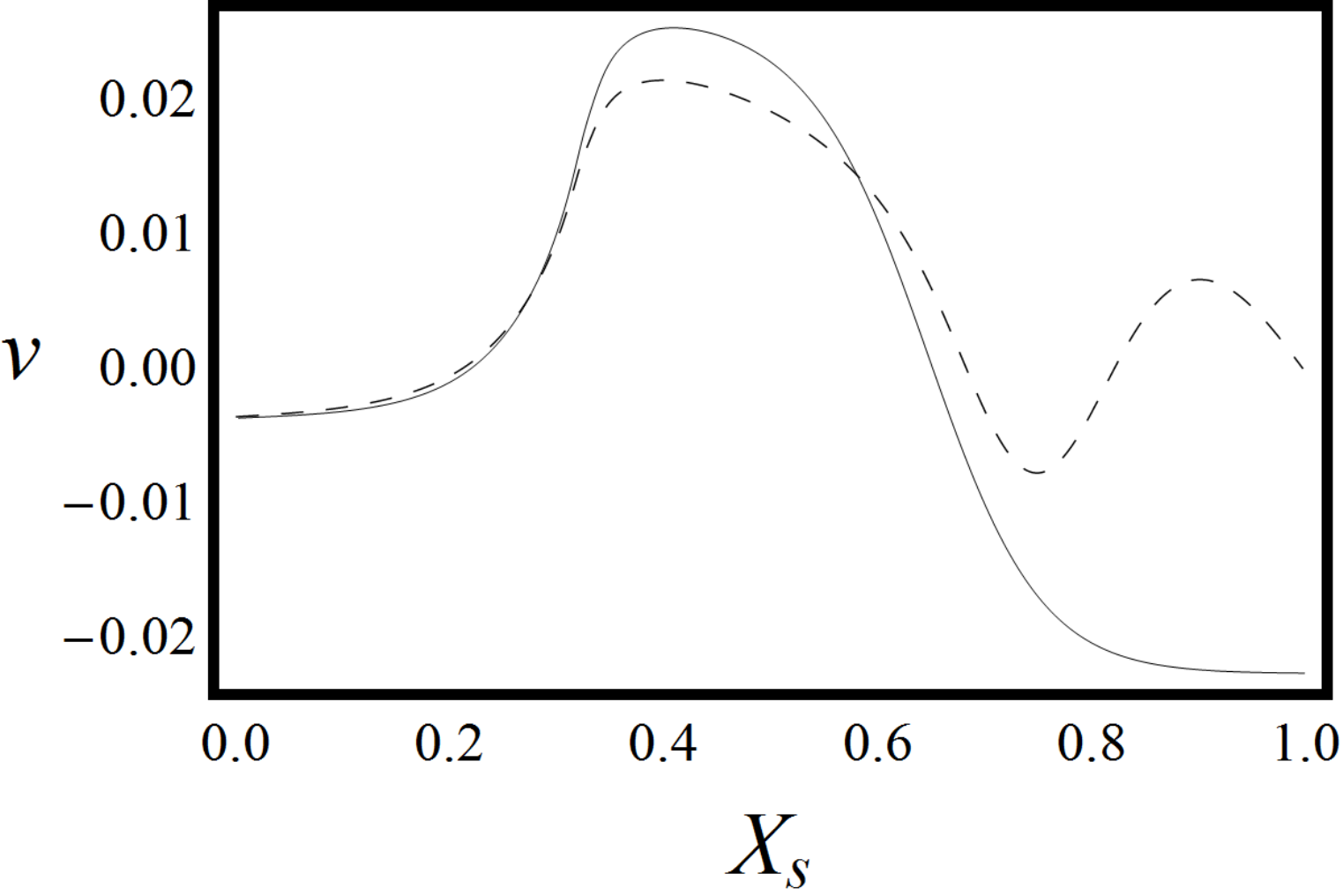}}}
}
\put(-40,100)
{
\resizebox{5.5cm}{!}{\rotatebox{0}{\includegraphics{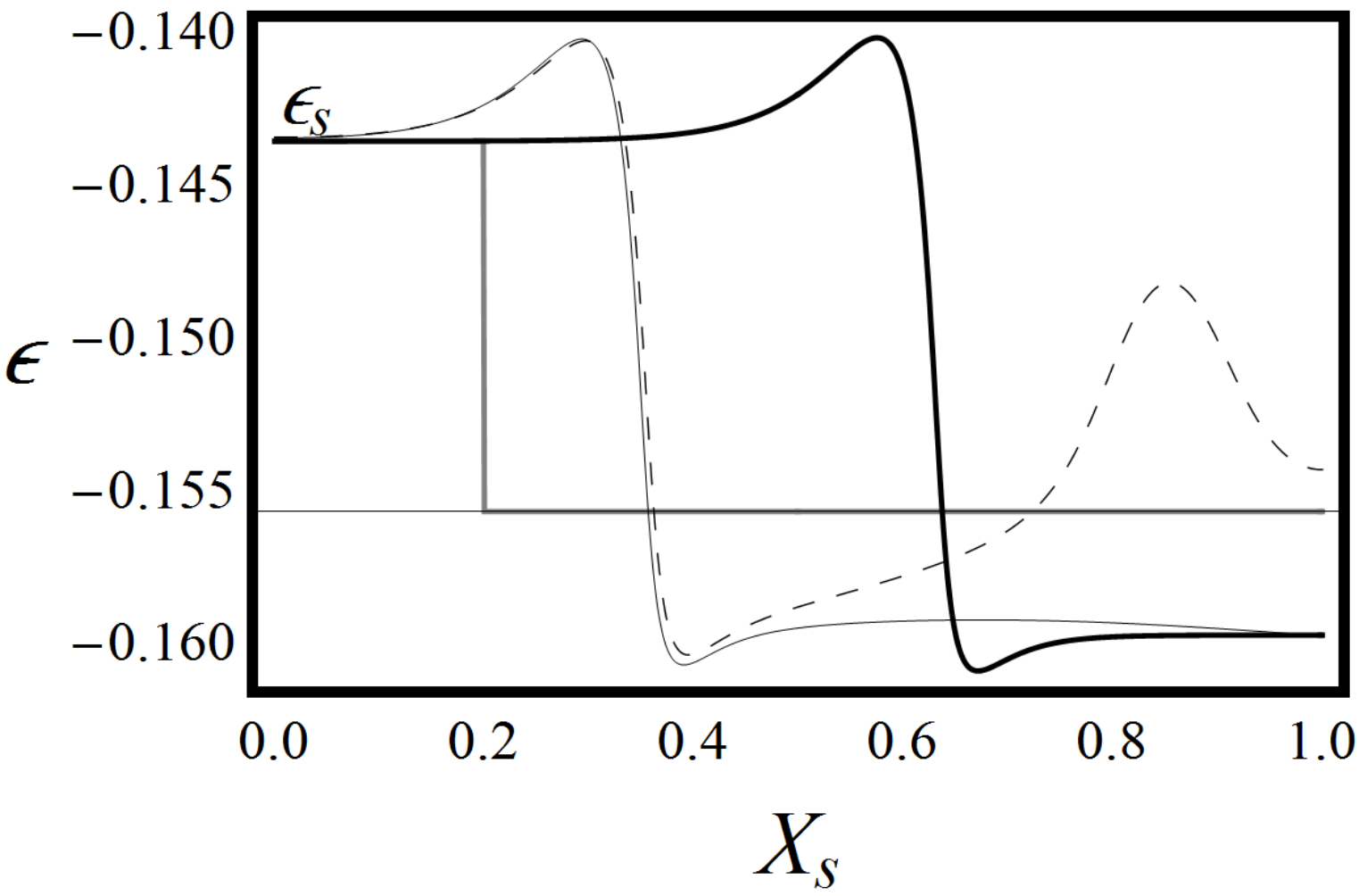}}}
}
\put(135,100)
{
\resizebox{5.5cm}{!}{\rotatebox{0}{\includegraphics{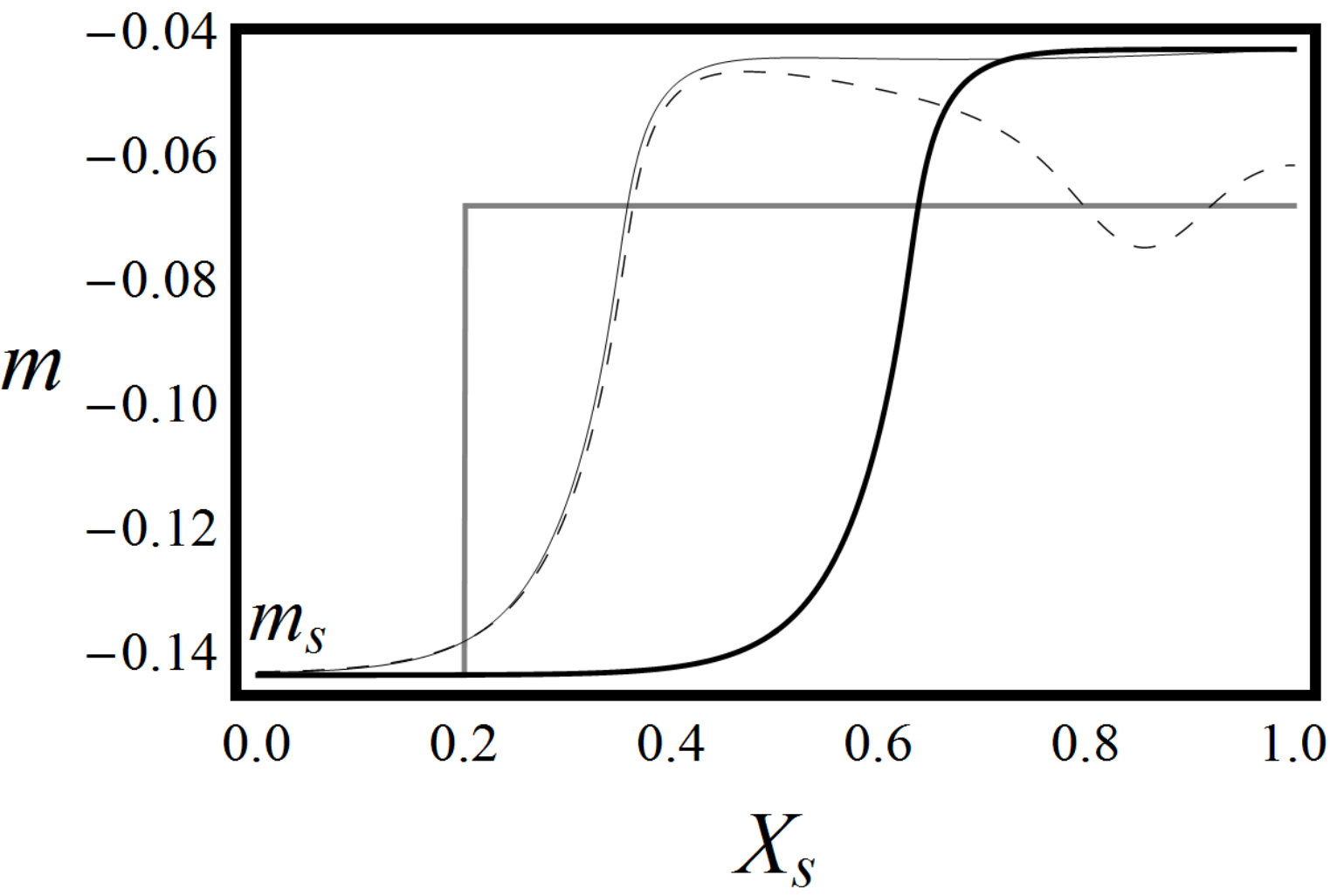}}}
}
\put(310,100)
{
\resizebox{5.5cm}{!}{\rotatebox{0}{\includegraphics{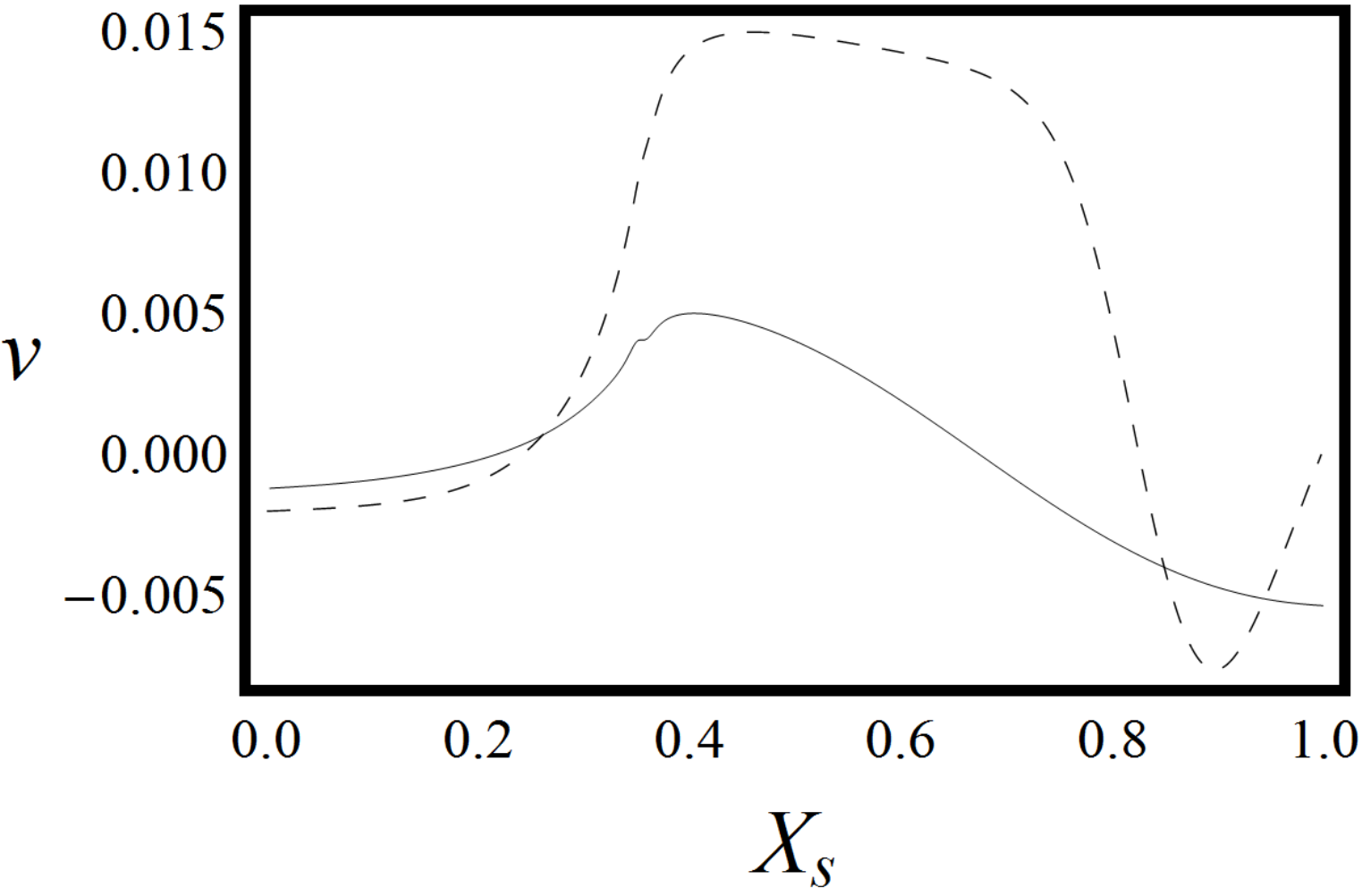}}}
}
\put(-40,-20)
{
\resizebox{5.5cm}{!}{\rotatebox{0}{\includegraphics{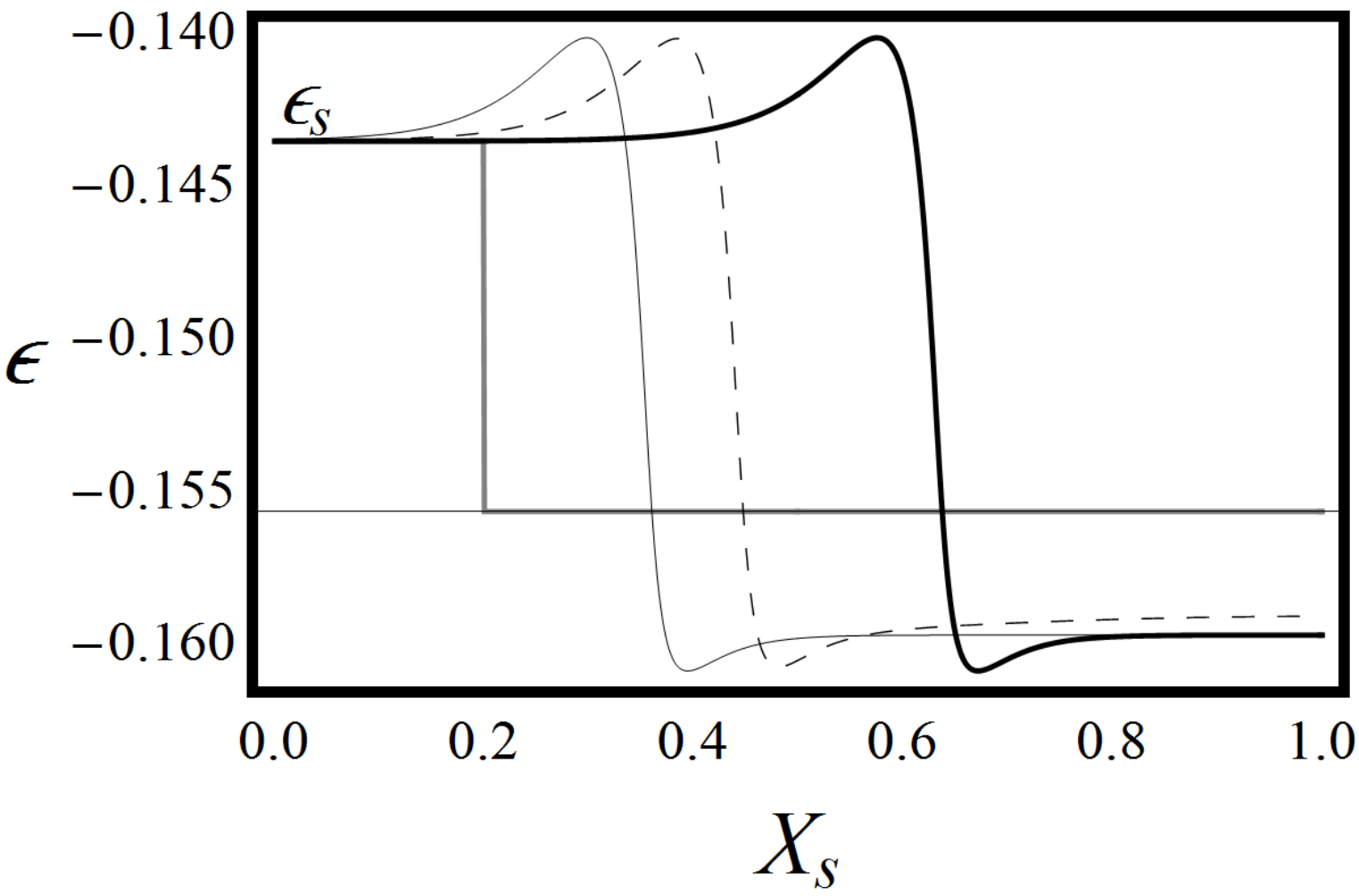}}}
}
\put(135,-20)
{
\resizebox{5.5cm}{!}{\rotatebox{0}{\includegraphics{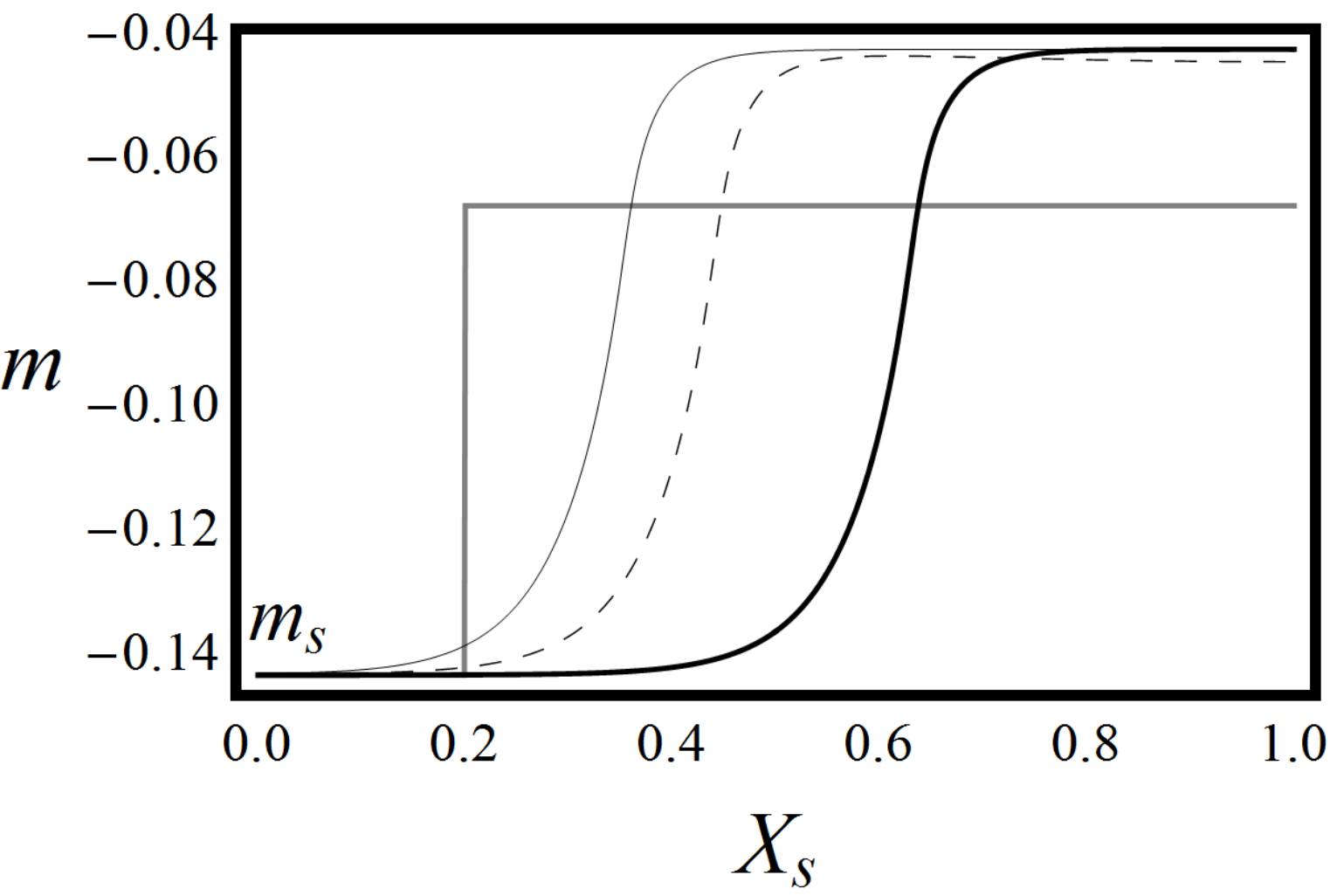}}}
}
\put(310,-20)
{
\resizebox{5.5cm}{!}{\rotatebox{0}{\includegraphics{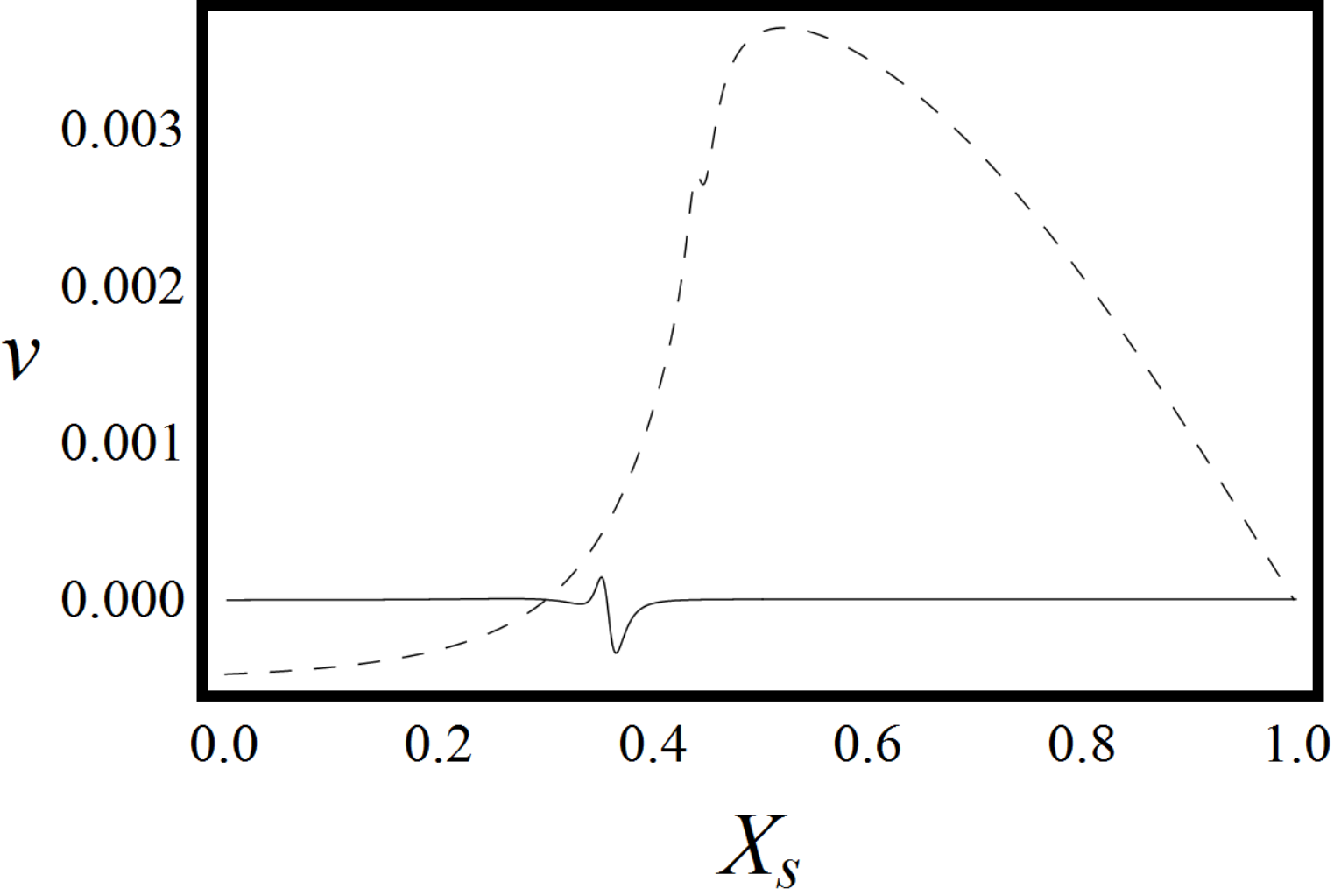}}}
}
\end{picture}
\vskip 0.8 cm
\centering
\caption{Solutions $\varepsilon(X_s,t)$, $m(X_s)$ and $v(X_s)$ for the zero chemical potential problem (thin black lines) and the one--side impermeable problem (dotted thin lines) obtained by solving 
Problem (\ref{problema-d}) and (\ref{problema-d-osi}) respectively, with the first initial condition (gray lines). We used Neumann boundary conditions $m'(0)=\varepsilon'(0)=m'(0)=m'(1)=0$ on the finite interval $[0,1]$, at the coexistence pressure for $a=0.5,\,b=1,\,\alpha=100,\,k_1=k_2=k_3=10^{-3}$. 
Profiles at times $t=0.04,\,t=0.15,\,t=0.3,\,t=1$ are in lexicographic order. The solid black lines represent interface--type stationary profiles.
}
\label{dinamica1}
\end{figure}

In figures \ref{dinamica1}, \ref{dinamica2} and \ref{dinamica3} we report the evolution of the system starting from three different piecewise initial states.
\\\indent
The first condition, corresponding to case $8$ of Table \ref{tabella1}, is an evolution for which the final state in both cases (zero--chemical and one--side--impermeable problem) is the same. In fact, as can be seen in Figure \ref{dinamica1}, in both cases the motion is divided in two steps, the same as the evolutions studied in \cite{ACS2014}, i.e. the formation of the $\varepsilon$ and $m$ profiles and the slow motion towards the not costant (interfaces) stationary solutions. The main reason of the similar evolution in both cases is the fact that the amount of water of the initial state is very high (see Section above), so that the formation of the interfaces is a direct consequence. Even if the final state of both the evolutions is the same, some differences can be observed in the first step of the evolutions. In the case without the impermeable wall, water can enter from the right boundary so that the formation of the profile is not only quicker but also more straightforward, as confirmed by the velocity profile, than the first step with the impermeable wall. In fact, in this last case, due to the presence of the wall, the fluid has to slowly accumulate in the neighborhood of the right boundary, giving rise to depression phenomena in the first moments of evolution (see the first three rows of Figure \ref{dinamica1}). 
\\\indent
The evolution depicted in Figure \ref{dinamica2} corresponds to case $7$ of of Table \ref{tabella1}. In this case, the dry portion of the medium being the same, the water level of the initial state is less with respect to case $8$ so that the evolution without an impermeable wall ends with the classical consolidation of a porous medium (the fluid flows out completely) and the final state is the standard poor--fluid state $(\varepsilon_s,m_s)$. On the other hand, the corresponding evolution with an impermeable wall has a different beahvoiur, actually the same behaviour of the previous case $8$ and that of the the evolutions studied in \cite{ACS2014}.
The way of fluid emptying in the zero--chemical potential evolution is first the accumulation of the initial state fluid in the center of the medium, thus giving rise to a double--interface--like profile (as can be observed by the velocity profile), then the symmetric slow release of the fluid through the boundaries.
Regarding the evolution in the one--side--impermeable case, the first step of dynamics is slight different from that of Figure \ref{dinamica1}; in fact in this case, due to the less amount of water of the initial state, the depression phenomenon before the fluid accumulation in the neighborhood of the roght boundary is much more appreciable. 
\\\indent
Finally we analyze the evolution of Figure \ref{dinamica3}, which refers to case $1$. The water level of the initial state is the same as that of case $7$, but the dry portion of the medium is much less. The evolution with the impermeability is quite different in this case. In fact, the final state corresponds to the fourth case of Figure \ref{figurastazionario1}, i.e. a double interface. The fluid amount of the initial state is not enough to produce a single--interface profile, but not so little to let all the water exit from the left boundary. The result is that the fluid remains trapped in the center of the medium instead of in its right part. In this case, the water depression is observed in the center of the porous medium. It is worth to note that the final velocity profile is the same of that of the cases studied in \cite{ACS2014}, but duplicated.
\\\indent
In this case the evolution of the system without impermeable wall is quite similar to that of Figure \ref{dinamica1}.

\subsection{Piecewiese continuous initial condition}
\begin{figure}[h!]
\vskip 2cm
\begin{picture}(200,600)(15,0)
\put(-40,550)
{
\resizebox{5cm}{!}{\rotatebox{0}{\includegraphics{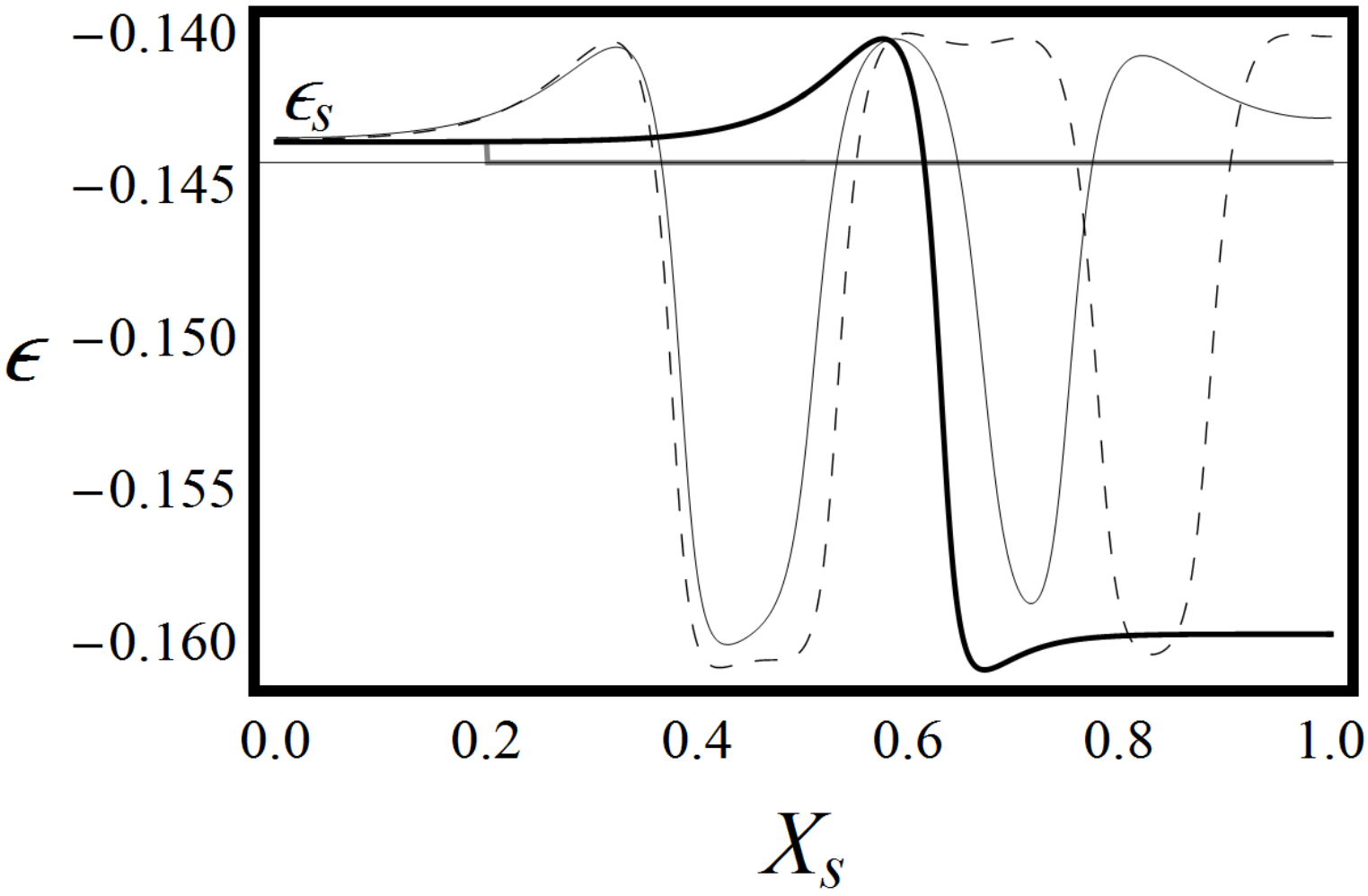}}}
}
\put(135,550)
{
\resizebox{5cm}{!}{\rotatebox{0}{\includegraphics{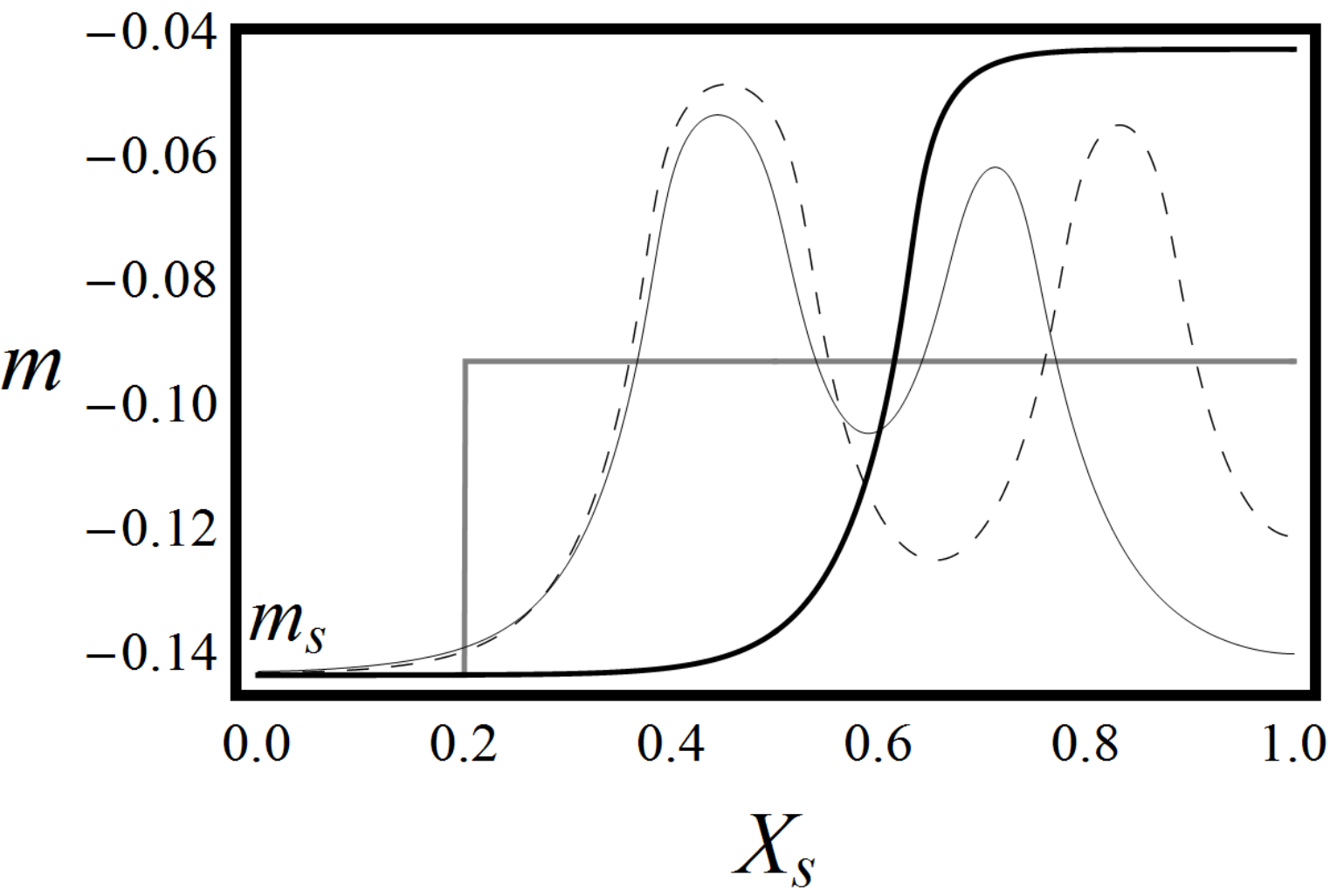}}}
}
\put(310,550)
{
\resizebox{5cm}{!}{\rotatebox{0}{\includegraphics{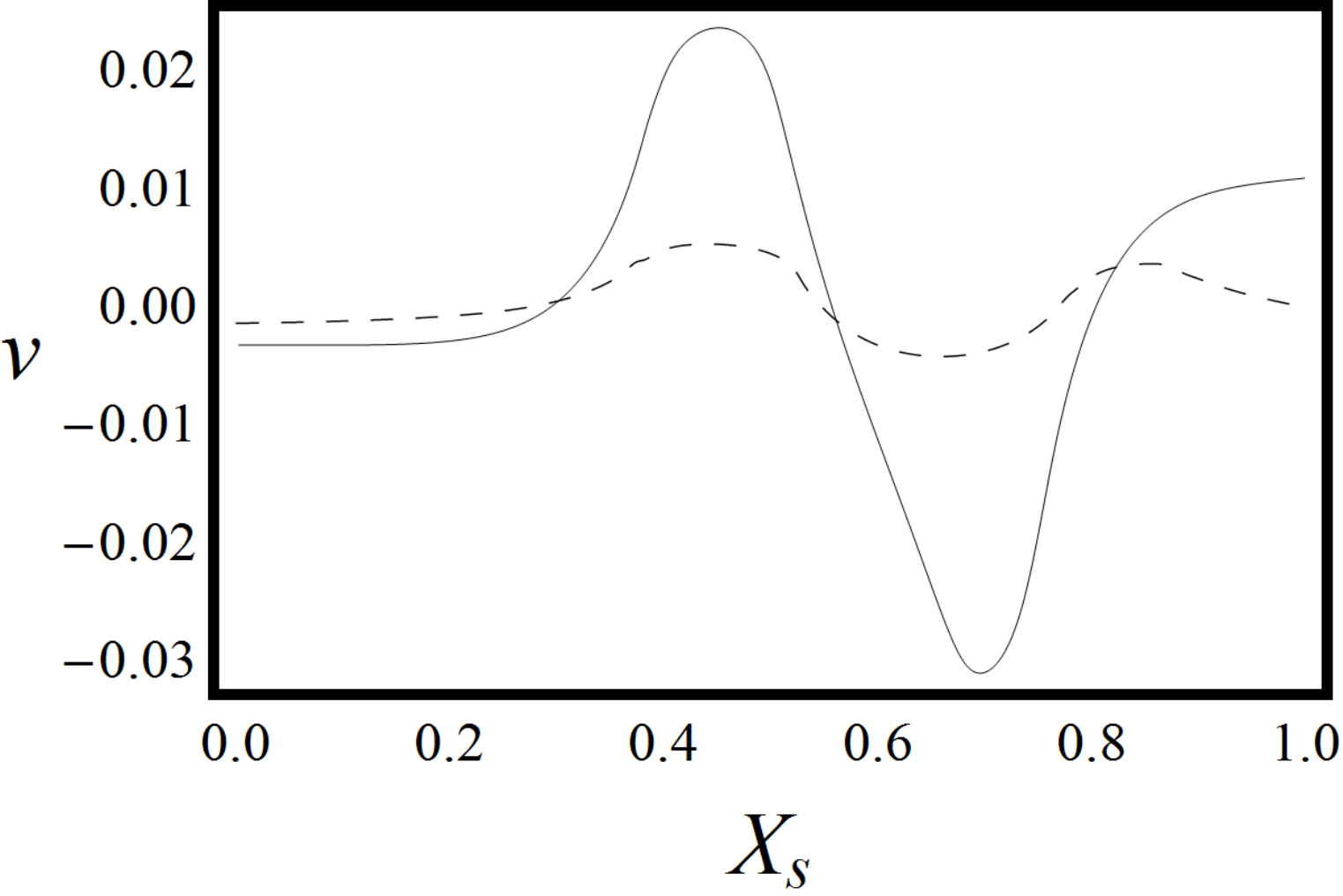}}}
}
\put(-40,450)
{
\resizebox{5cm}{!}{\rotatebox{0}{\includegraphics{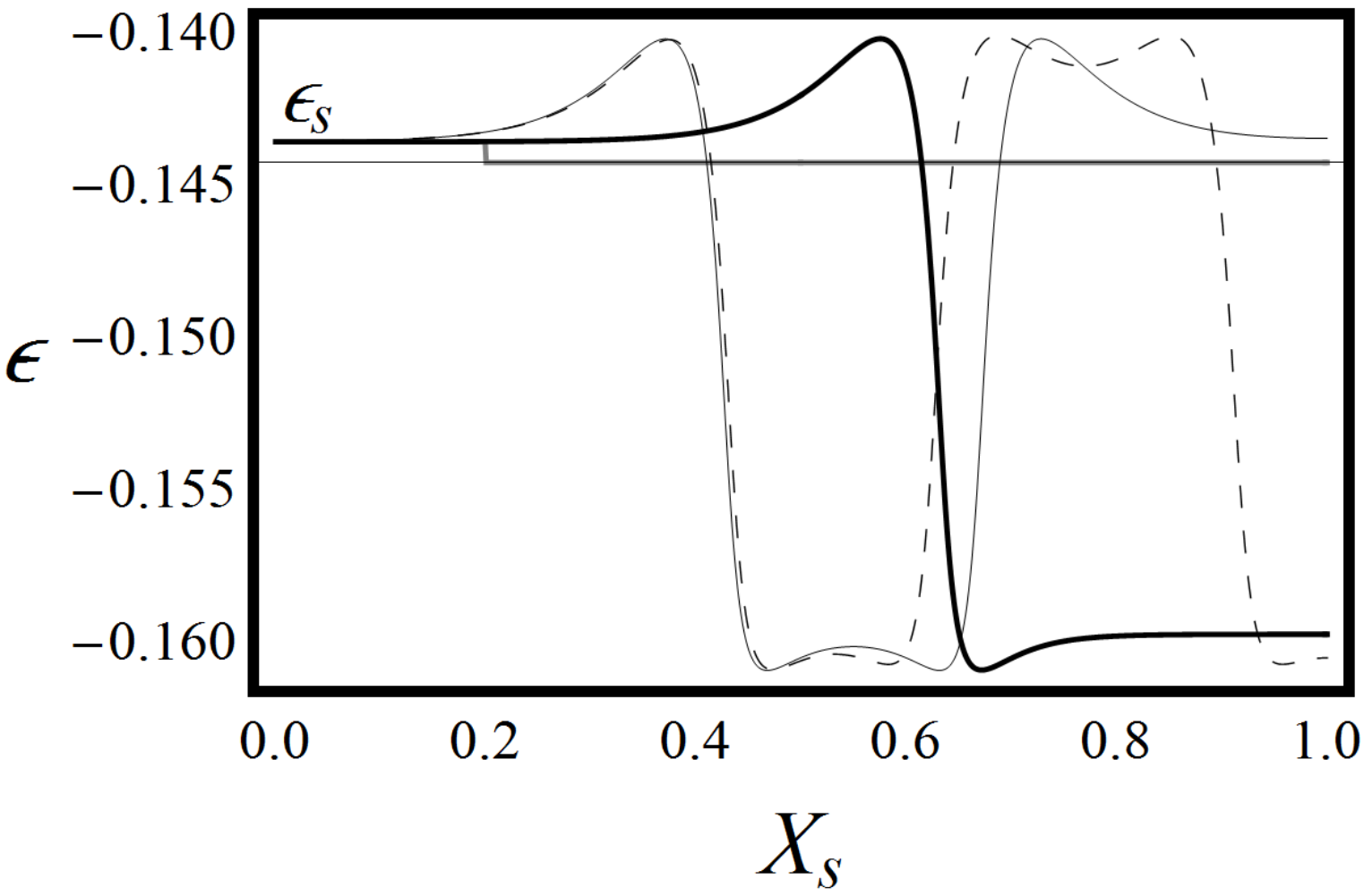}}}
}
\put(135,450)
{
\resizebox{5cm}{!}{\rotatebox{0}{\includegraphics{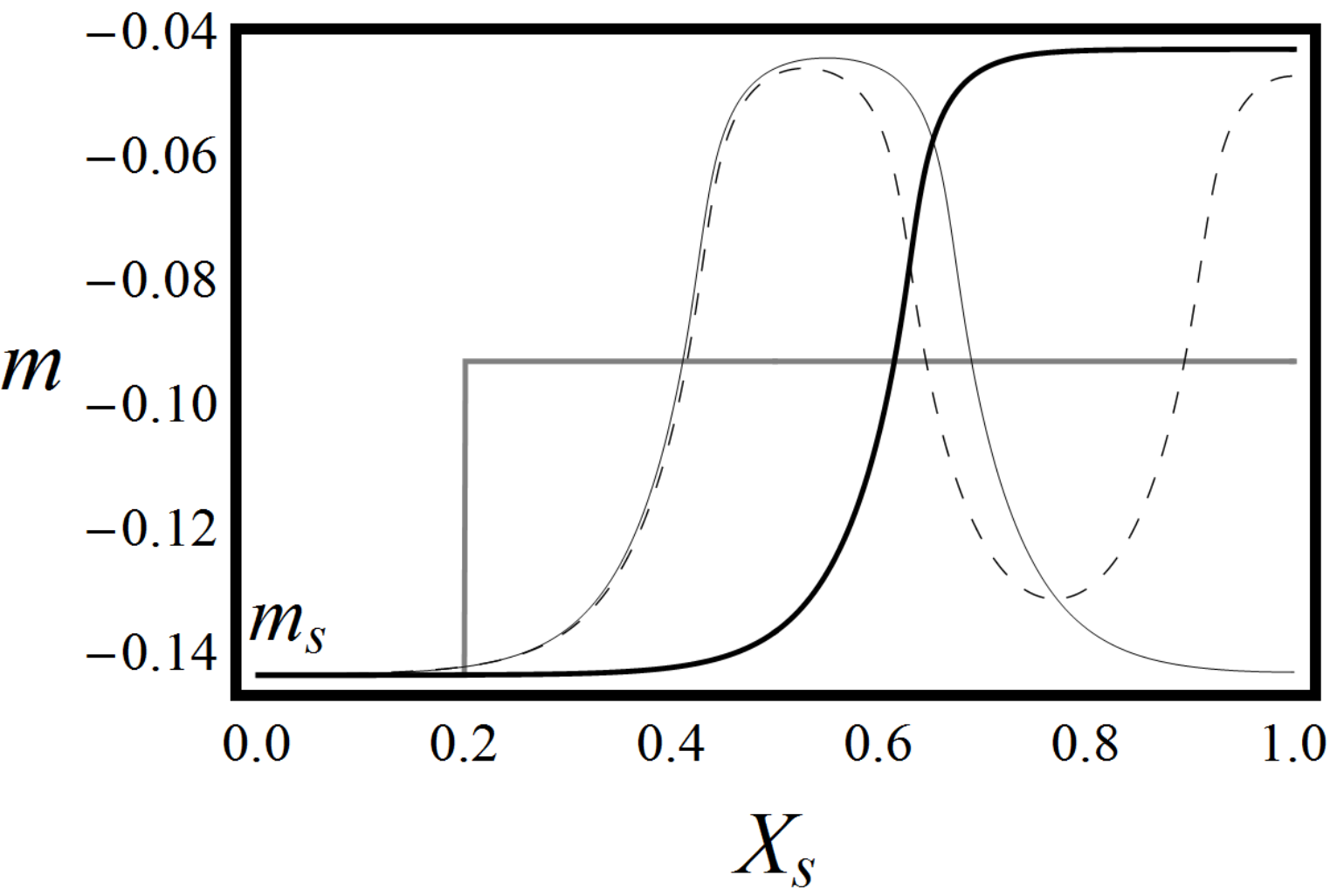}}}
}
\put(310,450)
{
\resizebox{5cm}{!}{\rotatebox{0}{\includegraphics{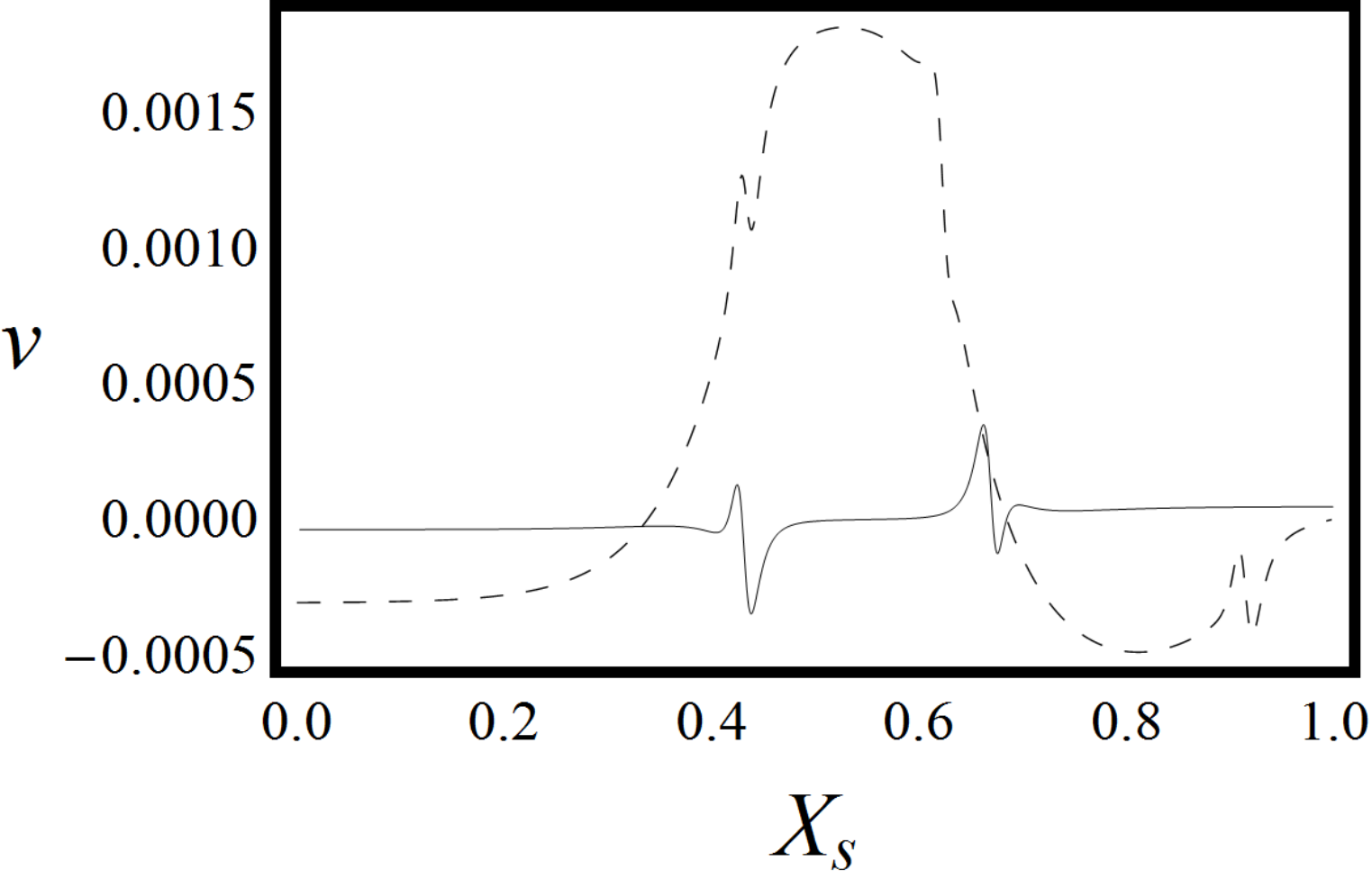}}}
}
\put(-40,350)
{
\resizebox{5cm}{!}{\rotatebox{0}{\includegraphics{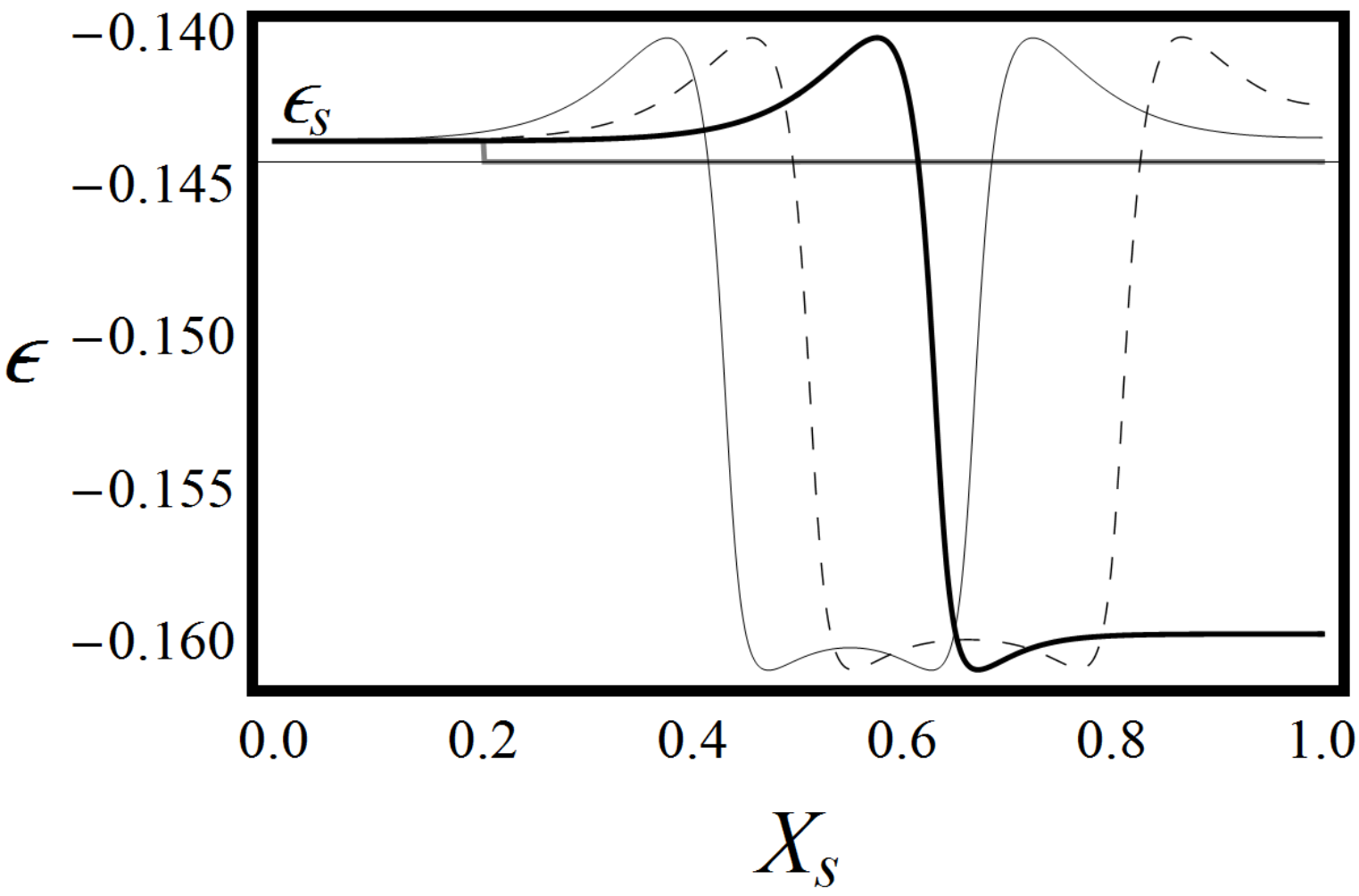}}}
}
\put(135,350)
{
\resizebox{5cm}{!}{\rotatebox{0}{\includegraphics{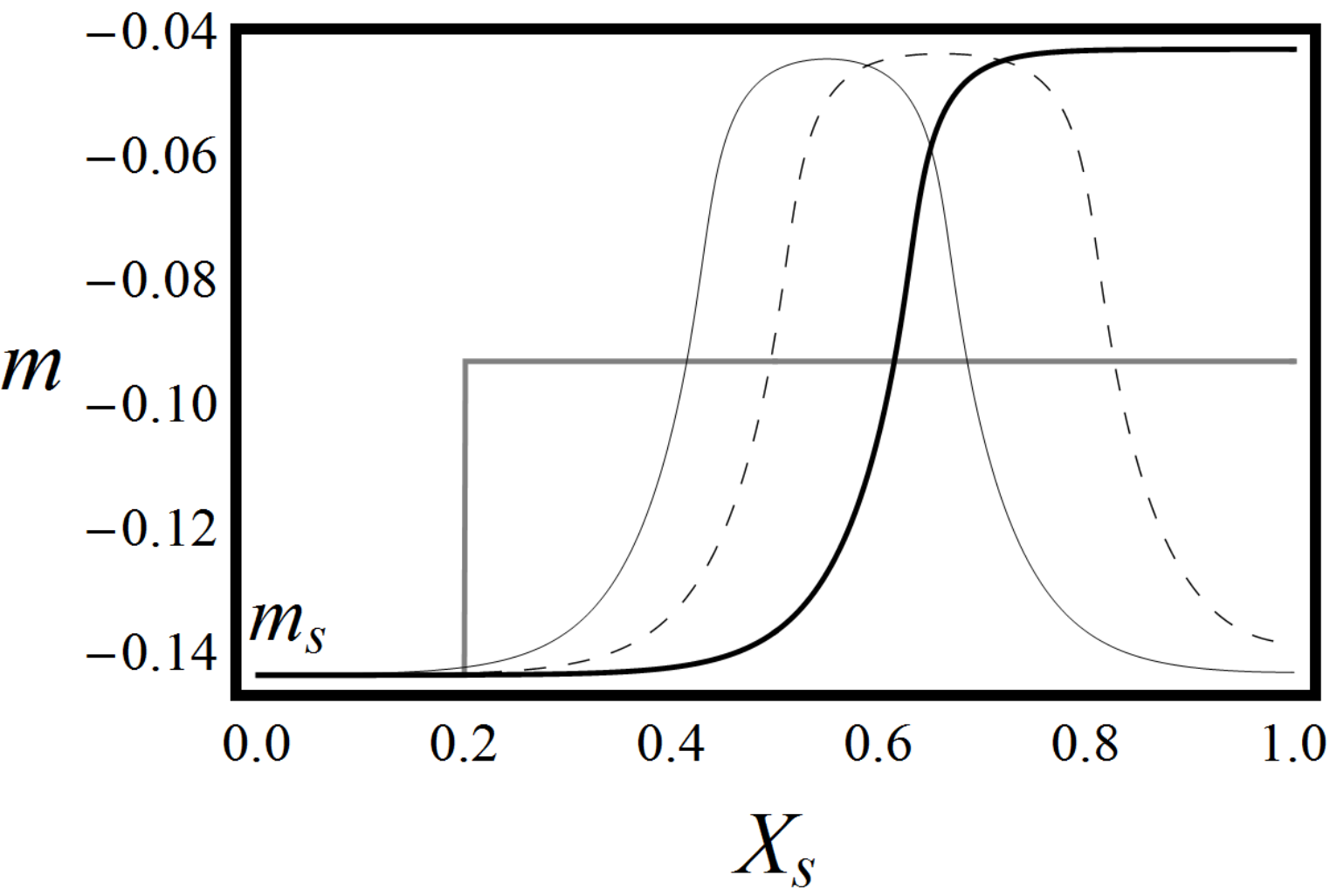}}}
}
\put(310,350)
{
\resizebox{5cm}{!}{\rotatebox{0}{\includegraphics{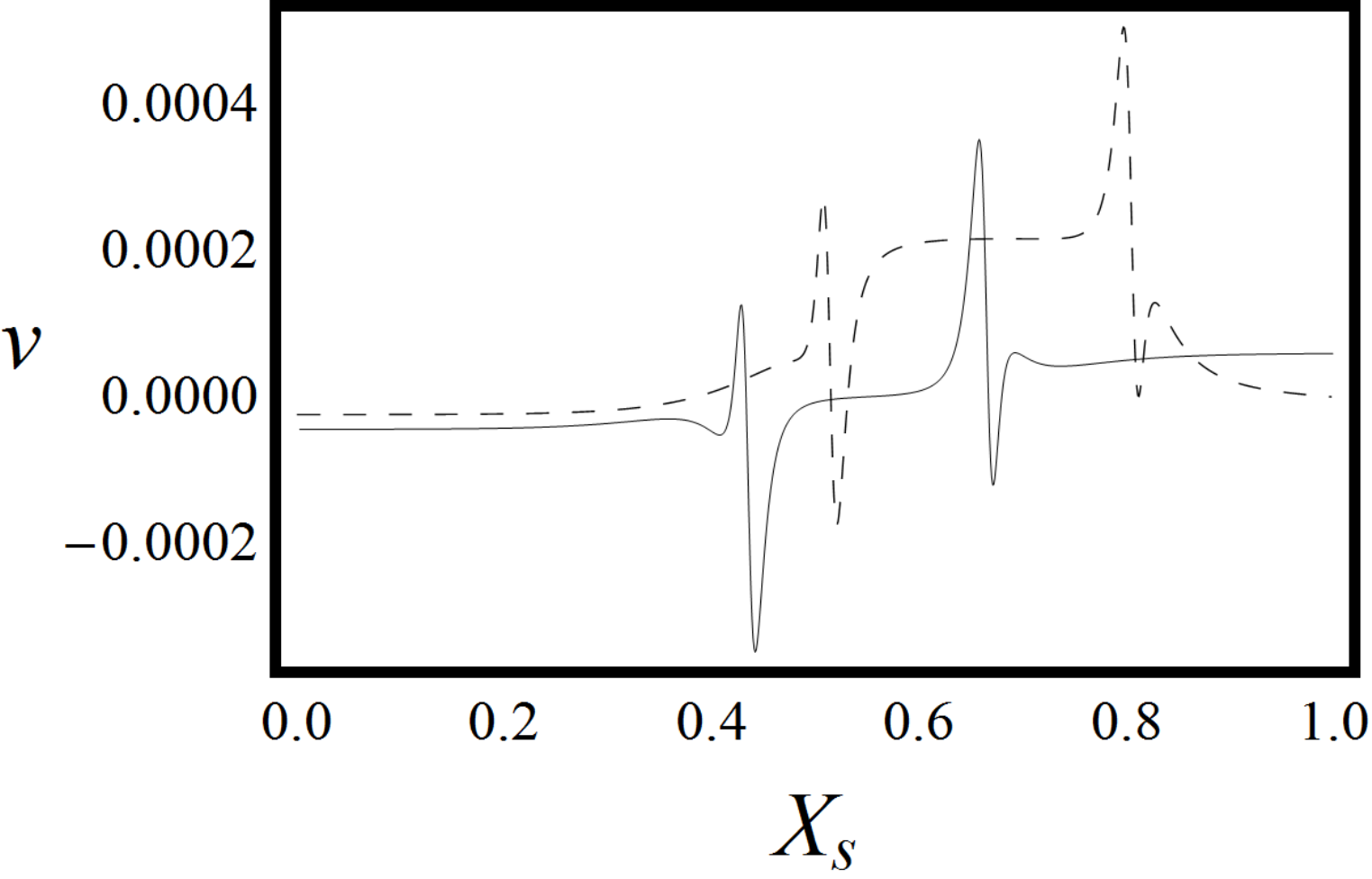}}}
}
\put(-40,250)
{
\resizebox{5cm}{!}{\rotatebox{0}{\includegraphics{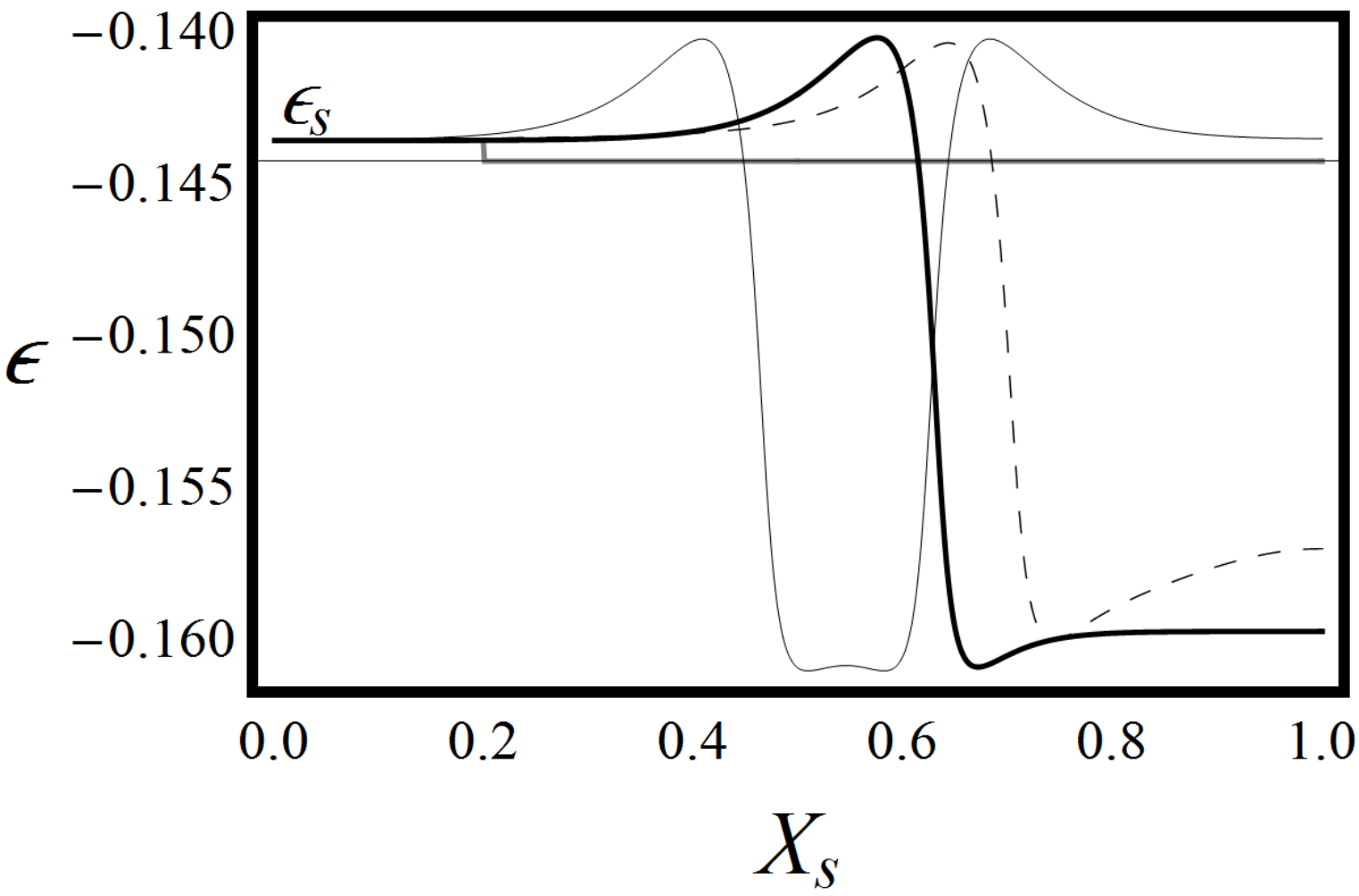}}}
}
\put(135,250)
{
\resizebox{5cm}{!}{\rotatebox{0}{\includegraphics{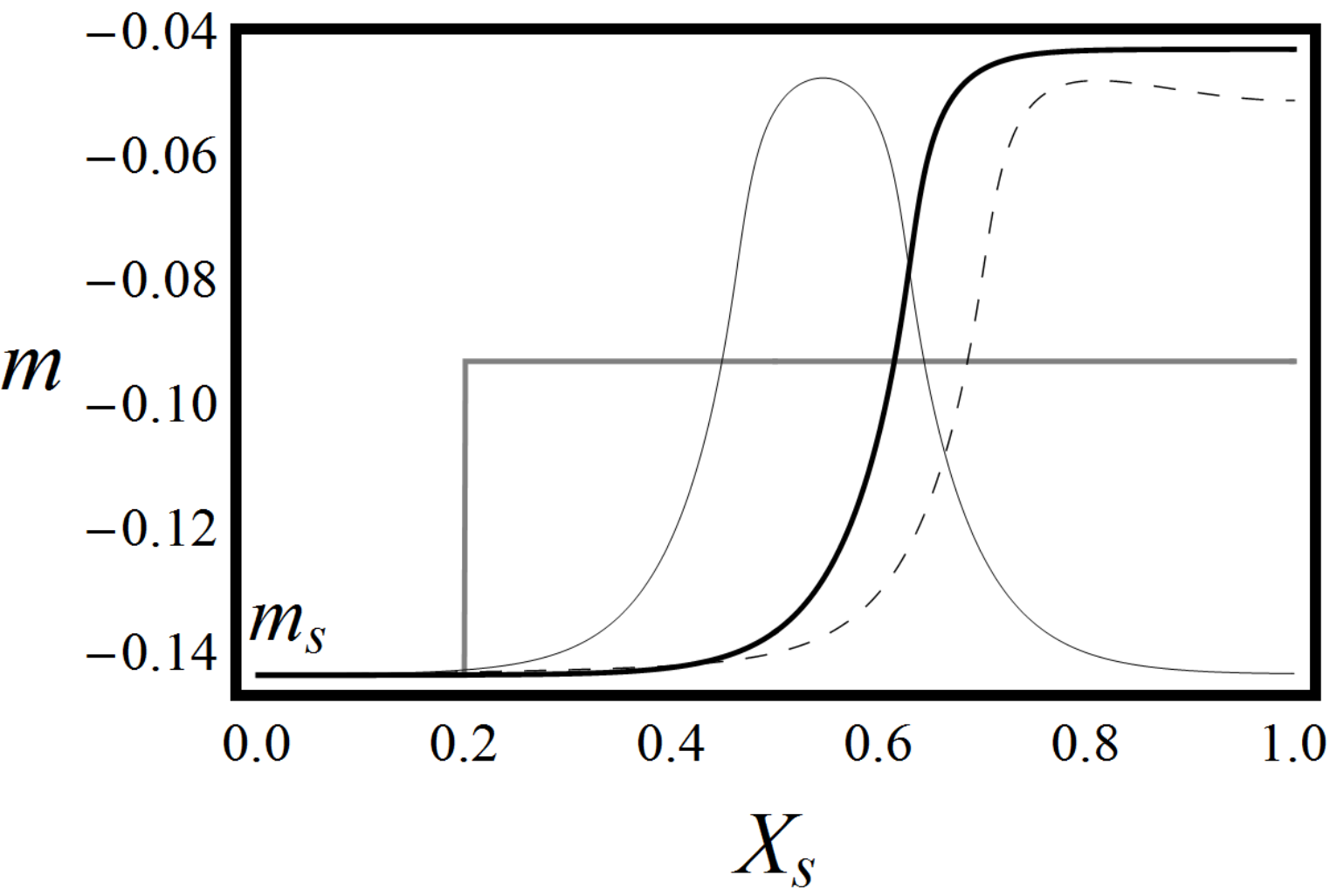}}}
}
\put(310,250)
{
\resizebox{5cm}{!}{\rotatebox{0}{\includegraphics{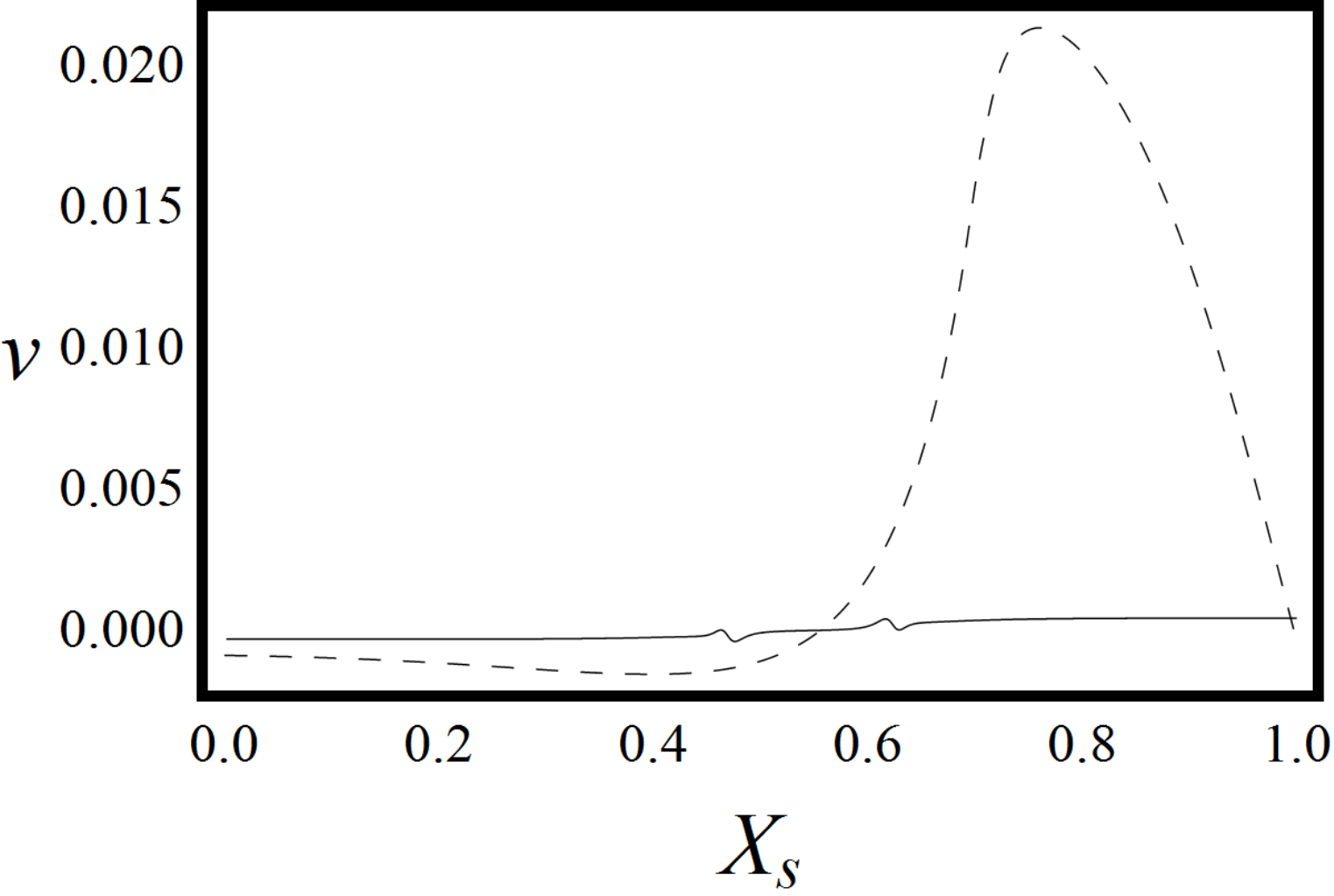}}}
}
\put(-40,150)
{
\resizebox{5cm}{!}{\rotatebox{0}{\includegraphics{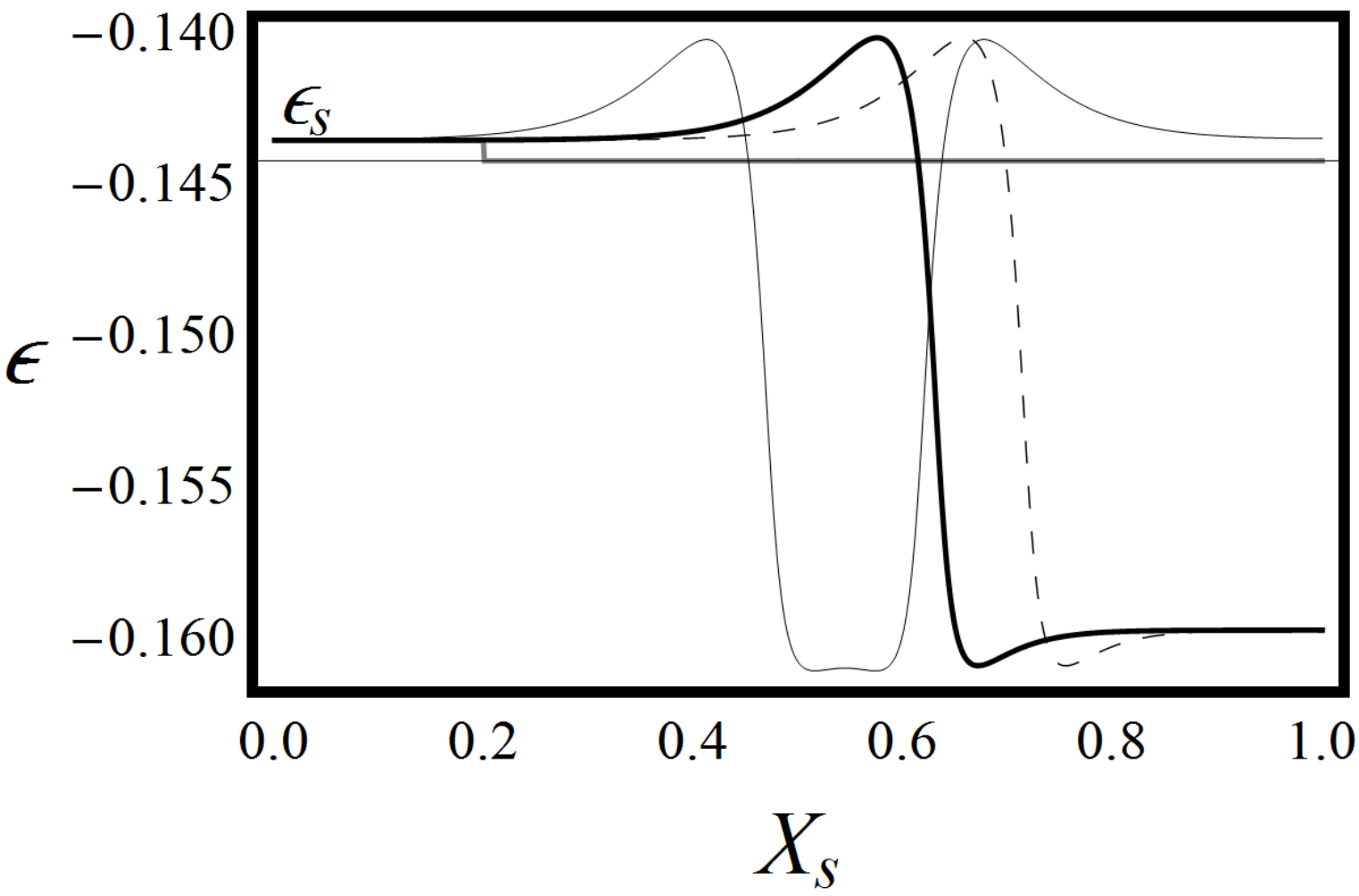}}}
}
\put(135,150)
{
\resizebox{5cm}{!}{\rotatebox{0}{\includegraphics{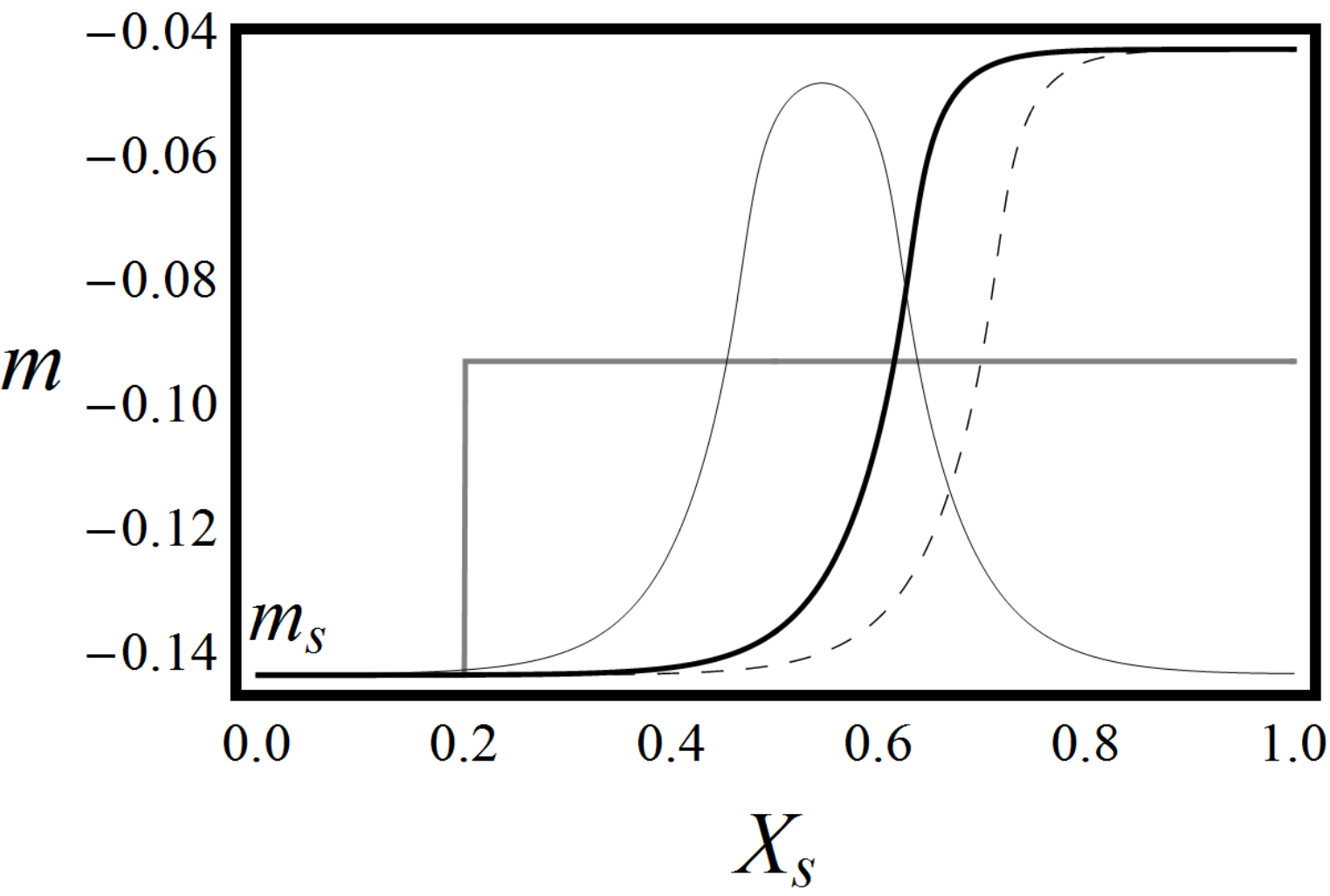}}}
}
\put(310,150)
{
\resizebox{5cm}{!}{\rotatebox{0}{\includegraphics{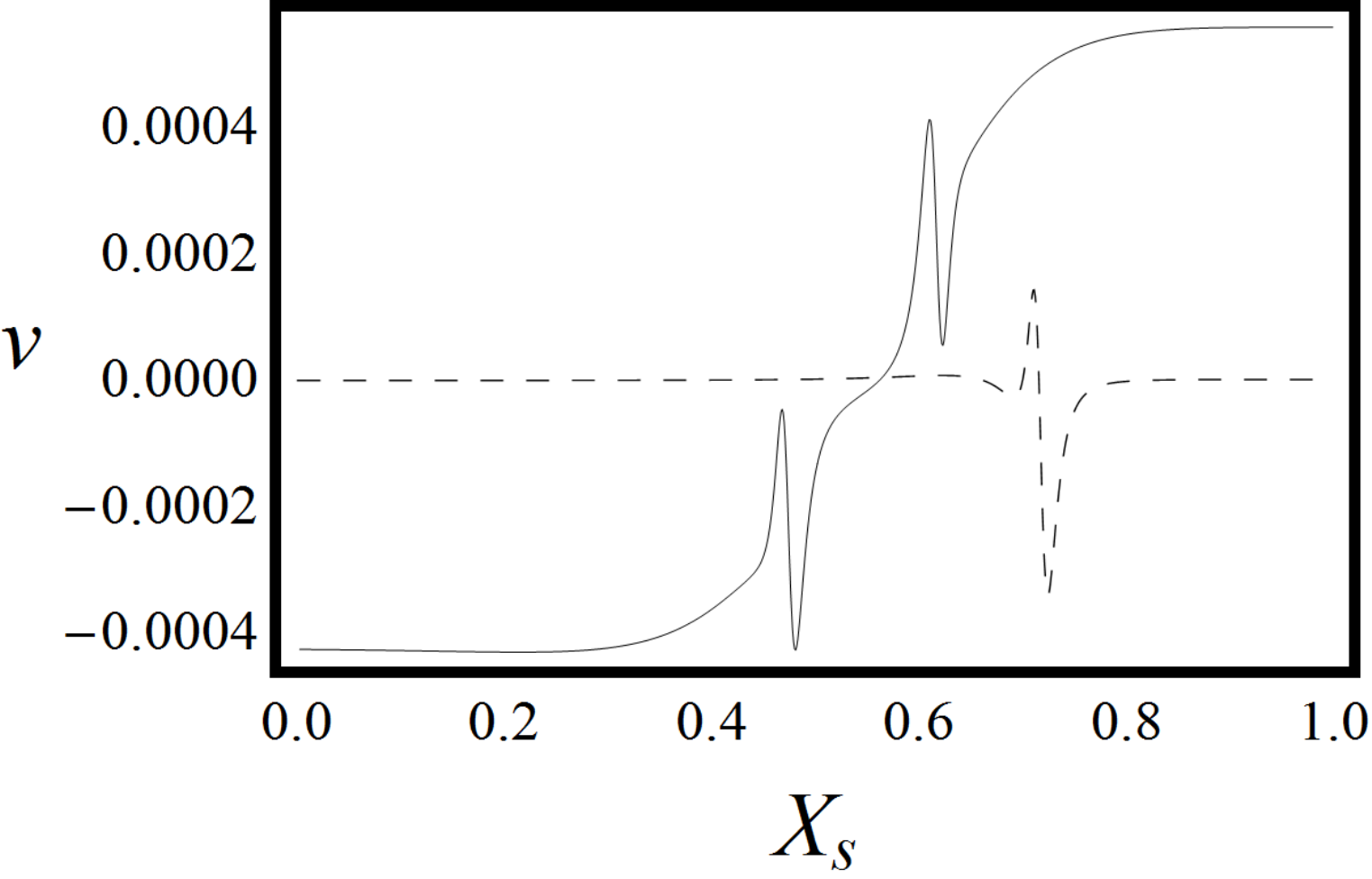}}}
}
\put(-40,50)
{
\resizebox{5cm}{!}{\rotatebox{0}{\includegraphics{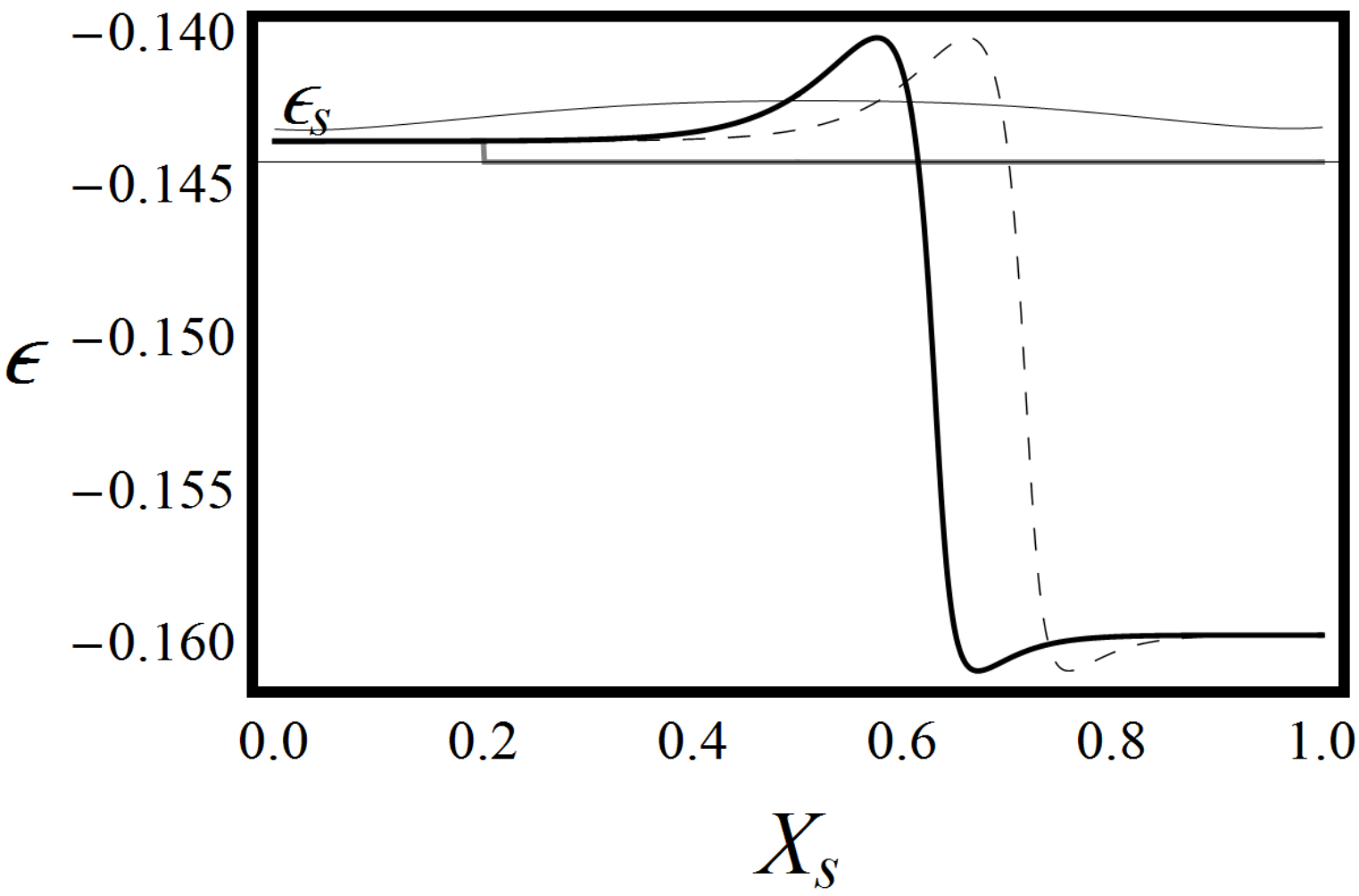}}}
}
\put(135,50)
{
\resizebox{5cm}{!}{\rotatebox{0}{\includegraphics{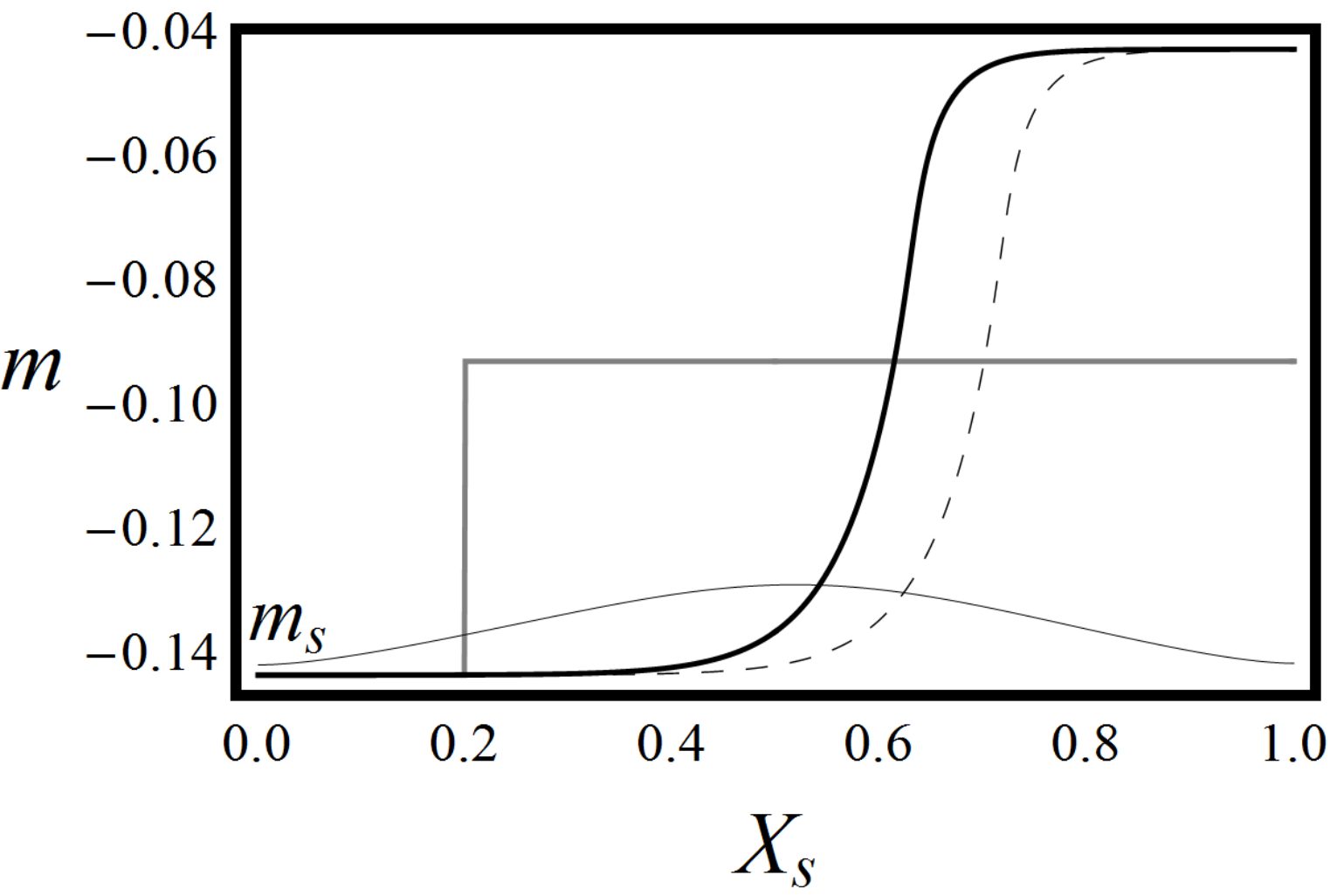}}}
}
\put(310,50)
{
\resizebox{5cm}{!}{\rotatebox{0}{\includegraphics{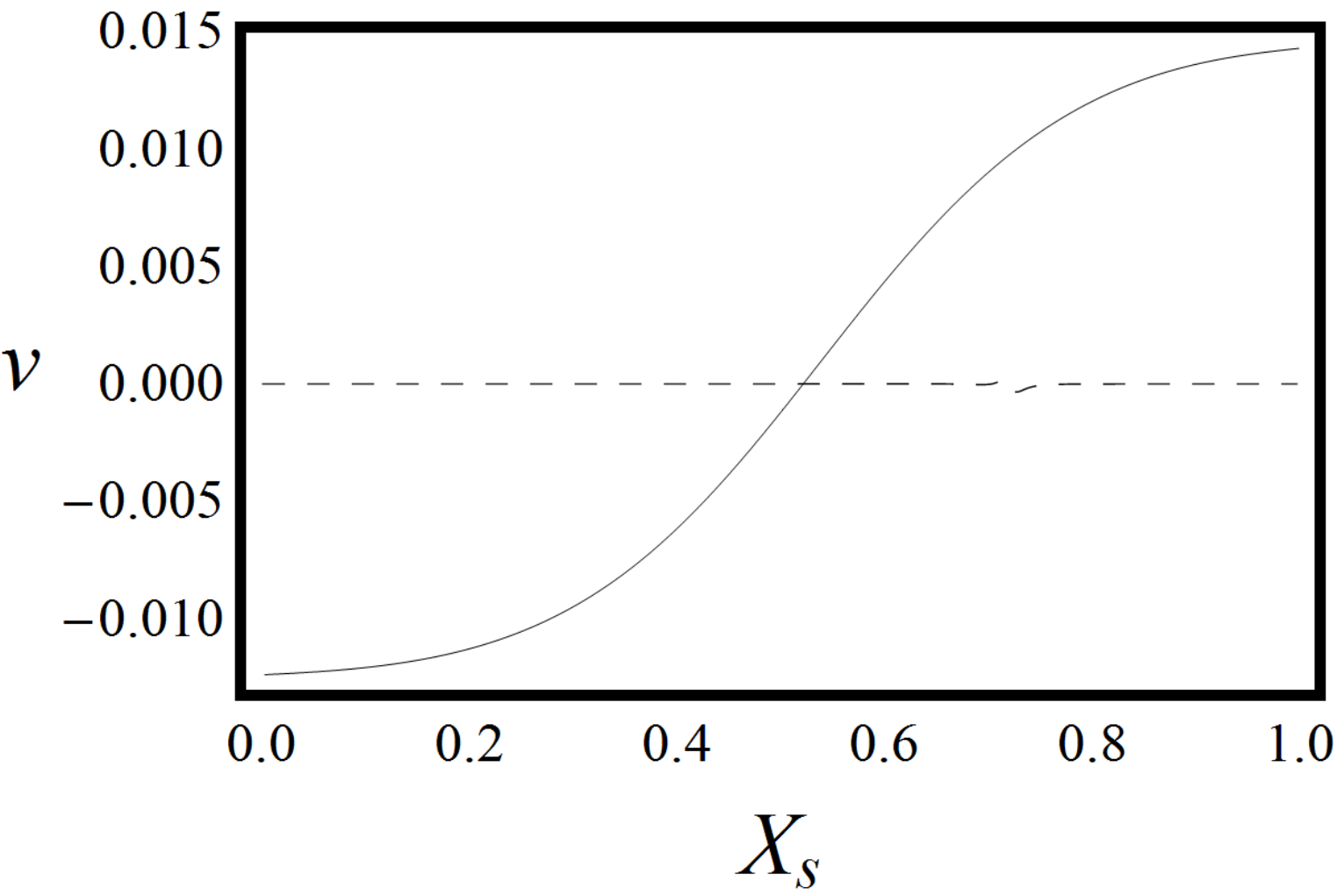}}}
}
\end{picture}
\vskip -1.5 cm
\centering
\caption{Solutions $\varepsilon(X_s,t)$, $m(X_s)$ and $v(X_s)$ for the zero chemical potential problem (thin black lines) and the one--side impermeable problem (dotted thin lines) obtained by solving 
Problem (\ref{problema-d}) and (\ref{problema-d-osi}) respectively, with the second initial condition (gray lines). We used Neumann boundary conditions $m'(0)=\varepsilon'(0)=m'(0)=m'(1)=0$ on the finite interval $[0,1]$, at the coexistence pressure for $a=0.5,\,b=1,\,\alpha=100,\,k_1=k_2=k_3=10^{-3}$. 
Profiles at times $t=0.201,\,t=2.245,\,t=5,\,t=21,\,t=22,\,t=25$ are in lexicographic order. The solid black lines represent interface--type stationary profiles.}
\label{dinamica2}
\end{figure}

\begin{figure}[h!]
\vskip 2cm
\begin{picture}(200,400)(15,0)
\put(-40,360)
{
\resizebox{5.5cm}{!}{\rotatebox{0}{\includegraphics{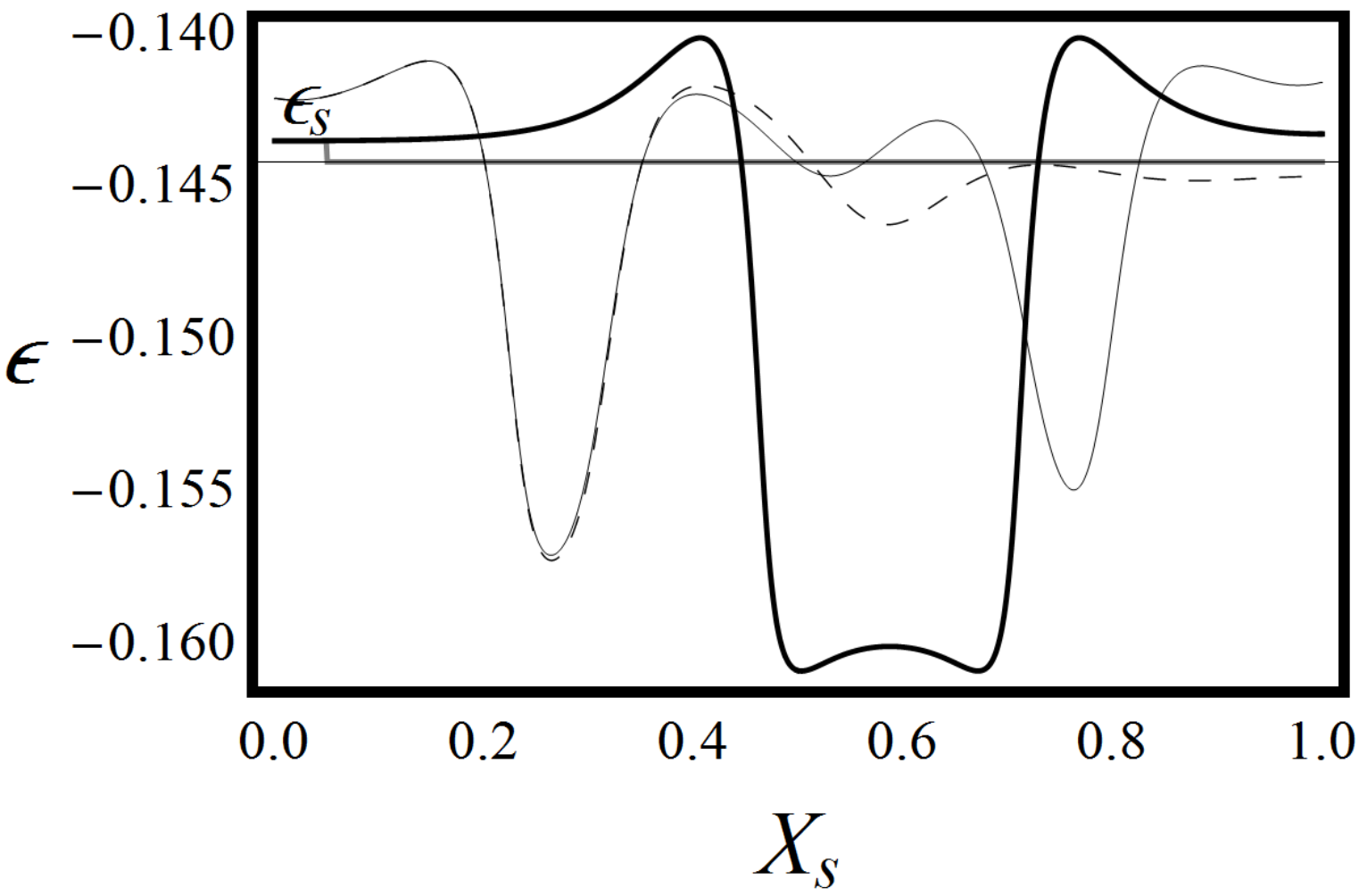}}}
}
\put(135,360)
{
\resizebox{5.5cm}{!}{\rotatebox{0}{\includegraphics{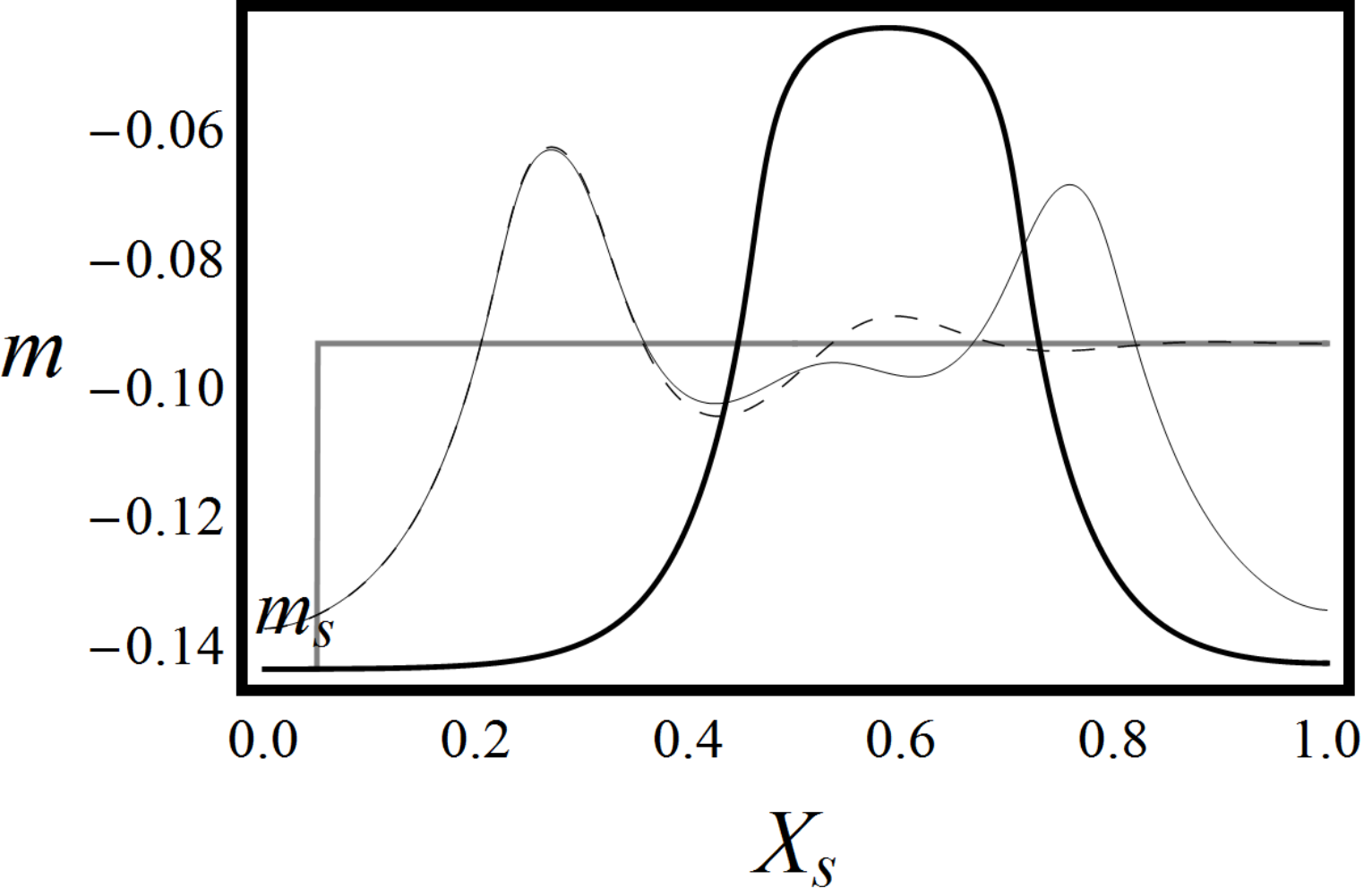}}}
}
\put(310,360)
{
\resizebox{5.5cm}{!}{\rotatebox{0}{\includegraphics{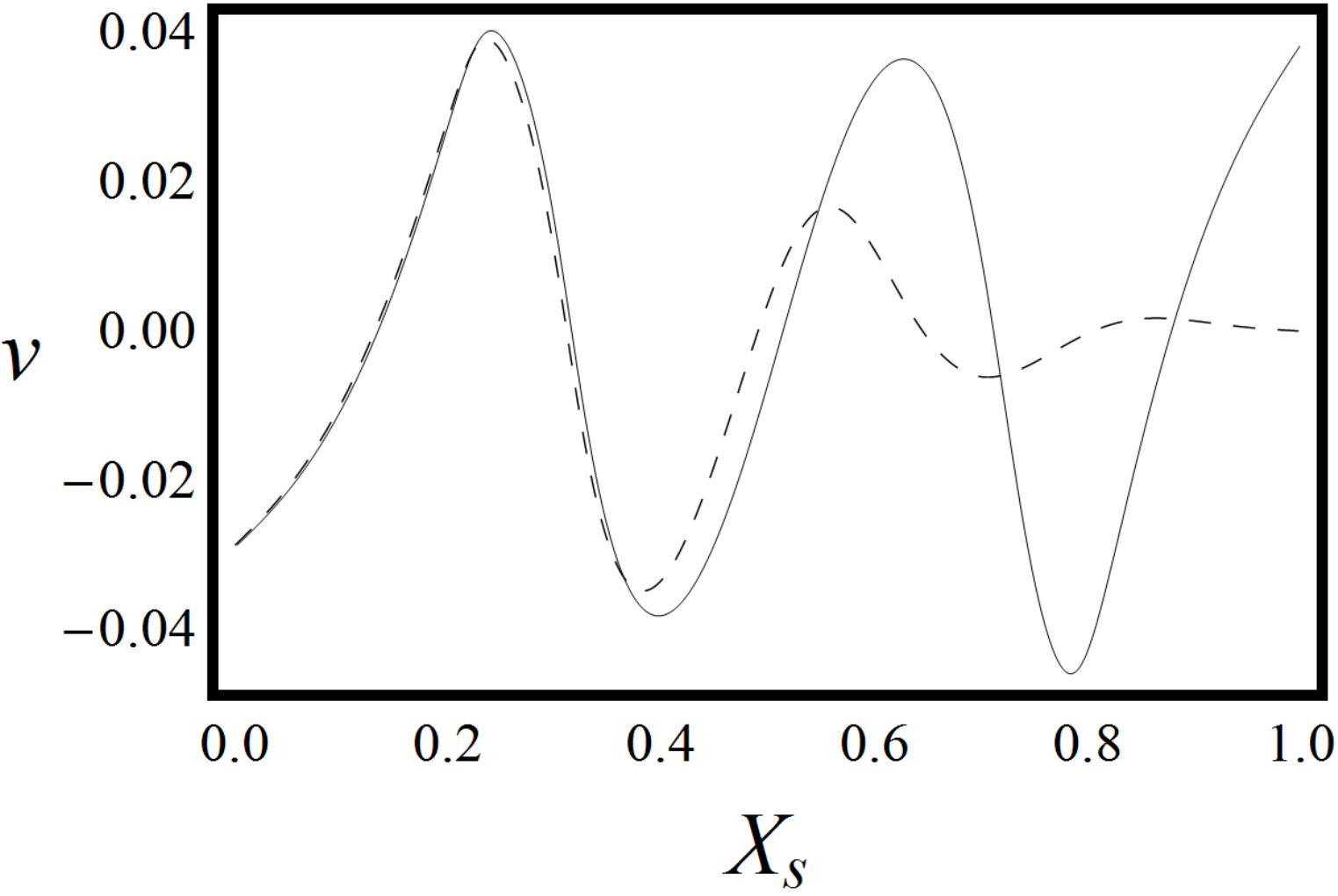}}}
}
\put(-40,240)
{
\resizebox{5.5cm}{!}{\rotatebox{0}{\includegraphics{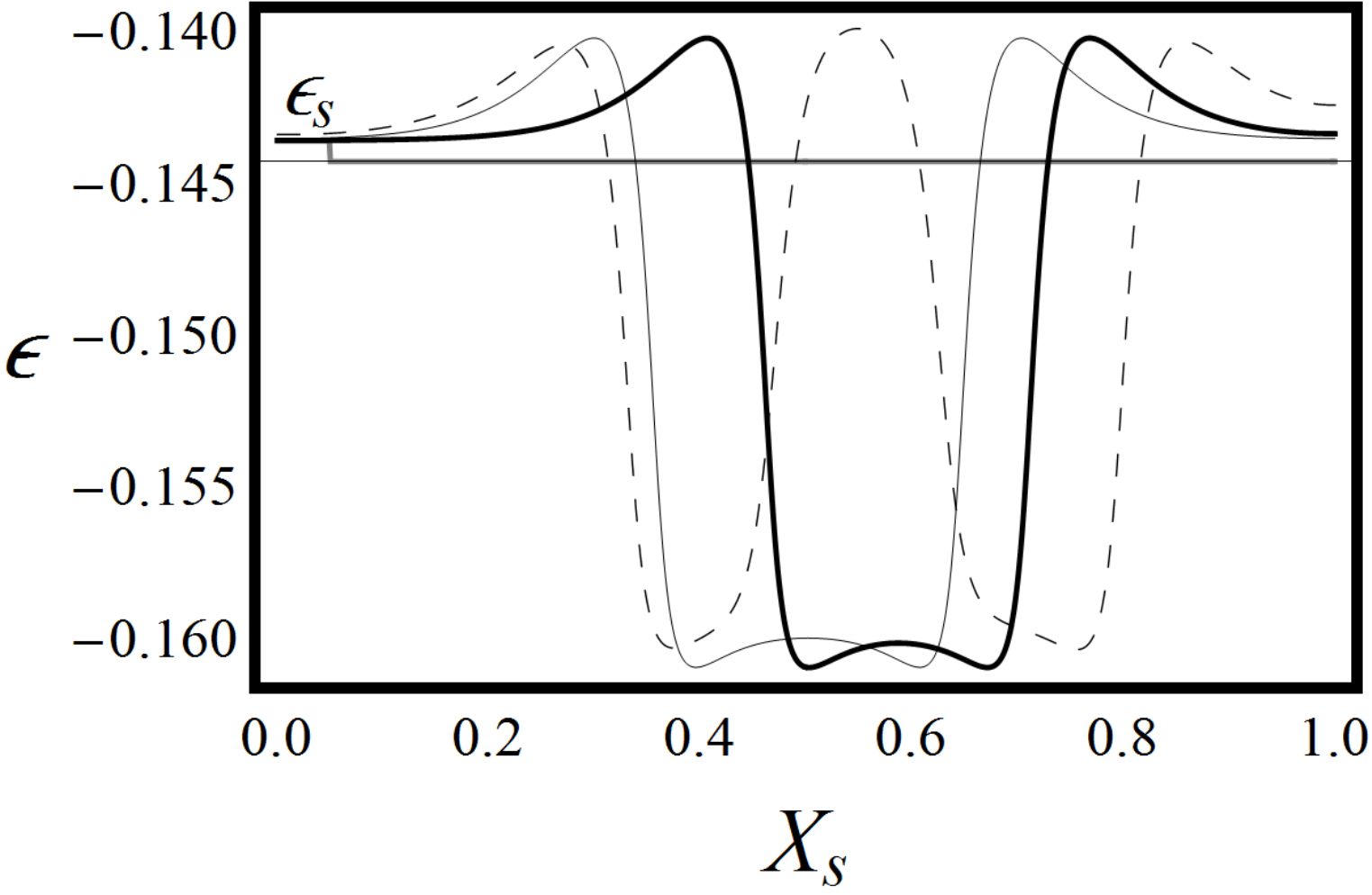}}}
}
\put(135,240)
{
\resizebox{5.5cm}{!}{\rotatebox{0}{\includegraphics{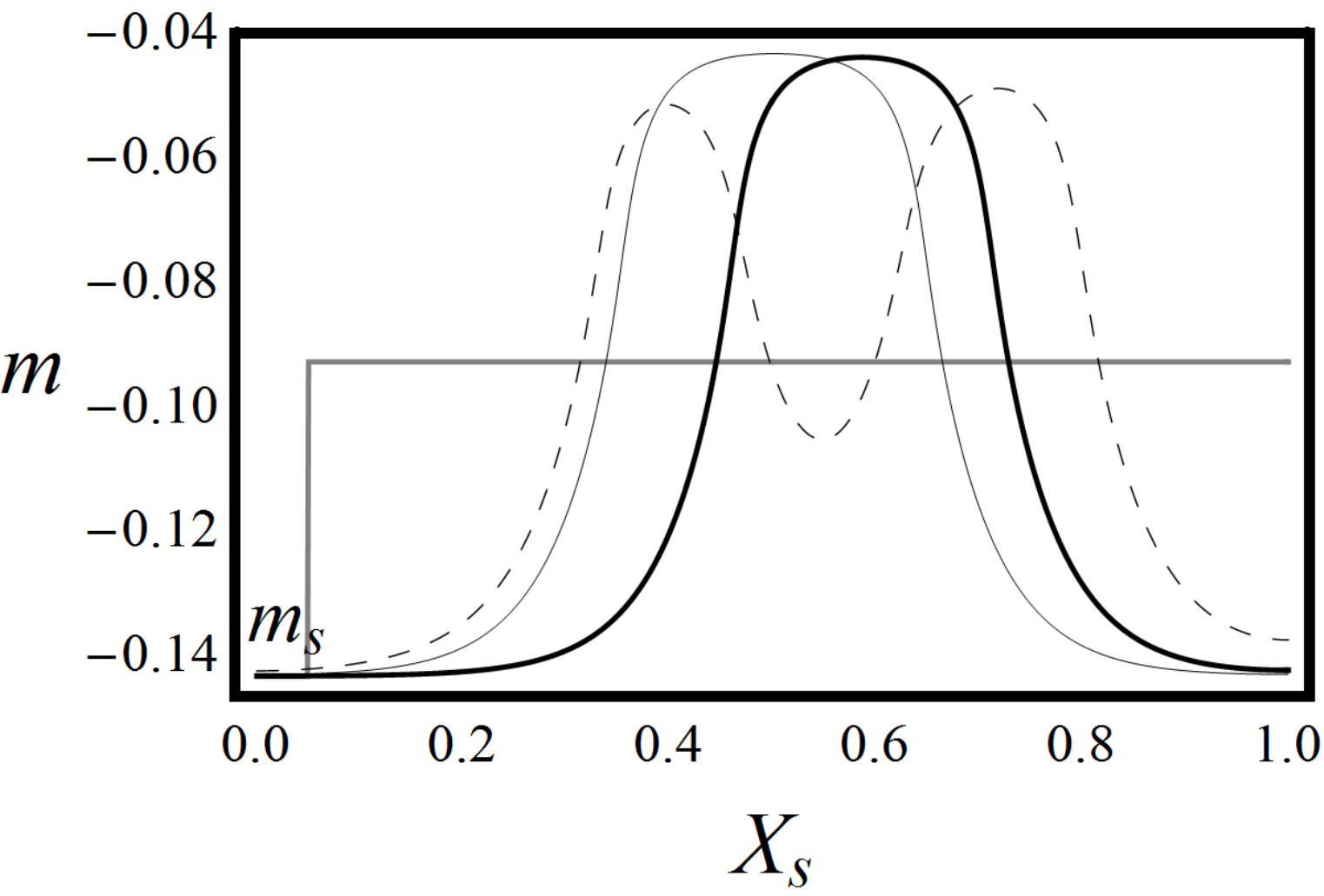}}}
}
\put(310,240)
{
\resizebox{5.5cm}{!}{\rotatebox{0}{\includegraphics{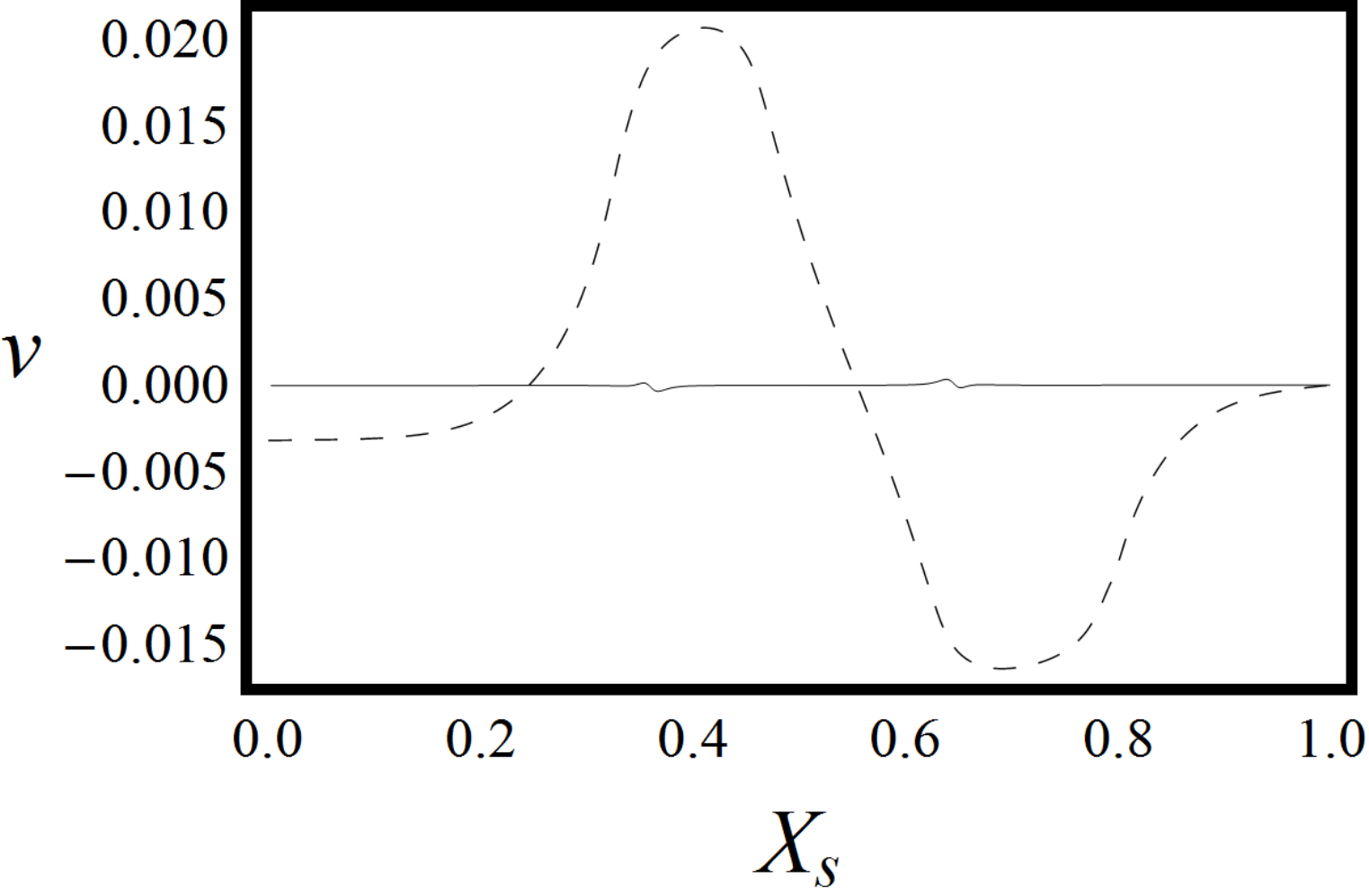}}}
}
\put(-40,120)
{
\resizebox{5.5cm}{!}{\rotatebox{0}{\includegraphics{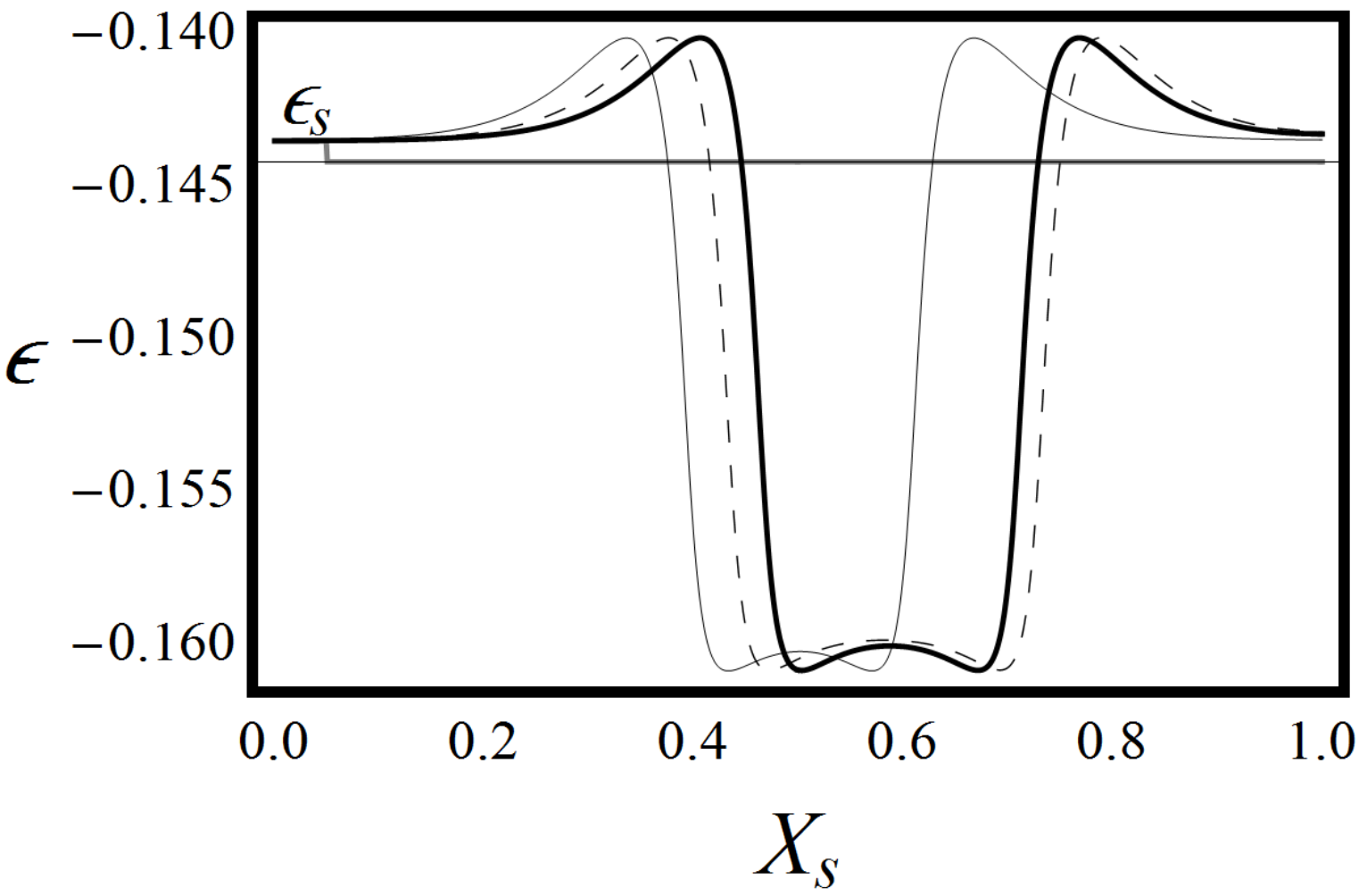}}}
}
\put(135,120)
{
\resizebox{5.5cm}{!}{\rotatebox{0}{\includegraphics{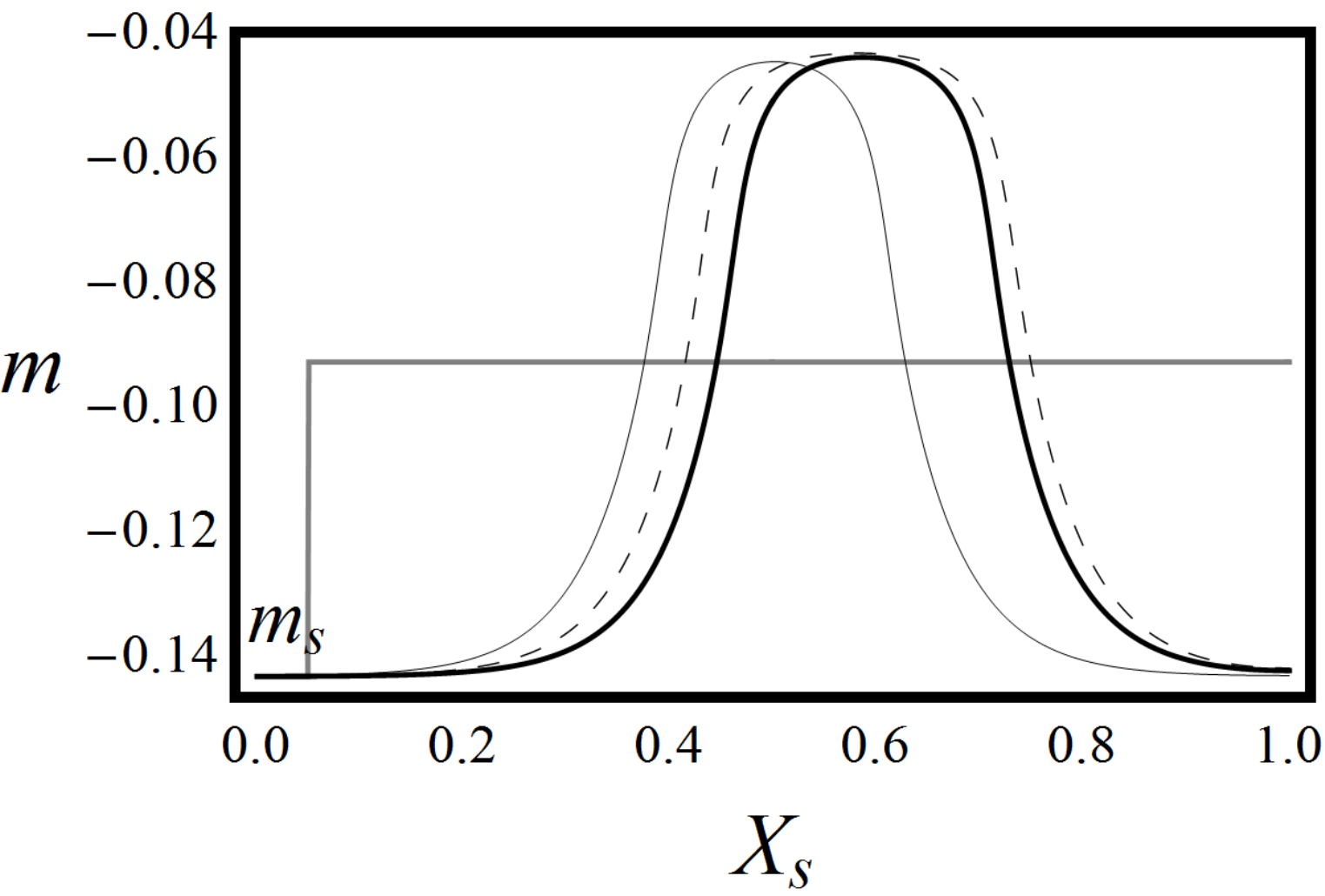}}}
}
\put(310,120)
{
\resizebox{5.5cm}{!}{\rotatebox{0}{\includegraphics{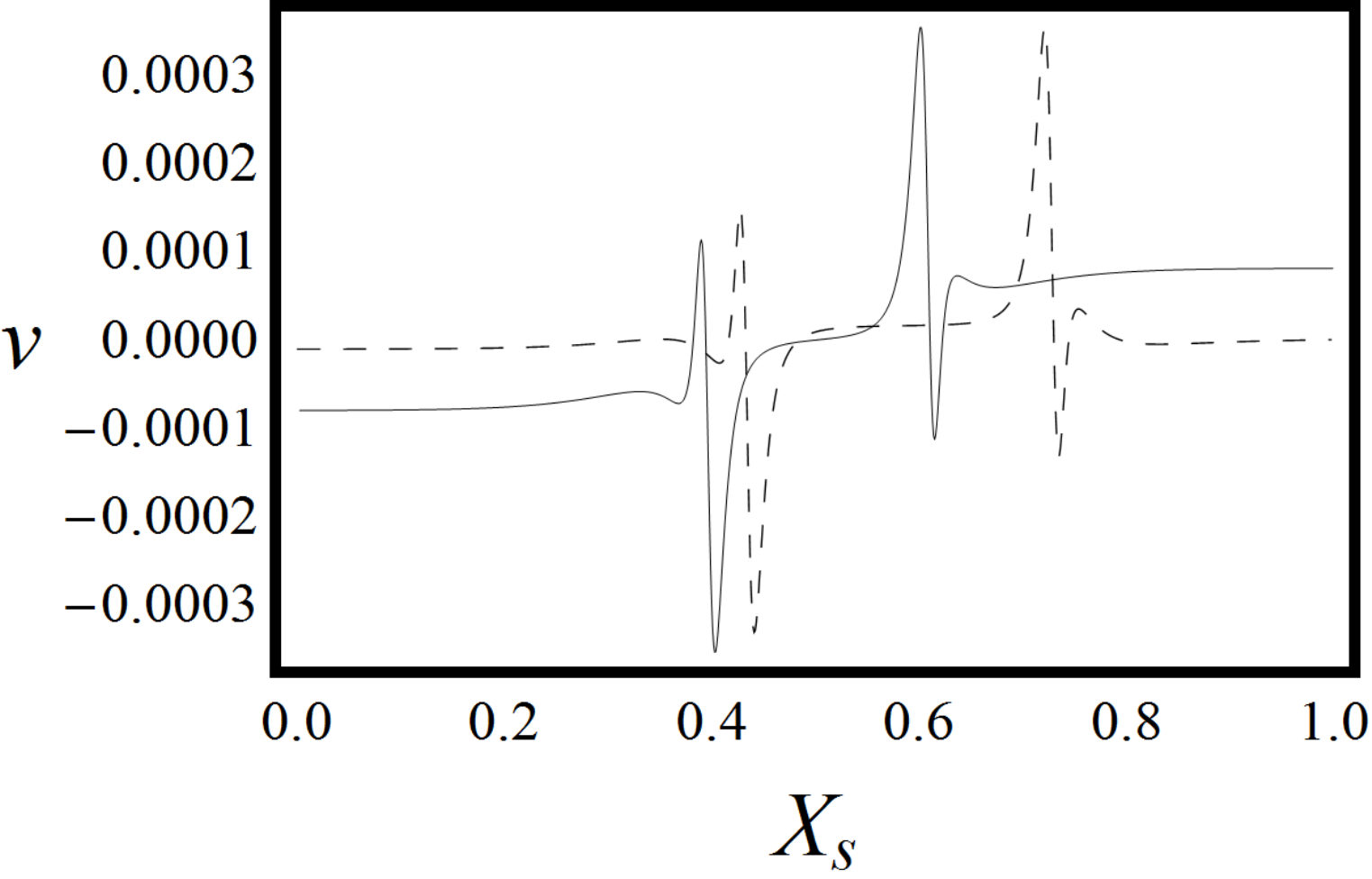}}}
}
\put(-40,0)
{
\resizebox{5.5cm}{!}{\rotatebox{0}{\includegraphics{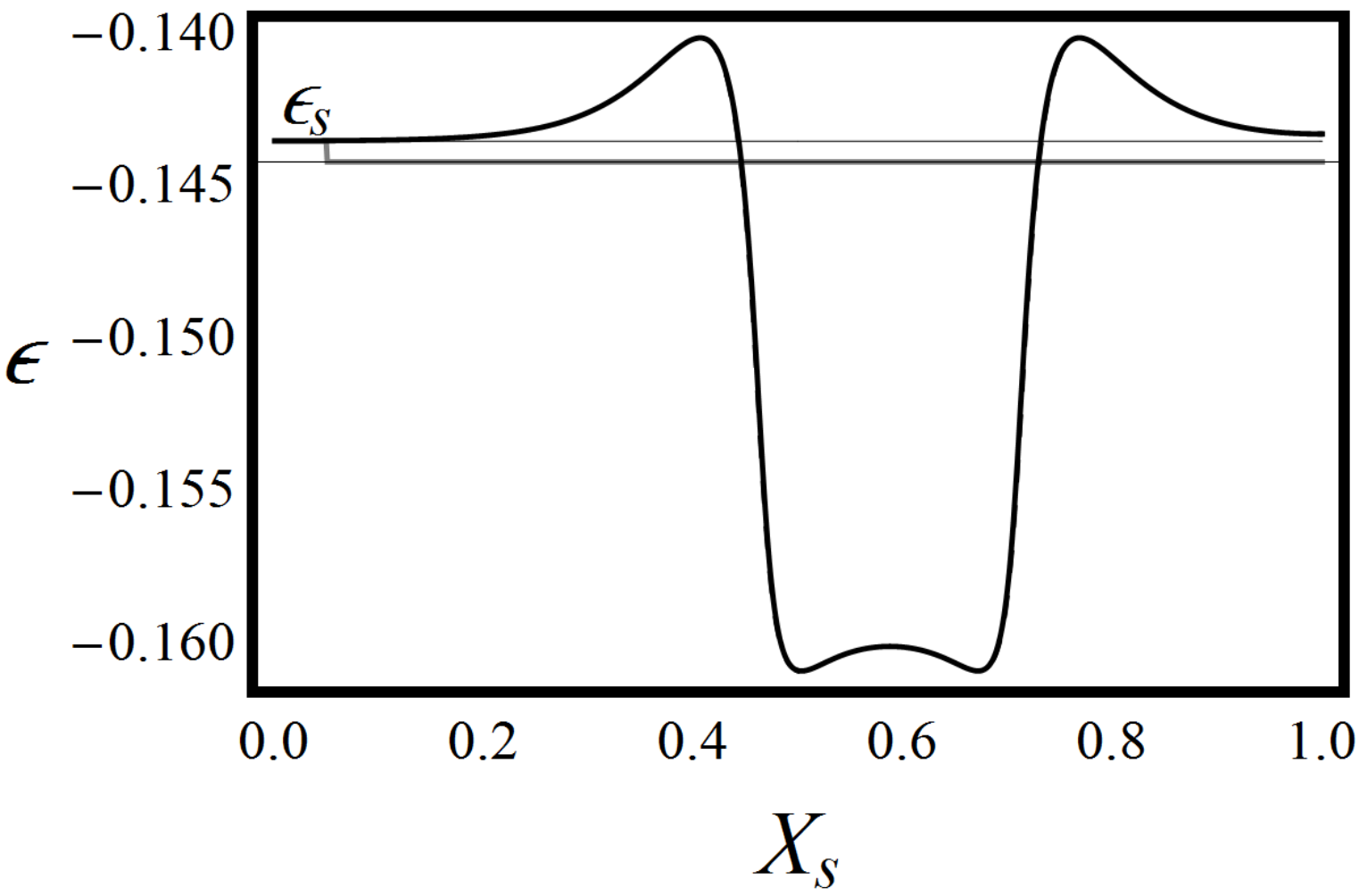}}}
}
\put(135,0)
{
\resizebox{5.5cm}{!}{\rotatebox{0}{\includegraphics{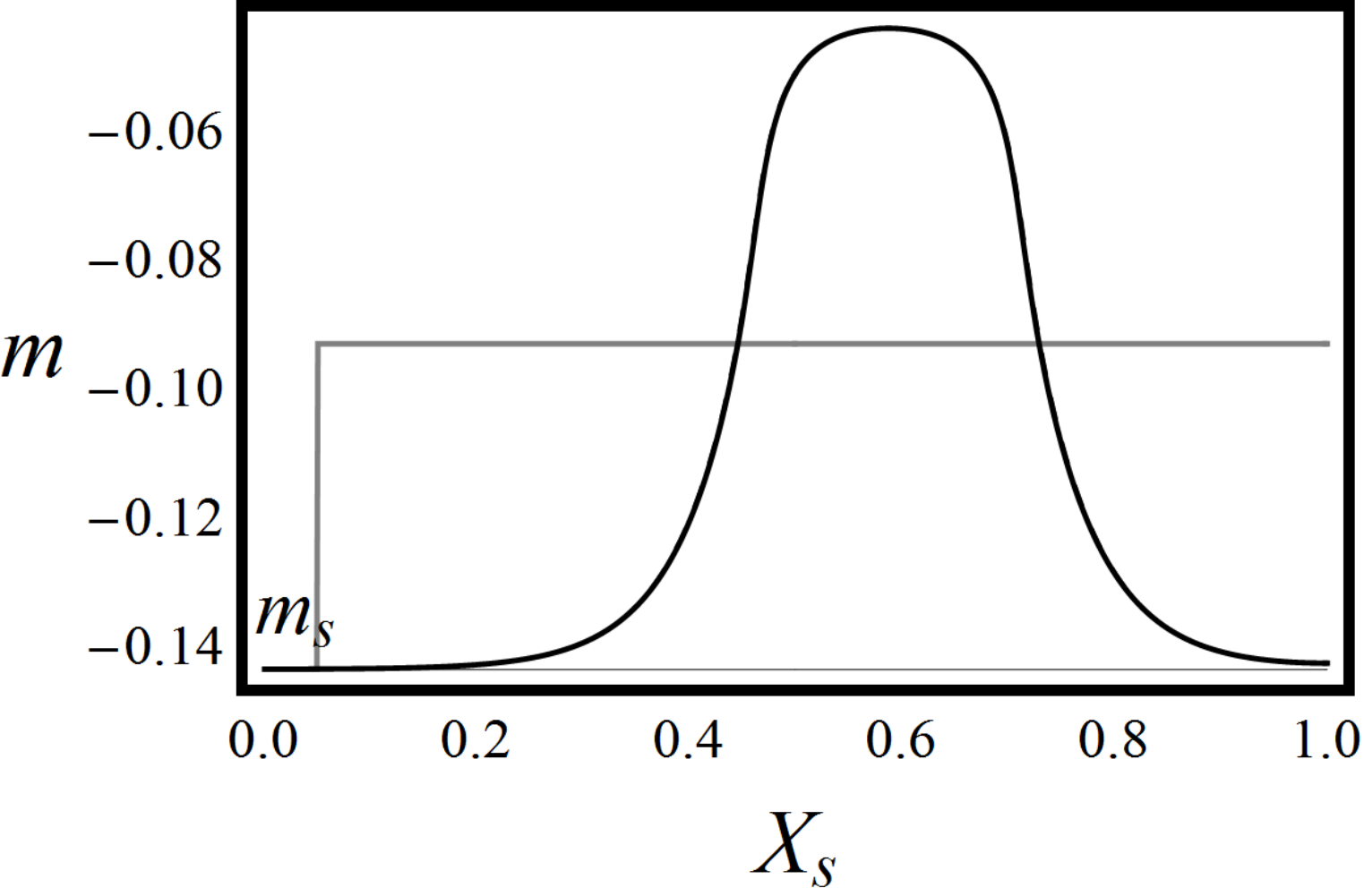}}}
}
\put(310,0)
{
\resizebox{5.5cm}{!}{\rotatebox{0}{\includegraphics{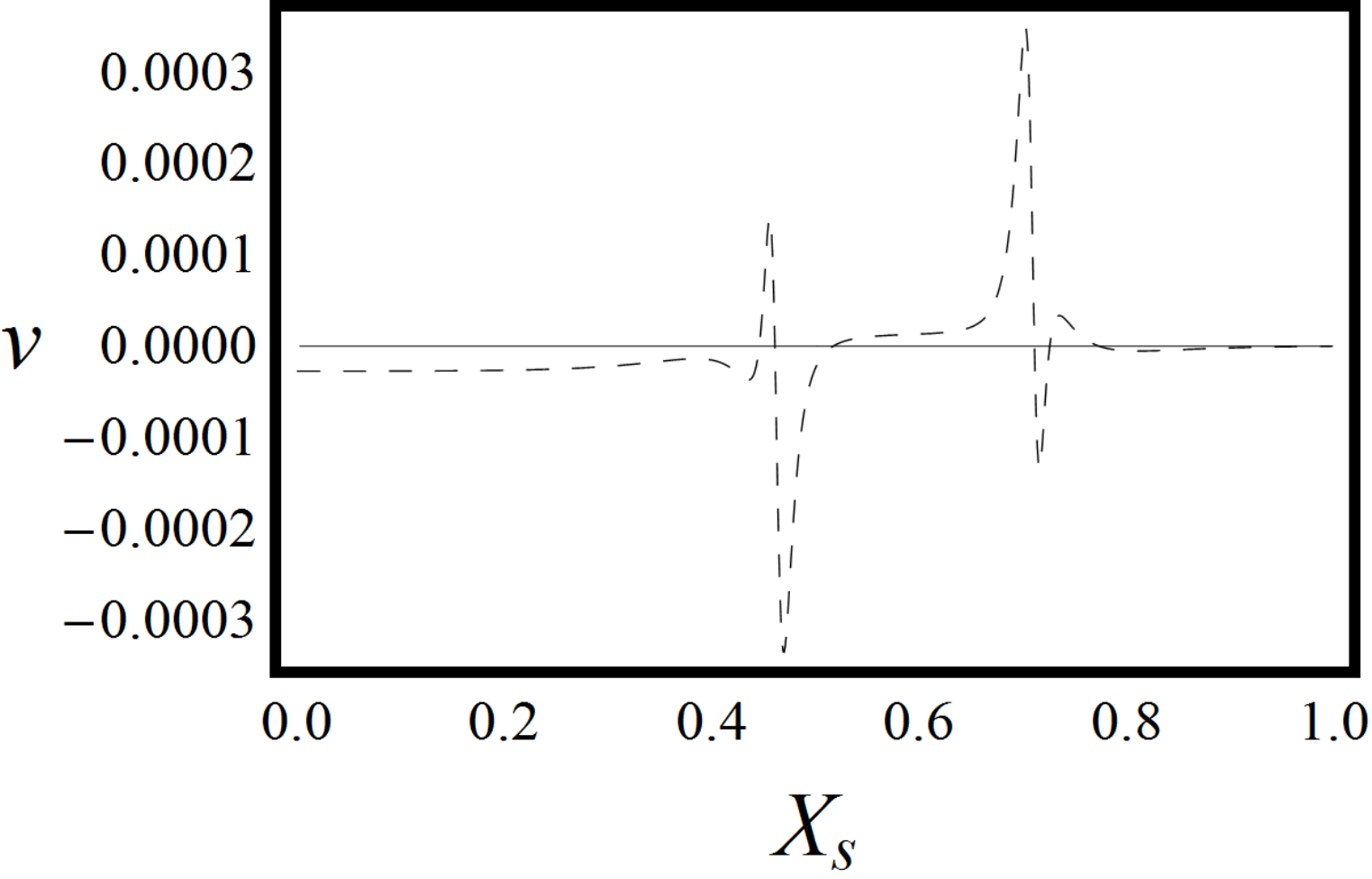}}}
}
\end{picture}
\vskip 1. cm
\centering
\caption{Solutions $\varepsilon(X_s,t)$, $m(X_s)$ and $v(X_s)$ for the zero chemical potential problem (thin black lines) and the one--side impermeable problem (dotted thin lines) obtained by solving 
Problem (\ref{problema-d}) and (\ref{problema-d-osi}) respectively, with the third initial condition (gray lines). We used Neumann boundary conditions $m'(0)=\varepsilon'(0)=m'(0)=m'(1)=0$ on the finite interval $[0,1]$, at the coexistence pressure for $a=0.5,\,b=1,\,\alpha=100,\,k_1=k_2=k_3=10^{-3}$. 
Profiles at times $t=0.052,\,t=1,\,t=28.333,\,t=50.038$ are in lexicographic order. The solid black lines represent double--interface--type stationary profiles.}
\label{dinamica3}
\end{figure}

\begin{figure}[h!]
\vskip 2cm
\begin{picture}(200,400)(15,0)
\put(-40,360)
{
\resizebox{5.5cm}{!}{\rotatebox{0}{\includegraphics{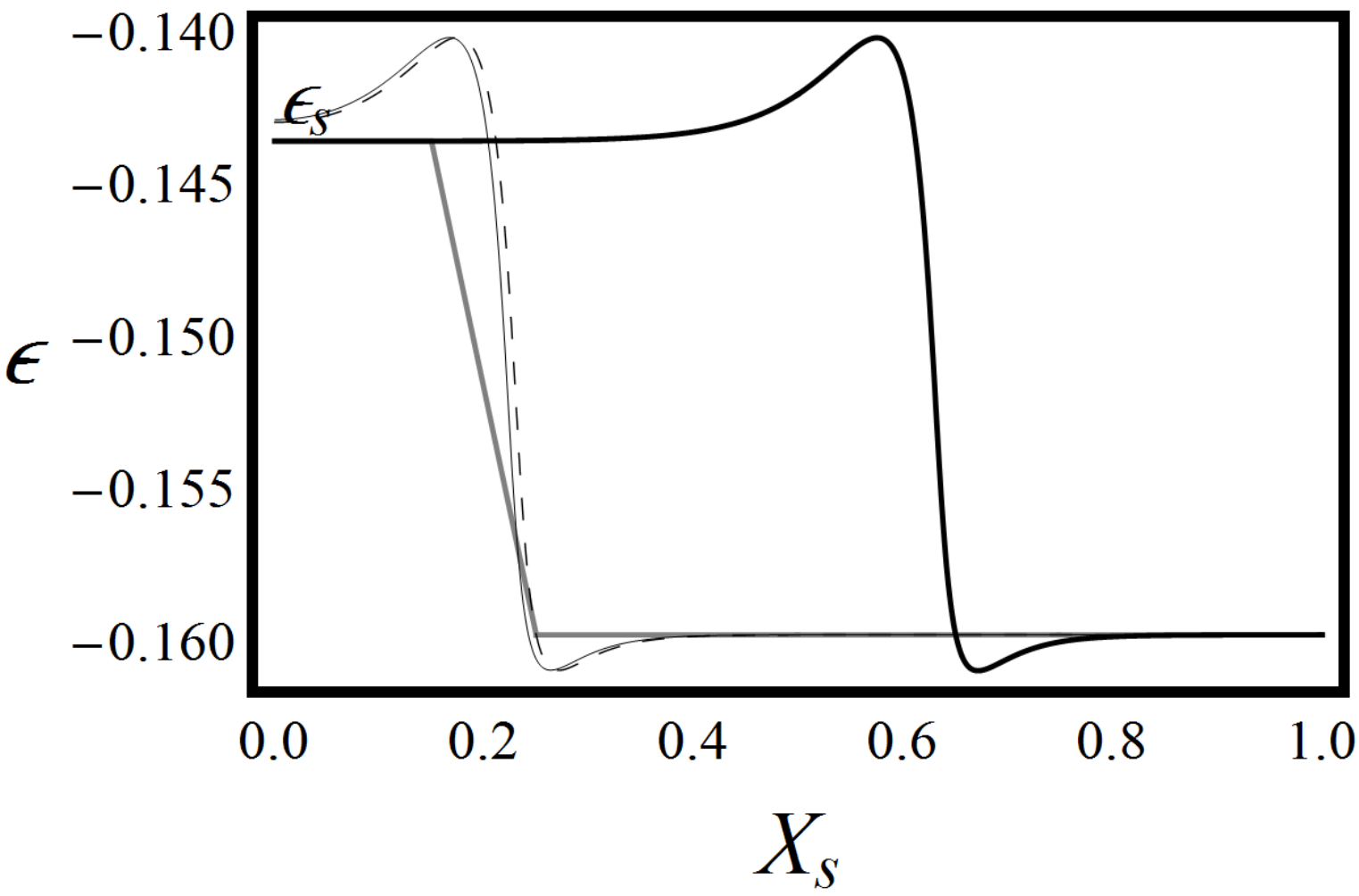}}}
}
\put(135,360)
{
\resizebox{5.5cm}{!}{\rotatebox{0}{\includegraphics{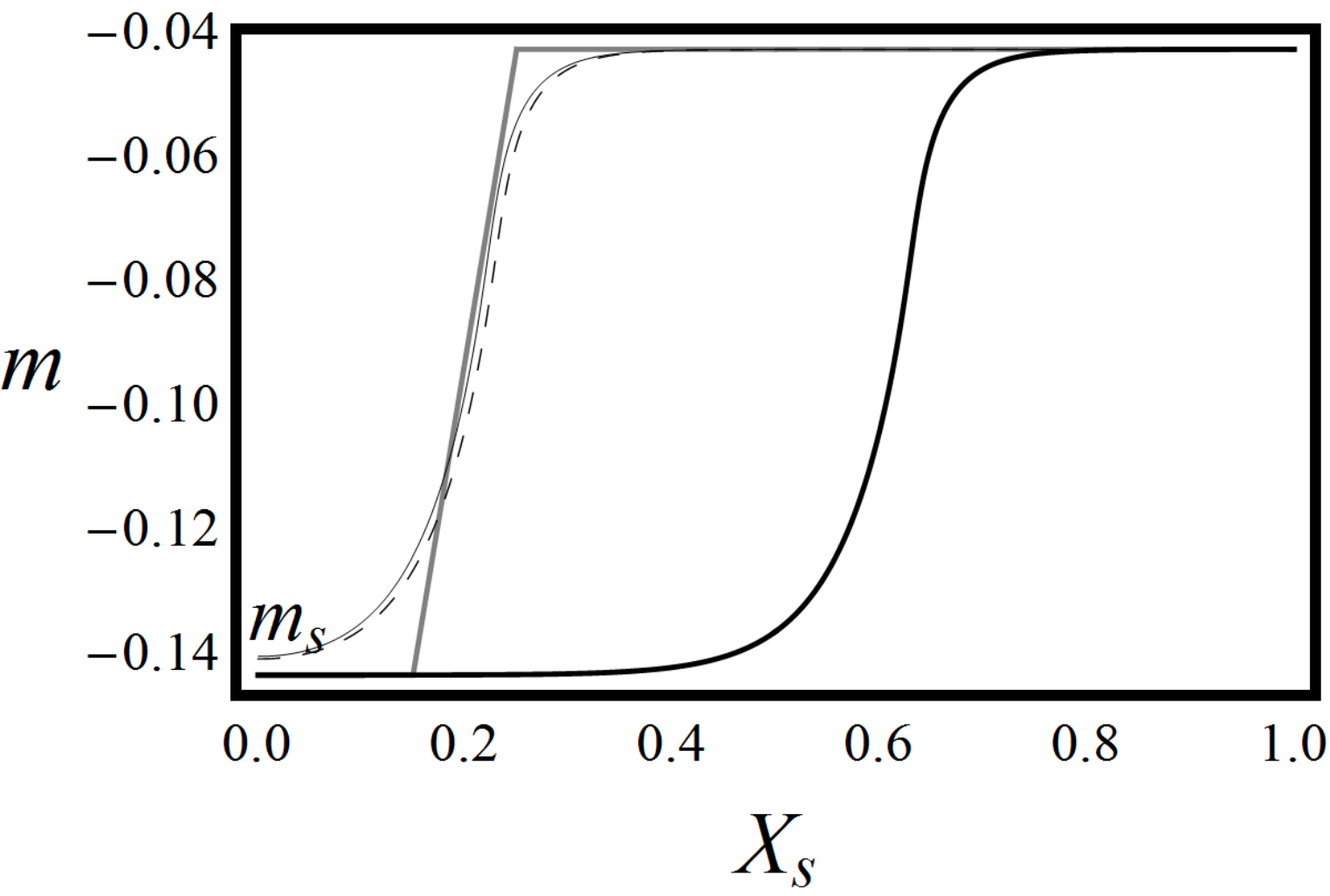}}}
}
\put(310,360)
{
\resizebox{5.5cm}{!}{\rotatebox{0}{\includegraphics{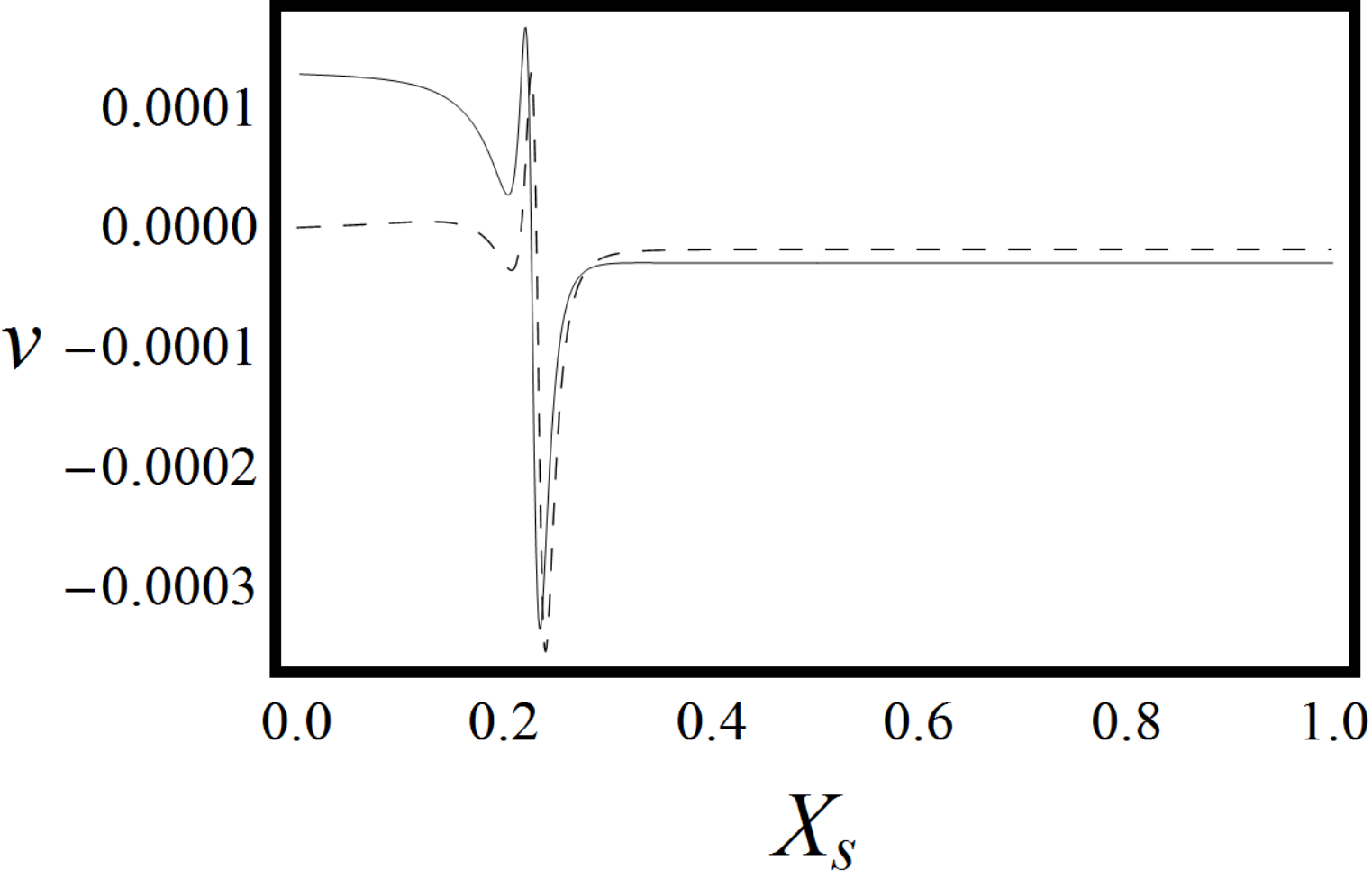}}}
}
\put(-40,240)
{
\resizebox{5.5cm}{!}{\rotatebox{0}{\includegraphics{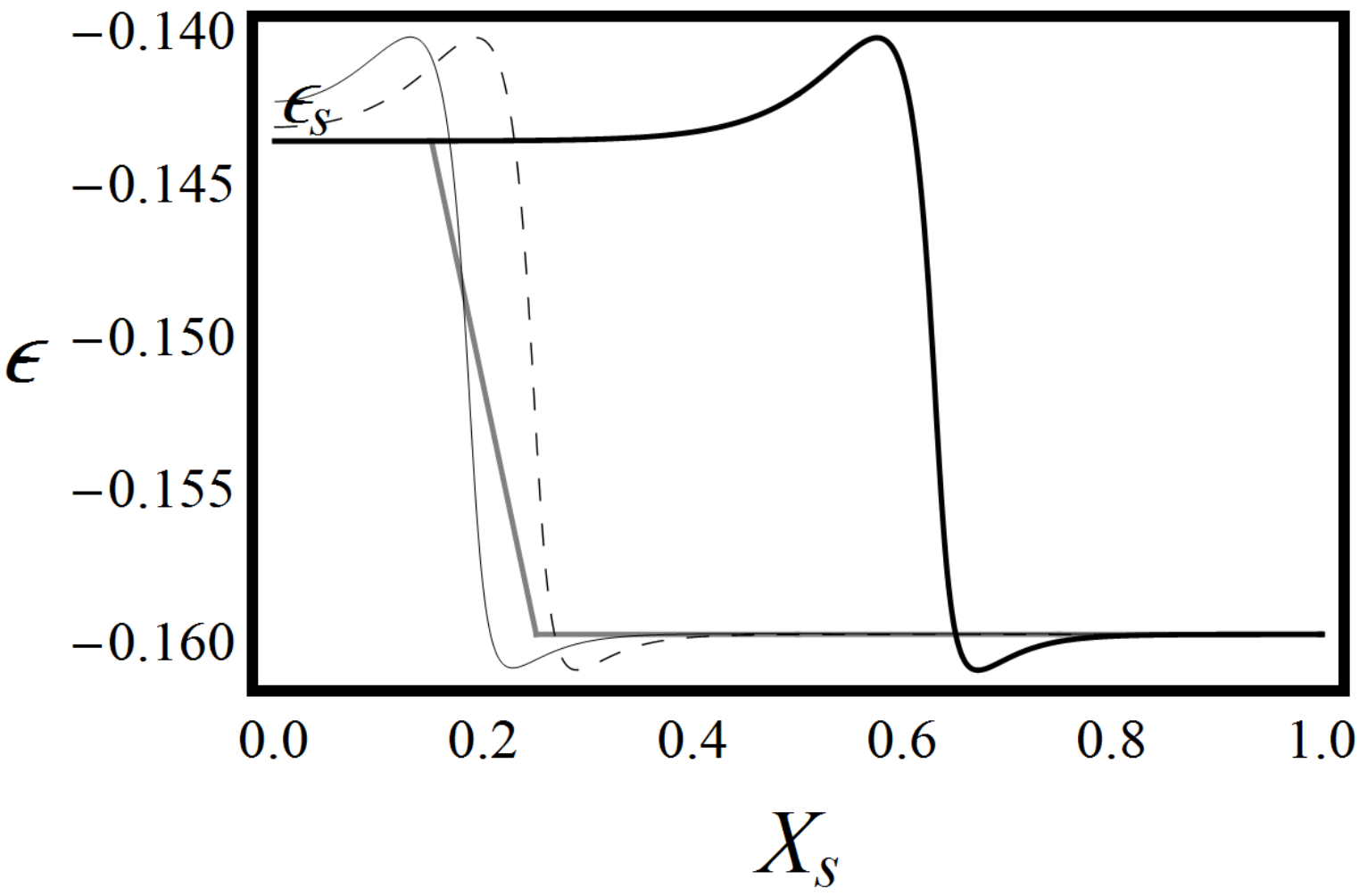}}}
}
\put(135,240)
{
\resizebox{5.5cm}{!}{\rotatebox{0}{\includegraphics{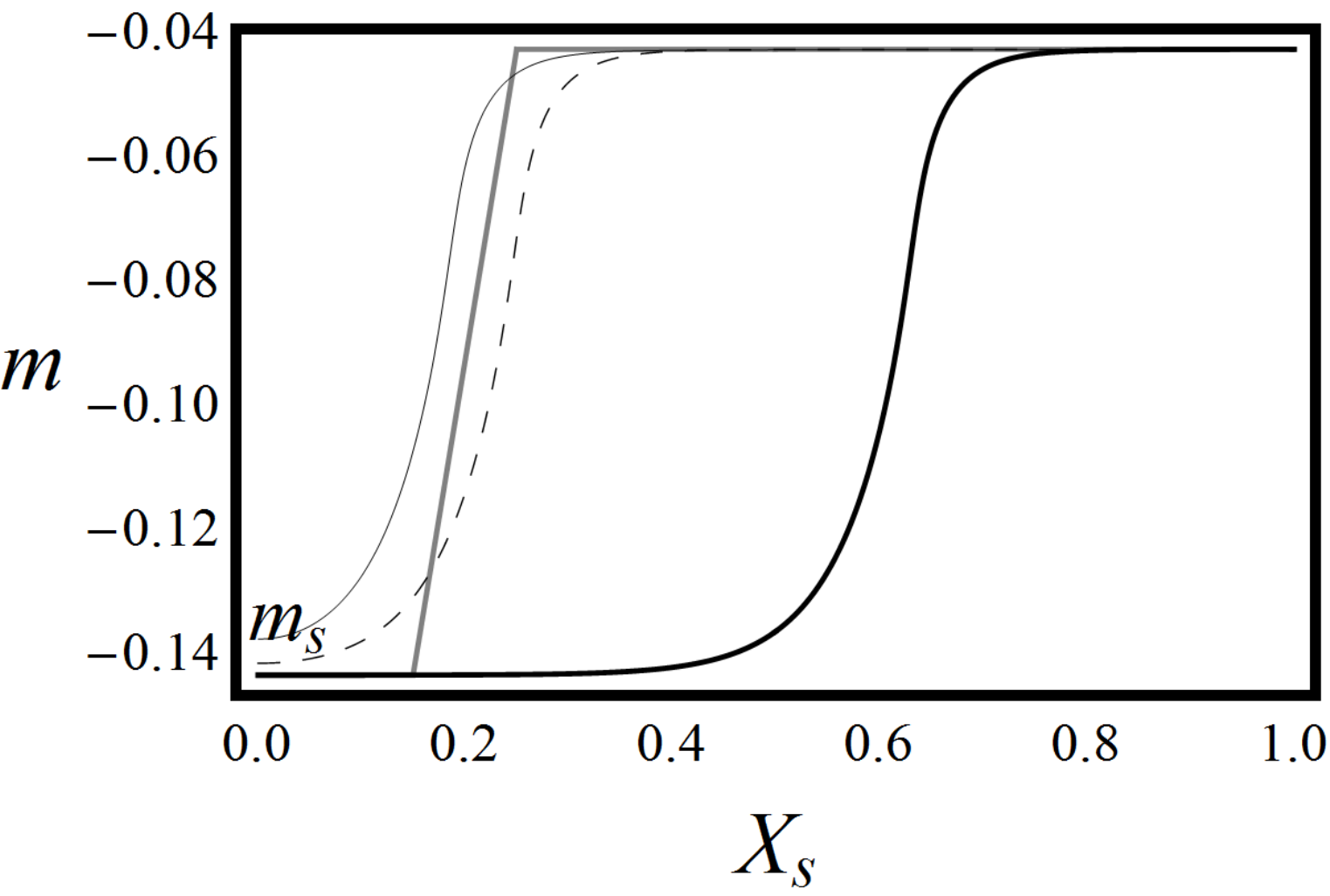}}}
}
\put(310,240)
{
\resizebox{5.5cm}{!}{\rotatebox{0}{\includegraphics{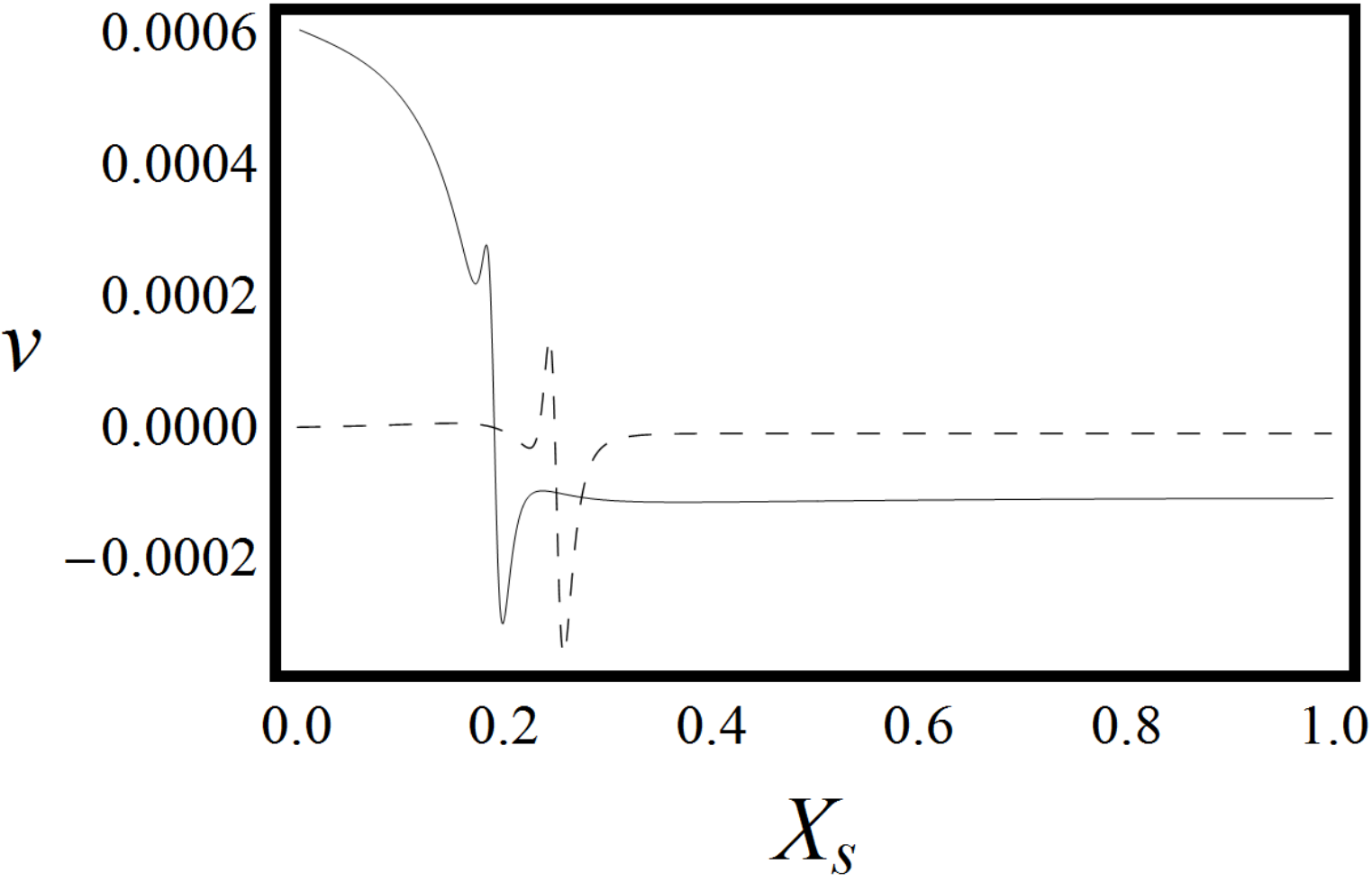}}}
}
\put(-40,120)
{
\resizebox{5.5cm}{!}{\rotatebox{0}{\includegraphics{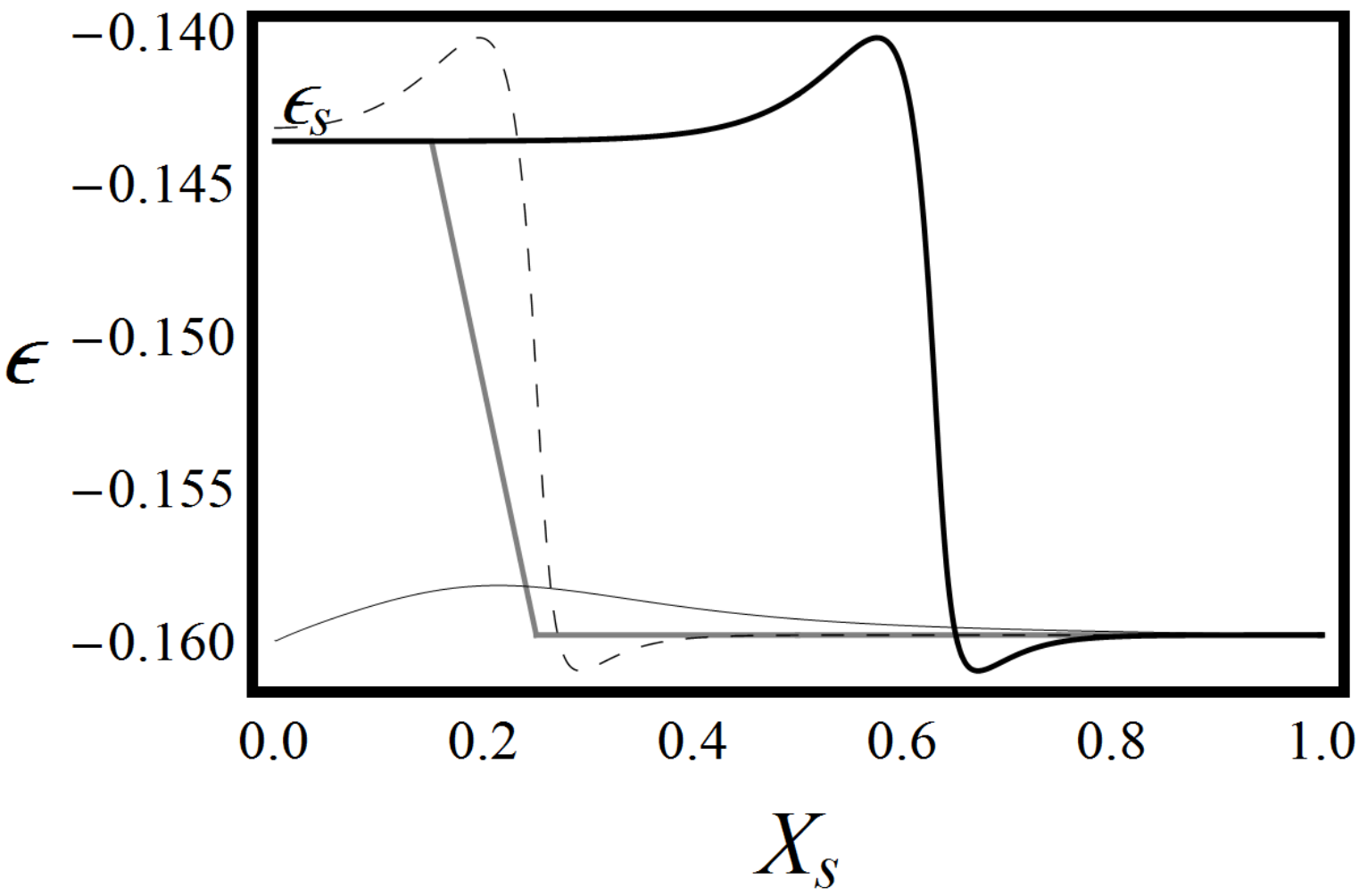}}}
}
\put(135,120)
{
\resizebox{5.5cm}{!}{\rotatebox{0}{\includegraphics{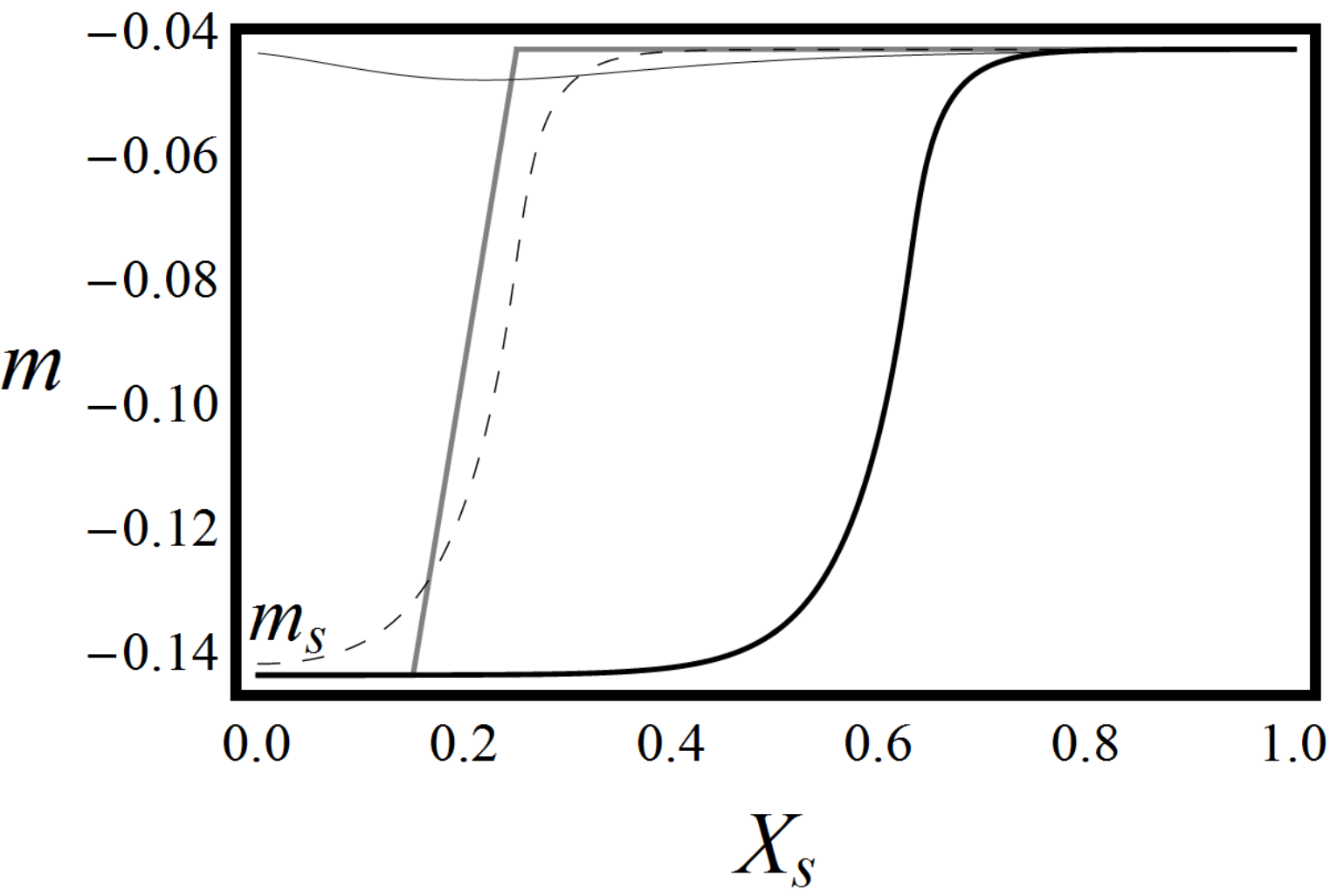}}}
}
\put(310,120)
{
\resizebox{5.5cm}{!}{\rotatebox{0}{\includegraphics{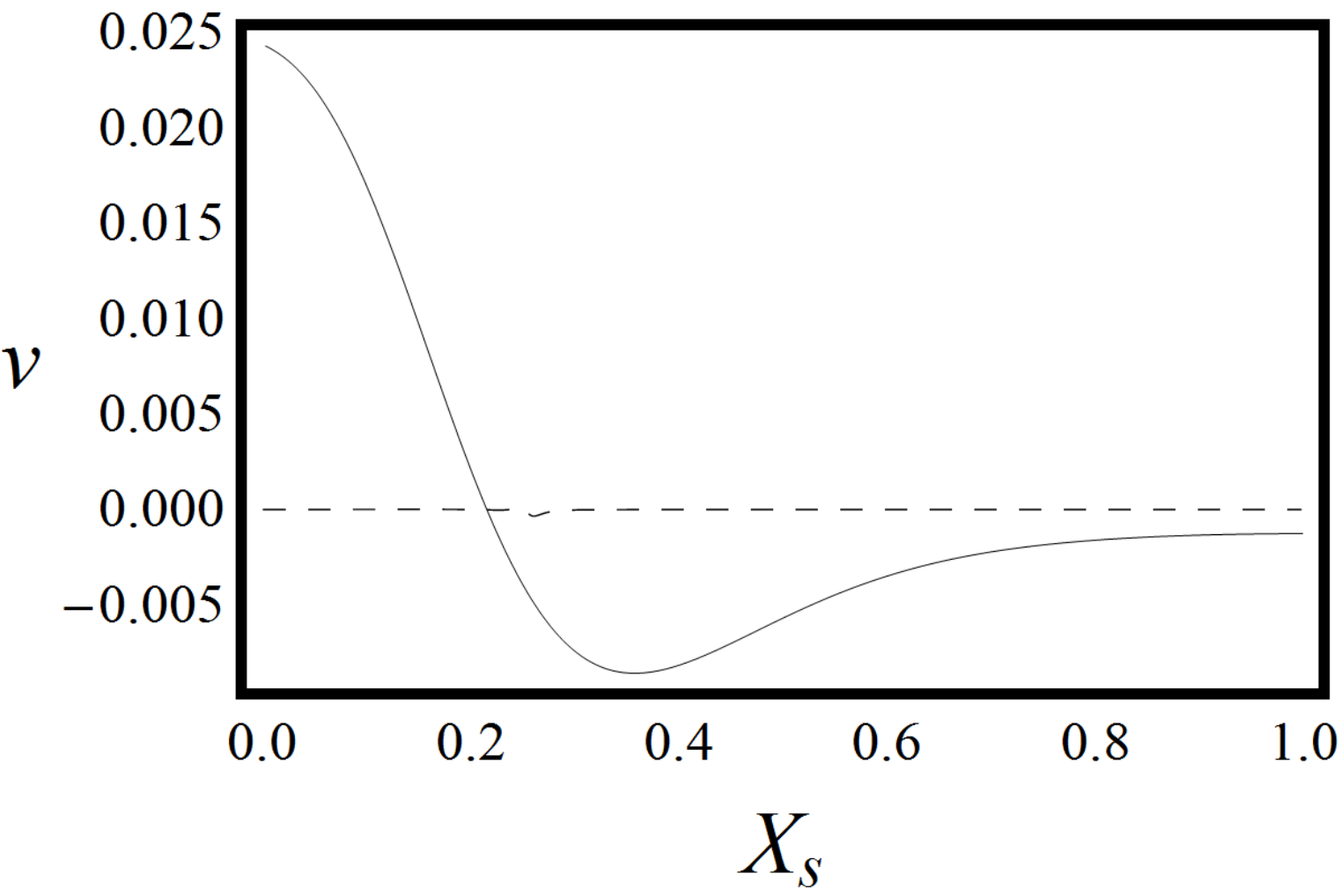}}}
}
\put(-40,0)
{
\resizebox{5.5cm}{!}{\rotatebox{0}{\includegraphics{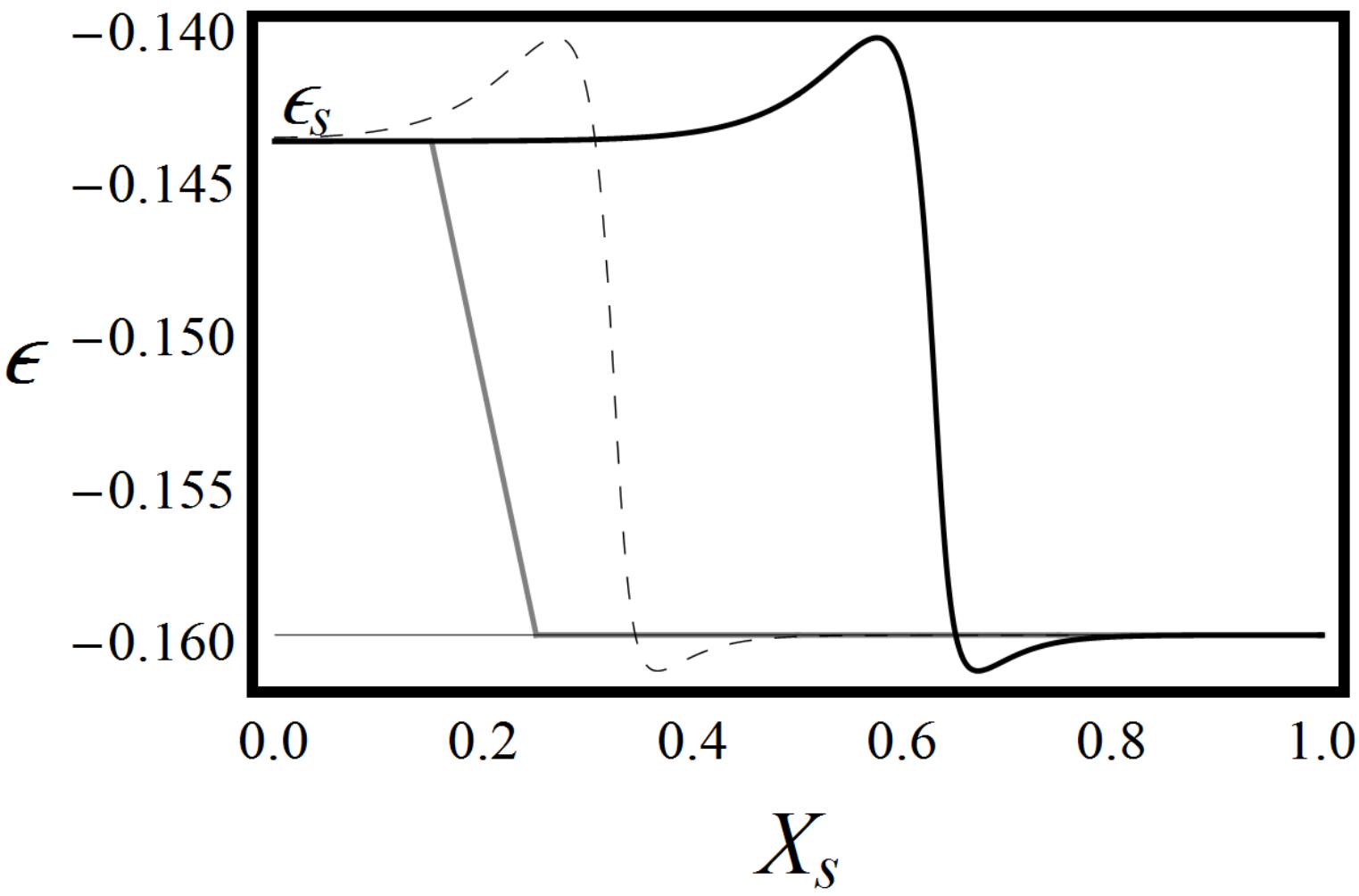}}}
}
\put(135,0)
{
\resizebox{5.5cm}{!}{\rotatebox{0}{\includegraphics{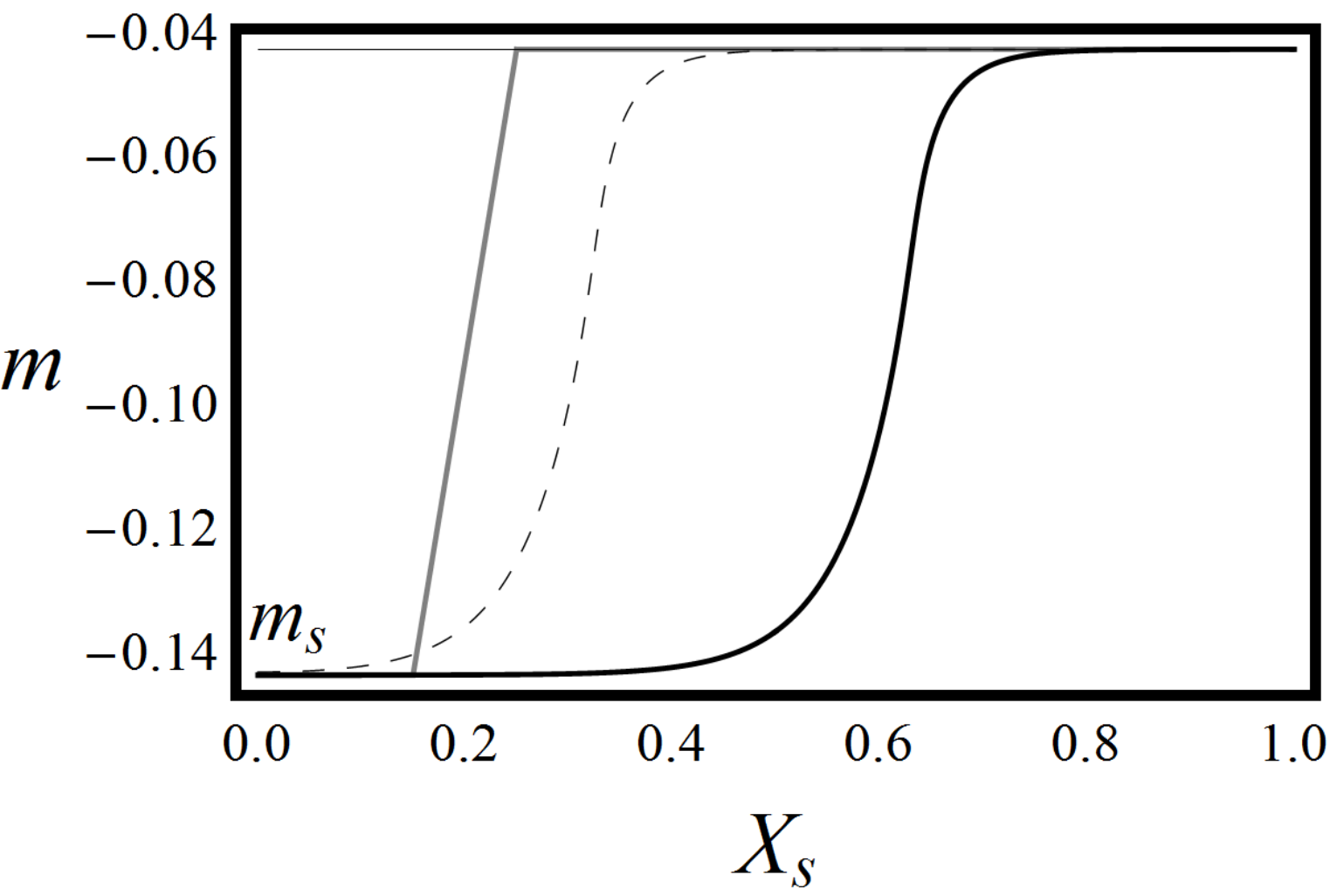}}}
}
\put(310,0)
{
\resizebox{5.5cm}{!}{\rotatebox{0}{\includegraphics{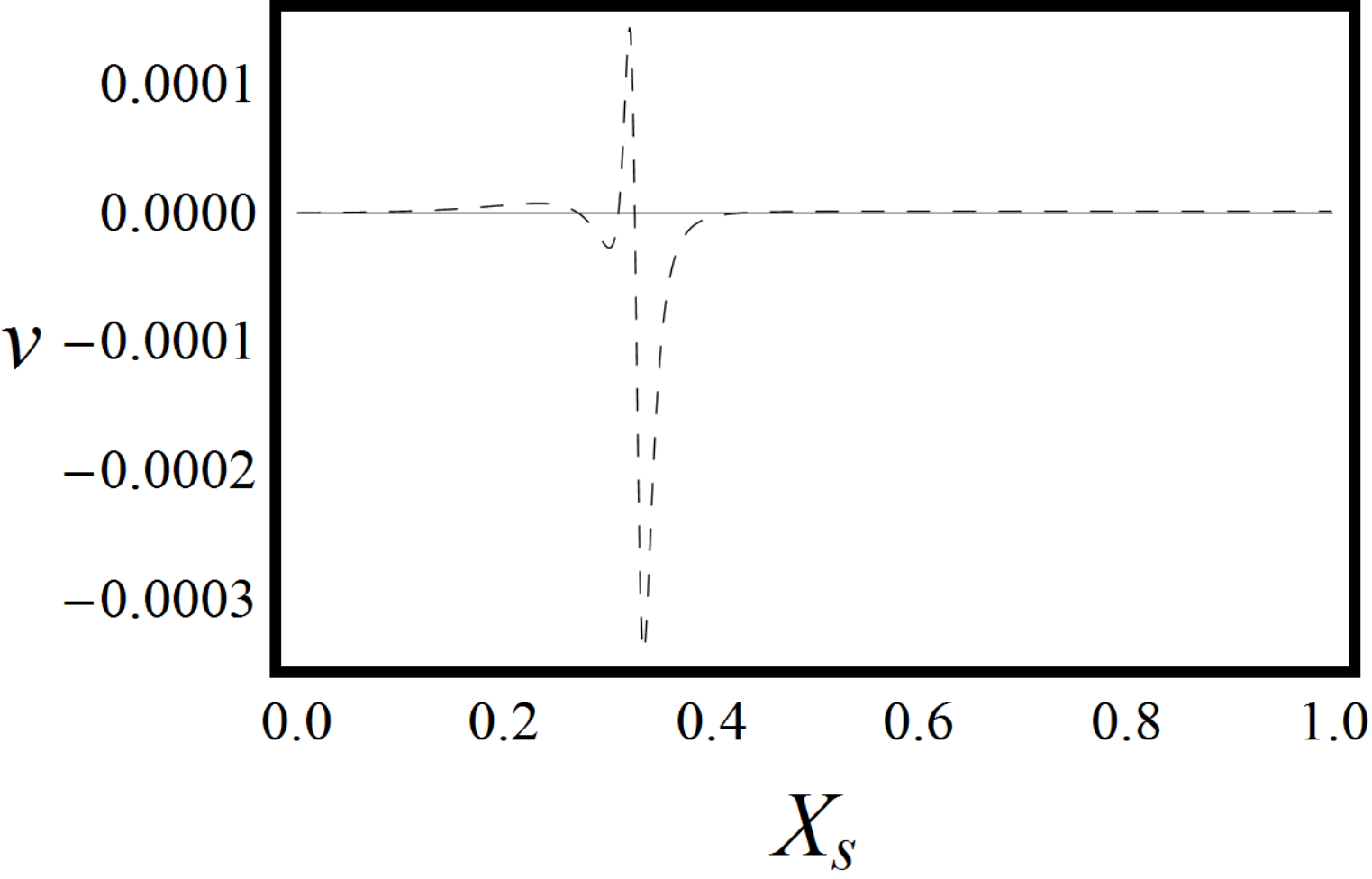}}}
}
\end{picture}
\vskip 1. cm
\centering
\caption{Solutions $\varepsilon(X_s,t)$, $m(X_s)$ and $v(X_s)$ for the zero chemical potential problem (thin black lines) and the one--side impermeable problem (dotted thin lines) obtained by solving 
Problem (\ref{problema-d}) and (\ref{problema-d-osi}) respectively, with the fourth initial condition (gray lines). We used Neumann boundary conditions $m'(0)=\varepsilon'(0)=m'(0)=m'(1)=0$ on the finite interval $[0,1]$, at the coexistence pressure for $a=0.5,\,b=1,\,\alpha=100,\,k_1=k_2=k_3=10^{-3}$. 
Profiles at times $t=1,\,t=21.6,\,t=25,\,t=101$ are in lexicographic order. The solid black lines represent interface--type stationary profiles.}
\label{dinamica4}
\end{figure}

The evolution depicted in Figure \ref{dinamica4} represents the dynamics described by Problems (\ref{problema-d}) and (\ref{problema-d-osi}) with the impermeabilty in $\ell_1$ instead of $\ell_2$. The initial state is a piecewise function with a continuous junction between the fluid--poor $(\varepsilon_s,m_s)$
and the fluid--rich phase $(\varepsilon_f,m_f)$. In this case, like those of Figures \ref{dinamica2} and \ref{dinamica3}, the final state of the evolution is different with the different natural boundary conditions. The evolution with the impermeability condition is very similar to that of the the evolutions studied in \cite{ACS2014}, as can be seen by the final shape of the velocity profile in Figure \ref{dinamica4}. The dynamics without the impermeable wall is just a process of fluid filling inside the medium so that, when the fluid--rich state is reached, the evolution ends and the velocity profile is identically zero.
\\\indent
The reason for this behaviour is that, at the beginning, the medium is almost filled with the fluid and the possibility of flowing trough the left boundary quickly fills the whole bar with water while, in the case of impermeability, a water depression in the neighborhood, remaains as a feature of the whole evolution, thus leading to the formation of the interface, which is the most close solution to the initial state.
\\\indent
The choice of this initial state is that of (numerically) investigating the occurence of different final states (with different boundary conditions) in the presence of a water depression in the neighborhood of the left boundary. 
\\\indent
With this choice we proved that, in the case of the one-side--impermeable problem, if the evolution starts with a water depression, due to the wall, the depression remains during the whole evolution.  
%%figura caso 1

%%figura caso 2
%figura caso 3
%figura caso 4
\clearpage
\section{Conclusions}
\label{s:conclusioni}
Here we studied the dissipative dynamics of a porous material under consolidation. Thanks to a 
second gradient total potential energy, see Section \ref{sec.phi}, we modeled heterogenous elastic strain and fluid density variations for the consolidating porous medium. The overall potential energy, introduced in previous works 
\cite{CIS2010,CIS2013} describes a kind of phase transition between a fluid--poor and a fluid--rich 
phase inside the porous network.

We studied the problem of fluid--segregation in the consolidating dynamics of the
porous medium with pure Darcy dissipation, considering two kinds of evolutions: the
first without an impermeable wall at one of the boundaries (zero--chemical potential
problem, see Subsection \ref{ss:zcp}) and the other one with the impermeable wall (one--side
impermeable problem, see Subsection \ref{s:osi}).  Thanks to  Neumann boundary conditions, thus without prescribing the formation of the interfaces between the two phases, we  found that the impermeable wall higly influences the
motion; in fact, the
dynamics with the impermeable wall ends with the fluid--segregation inside the
porous medium, whereas without the wall the dynamics goes toward the standard
phase solution and the soil consolidates.

The stationary problem is the same for both zero--chemical and one--side--impermeable problems. We were able
to prove that, with Neumann homogenous boundary conditions,
the position of the interface is the same of the  case of Dirichlet essential boundary
conditions. 
\\\\
\textbf{Acknowledgments}

I wish to thank E.N.M. Cirillo and G. Sciarra (Rome) for fruitful discussions.

%\bibliographystyle{plain} % BibTeX style
%\bibliography{biblio}

\end{document}